\documentclass[preprint2,times,tighten]{aastex6}
\pdfoutput=1 
\usepackage{amsmath,amstext}
\usepackage{color,soul}
\usepackage[figure,figure*]{hypcap}
\usepackage{newtxmath} 

\newcommand{\chandra}{{\em Chandra}}
\newcommand{\cygob}{Cygnus\,OB2}

\newcommand{\gaia}{{\em Gaia}}
\newcommand{\lk}{{\mathcal{L}}}
\newcommand{\lkm}{{\mathcal{L}}_{nn}}
\newcommand{\lm}{{\mathcal{L}}_{n}}
\newcommand{\lz}{{\mathcal{L}}_{0}}
\newcommand{\lp}{{\mathcal{L}}_{p}}
\newcommand{\lkp}{{\mathcal{L}}_{pp}}
\newcommand{\dx}{{\delta}}
\newcommand{\sig}{{\sigma}}
\newcommand{\ee}{{\mathcal{E}}}
\newcommand{\Av}{A_{\rm v}}
\newcommand{\us}{u}
\newcommand{\gs}{g}
\newcommand{\rr}{r}
\newcommand{\ii}{i}
\newcommand{\zz}{z}
\newcommand{\rp}{r_I}
\newcommand{\ip}{i_I}
\newcommand{\zp}{z_I}
\newcommand{\gp}{g_I}
\newcommand{\up}{u_I}
\newcommand{\Ha}{{\rm H}_\alpha}
\newcommand{\jj}{J}
\newcommand{\hh}{H}
\newcommand{\kk}{K}
\newcommand{\qs}{Q_{25}}
\newcommand{\qm}{Q_{50}}
\newcommand{\qh}{Q_{75}}
\newcommand{\hrat}{HR}

\newcommand{\Item}{\kern -8pt \item}

\shorttitle{CygOB2 Classification}
\shortauthors{Kashyap et al.}

\begin{document}

\title{Classification of Chandra X-ray Sources in Cygnus\,OB2}
\author{Vinay L.\ Kashyap\altaffilmark{1},
Mario G.\ Guarcello\altaffilmark{2},
Nicholas J.\ Wright\altaffilmark{3},
Jeremy J.\ Drake\altaffilmark{1},
Ettore Flaccomio\altaffilmark{2},
Tom L.\ Aldcroft\altaffilmark{1},
Juan F.\ Albacete Colombo\altaffilmark{4},
Kevin Briggs\altaffilmark{5},
Francesco Damiani\altaffilmark{2},
Janet E.\ Drew\altaffilmark{6},
Eduardo L.\ Martin\altaffilmark{7},
Giusi Micela\altaffilmark{2},
Tim Naylor\altaffilmark{8},
Salvatore Sciortino\altaffilmark{2}
}
\email{vkashyap@cfa.harvard.edu}
\altaffiltext{1}{Harvard-Smithsonian Center for Astrophysics, 60 Garden St.\ Cambridge MA 02138, USA}
\altaffiltext{2}{INAF - Osservatorio Astronomico di Palermo, Piazza del Parlamento 1, I-90134, Palermo, Italy}
\altaffiltext{3}{Astrophysics Group, Keele University, Keele, ST5 5BG, UK}
\altaffiltext{4}{Universidad de Rio Negro, Sede Atl\'antica - CONICET, Viedma CP8500, Argentina}
\altaffiltext{5}{Hamburger Sternwarte, Germany}
\altaffiltext{6}{University of Hertfordshire, Center for Astrophysics Research, Hatfield AL10 9AB, UK}
\altaffiltext{7}{CSIC-INTA, Centro de Astrobiologia, Madrid, Spain}
\altaffiltext{8}{Astrophysics Group, School of Physics, University of Exeter, UK}

\begin{abstract}
{We have devised a predominantly Naive Bayes based method to classify X-ray sources detected by \chandra\ in the \cygob\ association into members, foreground, and background objects. We employ a variety of X-ray, optical, and infrared characteristics to construct likelihoods using training sets defined by well-measured sources. Combinations of optical photometry from Sloan Digital Sky Survey (SDSS; $\rr\ii\zz$) and Isaac Newton Telescope Photometric H$\alpha$ Survey (IPHAS; $\rp\ip\Ha$), infrared  magnitudes from United Kingdom Infrared Telescope Deep Sky Survey and Two-Micron All Sky Survey (UKIDSS and 2MASS; $\jj\hh\kk$), X-ray quantiles and hardness ratios, and estimates of extinction $\Av$ are used to compute the relative probabilities that a given source belongs to one of the classes. Principal Component Analysis is used to isolate the best axes for separating the classes for the photometric data, and Gaussian component separation is used for X-ray hardness and extinction. Errors in the measurements are accounted for by modeling as Gaussians and integrating over likelihoods approximated as quartic polynomials. We evaluate the accuracy of the classification by inspection and reclassify a number of sources based on infrared magnitudes, the presence of disks, and spectral hardness induced by flaring. We also consider systematic errors due to extinction. Of the 7924 X-ray detections, 5501 have a total of 5597 optical/infrared matches, including 78 with multiple counterparts.  We find that $\approx$6100 objects are likely association members, $\approx$1400 are background, and $\approx$500 are foreground objects, with an accuracy of 96\%, 93\%, and 80\% respectively, with an overall classification accuracy of approximately 95\%.
}
\end{abstract}


\section{Introduction}

Nearby star forming regions provide opportunities for studying the characteristics of young stellar objects and the star formation process itself.
Growing realisation that exoplanets are very common in the Galaxy has also provided impetus to explore the sites of planet formation and how this process might be affected by astrophysical environments.
Star forming regions in the solar vicinity, such as those found along the Gould Belt within 500~pc or so
\citep[e.g.][]{Comeron+92},
have proven fruitful resources for exploitation and form much of the observational basis of our current picture of star and planet formation.
However, the Gould Belt represents fairly modest star formation activity, with its clusters typically containing only a few to tens of massive stars of early B or O spectral type.
In order to study truly massive sites of star formation we need to look further afield.

This is the aim of the {\em Chandra} \cygob\ Legacy Survey (Drake et al.\ this issue).
\cygob\ is one of the largest sites of recent star formation in our
Galaxy
\citep{Knodlseder:00, Hanson:03, Wright.Drake:09, Wright+15a},
hosting tens of O stars and hundreds of OB stars and with an estimated stellar mass of $\sim 3 \times 10^4 M_\odot$
\citep[e.g.][]{Massey.Thompson:91,Hanson:03,Drew+08, Wright+10}.
The Survey comprises a mosaic of \chandra/ACIS-I observations covering the central square degree of the \cygob\ Association.
X-ray observations and the X-ray source catalog are described in
{\citet{CygOB2cat.Wright+2014},}
while the survey sensitivity resulting completeness in terms of 
X-ray luminosity and stellar mass are discussed by
\citet{Wright+15b}.

The X-ray survey aims to exploit the comparative X-ray brightness of young low-mass stars in the T~Tauri phase as a means of
distinguishing the true association members from
a plethora of foreground and background objects in the Galactic plane.
{7924} 
X-ray point sources were detected
{\citep{CygOB2cat.Wright+2014}}
and while the majority are expected to be in \cygob\ itself, a significant population of interlopers comprising mostly background Active Galactic Nuclei (AGN) and active late-type stars in the foreground are expected.
In order to be successful, we must solve the problem of separating out these populations and correctly identifying the sources located in the association.
This problem is especially challenging for \cygob, lying as it does behind the Great Cygnus Rift and being subject to significant extinction ($E(B-V) > 1.2)$) that varies considerably across the field
\citep[e.g.][]{Massey.Thompson:91,Hanson:03,Guarcello+15a,Wright+15a}.
\citet{Guarcello+15a}
have correlated the X-ray catalog with new optical and existing infrared (IR) photometric catalogs, such that we have a wealth of multiwavelength data on which to draw.
Here, we exploit this data in order to form a statistical basis for classification of survey objects as \cygob\ members or as foreground or background interlopers.

We describe the structure of our approach in Section~\ref{s:nbc}.
In Section~\ref{s:data}, we describe the various information streams we use in the classification, and in Section~\ref{s:likeli}, the likelihoods we develop based on the data.
We discuss the limitations and features of our method, and present the classifications, in Section~\ref{s:discuss}, and summarize in Section~\ref{s:summary}.
We provide a detailed description of how measurement error bars are incorporated into the analysis in Appendix~\ref{s:errorbars}, and a discussion of the different choices of extinction and its effect on the classification in Appendix~\ref{s:FF}.

\section{Naive Bayes Classification}
\label{s:nbc}

There are several methods available to cluster multi-dimensional datasets: machine learning methods like $k$-means clustering
\citep[e.g.,][]{Stein+15},
gaussian process modeling
\citep[e.g.,][]{Gopalan+15},
neural-network based deep-learning
\citep[see][and references therein]{Benavente+17}
probabilistic methods like multivariate Gaussian clustering
\citep[e.g.,][]{2019MNRAS.485.1085S},
and Naive Bayes
\citep[e.g.,][]{Broos+11}.
We have chosen the Naive Bayes approach since it allows us to incorporate relevant information and account for measurement uncertainties in a straightforward manner.
It allows us to construct and take advantage of scientifically meaningful likelihoods, generated using training sets in a manner similar to that used with neural nets, but with the advantage of their operational mechanism not being hidden.
It is well-suited for {\sl wide} datasets (those with large numbers of variables but with few categories, as here, in contrast to {\sl tall} datasets, which have small numbers of observations but large numbers of categories): the manner of construction naturally circumvents the curse of dimensionality when observations in a given category are expected to be correlated.
The value of the Naive Bayes approach has been previously demonstrated by its application in the MYStIX survey
\citep{Kuhn+13}.
{Note that as in typical machine learning (ML) and neural network (NN) methods, our procedure also relies on setting up and using training sets to define the likelihoods for classification.  However, the likelihoods are set up to be easily interpretable as being physically meaningful, and furthermore, unlike ML and NN, Bayesian methods are not subject to arbitariness in deciding how many iterations to run, or how many layers to include, or which activation functions to use.}

We compute the classification of X-ray sources detected in the \chandra\ \cygob\ field -- whether they are in the foreground, are association members, or in the background -- by computing the likelihood that they belong to each class and choosing the class for which the probability $>0.5$.
In principle, this excludes weak classifications where the sum of the probabilities for two classes is greater than the class with the highest probability, but as a practical matter that situation is never encountered in our analysis.
For a description of the application of Naive Bayes Classification (NBC) to an astronomical survey dataset, see
\citet{Broos+11}.
The method hinges on computing the probability that an object is of a given class given the associated data available for that object.
Formally, if the class is represented by $\theta=$\{foreground, member, background\}, and $D$ are various datasets that range from X-ray fluxes to optical magnitudes, we seek to compute the probability of each of the classes $\theta$ given the data $D$, $p(\theta|D)$.
By Bayes' Theorem,
\begin{equation}
p(\theta|D) \propto p(D|\theta) p(\theta)
\label{e:Bayes}
\end{equation}
where $p(D|\theta)$ are the likelihoods, that is, the probability of observing the data $D$ for a given class $\theta$, and $p(\theta)$ codify our prior belief about the relative fractions of the classes.

We describe our choices of the data, $D$, the forms of the likelihoods, and choices of priors in \S\ref{s:likeli}.

\section{Data}
\label{s:data}

\subsection{X-ray}
\label{s:xraydata}

The X-ray observations that make up the \chandra\ \cygob\ Legacy Survey consist of a grid of $6 \times 6$ {\it Chandra} ACIS-I pointings, offset from each other by half the width of the ACIS-I field of view.
In addition, two previous \chandra\ observations
\citep{AlbaceteColombo+07,Wright.Drake:09}
are included in the data processed.
The data were processed following standard {\it Chandra} data reduction procedures, including source detection, photon extraction, and background subtraction, to generate a catalog of 7924 X-ray sources 
{\citep{CygOB2cat.Wright+2014}.}
An analysis of the completeness of the observations and the resulting catalog is presented by 
\citet{Wright+15b}.

For each detected source, we calculate several measures of spectral shape.
First, we compute spectral quartiles ($\qs$, $\qm$, $\qh$ -- the energies corresponding to the $25^{th}$, $50^{th}$, and $75^{th}$ cumulative percentiles in the spectrum; 
\citep[see][]{Hong+04}.
Based on the combined counts obtained in source and background regions in the Soft ($S:\frac{1}{2}-2$~keV) and Hard ($H:2-7$~keV) passbands, we also compute the fractional hardness ratio ($HR=\frac{H-S}{H+S}$) and X-ray colors ($C=\log{S/H}$) for each source 
\citep{Park+06}.
These measures are used to supplement the optical and IR photometric measurements (see \S\ref{s:oirdata} below) to classify sources.

\subsection{OIR}
\label{s:oirdata}

{


The optical-infrared (OIR) catalog compiled for the \chandra\ \cygob\ Legacy Survey counts 328540 sources across the region of the survey, for which photometry from the following catalogs is available:

\begin{itemize}
\item 65349 sources with photometry in $\rr,\,\ii,\,\zz$ across the central $41^{\prime}\times 41^{\prime}$ region
\citep{Guarcello+12}
from specific observations with the Optical System for Imaging and low Resolution Integrated Spectroscopy (OSIRIS), mounted on the $10.4\,$m Gran Telescopio CANARIAS (GTC) of the Spanish Observatorio del Roque de los Muchachos in La Palma 
\citep{Cepa+00};

\item 24072 sources with photometry in $\rp,\,\ip,\,$H$\alpha$ bands from the second release of the INT Photometric H$\alpha$ Survey
catalog obtained from observations with the Wide Field Camera (WFC) on the $2.5\,$m Isaac Newton Telescope (INT)
(IPHAS,
\citealp{Drew+05,BarentsenFDG2014});

\item 27531 sources from the SDSS catalog (DR8, 
which covers the \chandra~FOV fully;
\citealp{Aihara+11})
with photometry in $\us,\,\gs,\,\rr,\,\ii,\,\zz$ bands;

\item 273473 sources with photometry in the $JHK$ bands from the United Kingdom Infrared Telescope Deep Sky Survey's Galactic Plane Survey (UKIDSS/GPS) catalog 
\citep{Hewett+06,Lucas+08},
from observations taken with the Wide Field Camera (WFCAM, 
\citealp{Casali+07}
on the United Kingdom InfraRed Telescope (UKIRT), and compiled
using
a new photometric procedure 
\citep{King+13}
based on the UKIDSS images 
\citep{Dye+18}; 

\item 43485 sources with photometry in $JHK$ from the Two-Micron All Sky Survey point source catalog (2MASS/PSC) catalog 
(\citealp{Cutri+03});

\item 149381 sources from the Spitzer/IRAC catalog with photometry in the $3.6,\,4.5,\,$ $5.8,\,8.0\,\mu$m and MIPS $24\mu$m bands, from the Spitzer Legacy Survey of the Cygnus~X region Spitzer 
\citep{BeererKHG2010}.

\end{itemize}

These catalogs have been combined into the OIR catalog in 
\citet{Guarcello+13},
adopting a matching procedure that can be divided into three steps.
First, a combined optical catalog was produced by matching the OSIRIS, IPHAS, and SDSS catalogs
pairwise for all combinations.
Second, an infrared catalog was created similarly by matching UKIDSS, 2MASS, and Spitzer data.
Each pair of catalogs were combined by using a close-neighbour method with specific matching radii defined in order to minimize the expected number of spurious coincidences and maximize the matched real pairs
\citep[see][for details]{Guarcello+13}.
In the last step, these two catalogs were merged into an unique OIR catalog.
All the data used here, except those from OSIRIS, are available over the entire area surveyed with \chandra/ACIS-I.
}


\subsection{X-ray/OIR Matching}

The adopted matching procedure between the X-ray and the multi-wavelength OIR catalog resulted in 2433 X-ray sources ($\approx30$\%) with no OIR counterparts 
\citep{Guarcello+15a}.
While we expect most background X-ray source AGNs to indeed have no OIR counterparts due to the large extinctions in this direction, some deeply embedded members of \cygob\ are also likely to have no OIR counterparts.
We discuss their effect on the prior and classification in Sections~\ref{s:priors} and \ref{s:catalog} below, and consider the possible presence of false negatives in more detail in Appendix~\ref{s:nomatch}.

\section{Likelihoods}
\label{s:likeli}

A large variety of measurements are available to use to determine the data vector $D$, and we limit ourselves at the outset to 
\begin{eqnarray}
D=\{&~~~~~~~~& \nonumber \\
    & \rr & \equiv {\rm SDSS}~r~{\rm band~magnitude},  \nonumber \\
    & \ii & \equiv {\rm SDSS}~i~{\rm band~magnitude},  \nonumber \\
    & \zz & \equiv {\rm SDSS}~z~{\rm band~magnitude},  \nonumber \\
    & \Ha & \equiv {\rm IPHAS~H}_{\alpha}~{\rm magnitude},  \nonumber \\
    & \rp & \equiv {\rm IPHAS}~r~{\rm band~magnitude},  \nonumber \\
    & \ip & \equiv {\rm IPHAS}~i~{\rm band~magnitude},  \nonumber \\
    & \jj & \equiv {\rm UKIDSS/2MASS}~J~{\rm band~magnitude},  \nonumber \\
    & \hh & \equiv {\rm UKIDSS/2MASS}~H~{\rm band~magnitude},  \nonumber \\
    & \kk & \equiv {\rm UKIDSS/2MASS}~K~{\rm band~magnitude},  \nonumber \\
    & \qs & \equiv {\rm X-ray~25^{th}~percentile~quantile},  \nonumber \\
    & \qm & \equiv {\rm X-ray~50^{th}~percentile~quantile},  \nonumber \\
    & \qh & \equiv {\rm X-ray~75^{th}~percentile~quantile},  \nonumber \\
    & \hrat & \equiv {\rm X-ray~hardness~ratio}~\frac{H-S}{H+S},  \nonumber \\
    & C & \equiv {\rm X-ray~color}~\log{(S/H)},  \nonumber \\
    & \Av & \equiv {\rm absorption~estimate}  \nonumber \\
\}&& \,.
\label{e:datavector}
\end{eqnarray}
This enables us to use IRAC photometry for verification of the classification and avoids confusion with the analysis of disk-bearing stars (see \S\ref{s:reclassify}).

A general assumption made in this type of analysis is that the likelihoods are independent of each other.
That is, given any two data components, say $D_1,D_2$, their individual likelihoods are independent of the other, i.e., $p(D_1|D_2,\theta) = p(D_1|\theta)$, and $P(D_2|D_1,\theta)=p(D_2|\theta)$.
This is not strictly true, as trends and correlations in the data components do exist and conditional independence is not assured, and classifications made using such systems will not be optimal.
In practice, however, even if full independence is not achieved, the Naive Bayes classifier is highly tolerant of attribute dependences \citep{Domingos.Pazzani.1997}.
Nevertheless, in order to minimize cross-talk between components, we carry out Principal Component Analysis (PCA) on several subgroups of data attributes.
{Several efforts have been made to use PCA-type methods to reduce the complexity of astronomical spectra \citep[see, e.g.,][]{2000MNRAS.317..965H,Hojnacki+2007,2016MNRAS.460..373S,2016MNRAS.461.2044S,2017ApJ...846...59D,2020MNRAS.498.5207W,2022ApJ...926...51P}.  Here, we consider principal components of the attributes }
$\{\mathbf{d}\} \equiv$ $\{\rr,\ii,\zz\}$, $\{\hh,\kk,\jj\}$, $\{\qs,\qm,\qh\}$, $\{\rp,\ip,\Ha\}$ (see Table~\ref{t:pca_project}; $\Av$ and {$C$ (the latter based on $\hrat$)} are used separately and by themselves) and select 7 principal components (PCs) that provide the best discriminatory power (note that in no case does a subgroup contribute more PCs to the classification stream than there are attributes).
PCA has been used often in astronomical analysis, though usually as an empirical classification technique \citep{Hojnacki+2007} or a compression technique \citep{Lee+2011,Xu+2014}.
Here we use it primarily as a scheme to find linear transformations of data subgroups that allows us to efficiently separate the foreground, member, and background sources.

{We create vector spaces
$$\mathbf{d}_{2i} = \left\{ \frac{X_i-\bar{X}}{\sqrt{\sum_i(X_i-\bar{X})^2}}, \frac{Y_i-\bar{Y}}{\sqrt{\sum_i(Y_i-\bar{Y})^2}} \right\}$$
of doublets, or
$$\mathbf{d}_{3i} = \left\{ \frac{X_i-\bar{X}}{\sqrt{\sum_i(X_i-\bar{X})^2}}, \frac{Y_i-\bar{Y}}{\sqrt{\sum_i(Y_i-\bar{Y})^2}}, \frac{Z_i-\bar{Z}}{\sqrt{\sum_i(Z_i-\bar{Z})^2}} \right\}$$
of triplets, where $X,Y,Z$ represent magnitudes or colors for a given object $i$.  By analyzing the correlation matrix, we then obtain principal component projections as} $v_i^{(k)} = \sum_j c_{jk} d_{ji}$, where the summation is over the dimensions of the subspace, carried out separately for each object (see Table~\ref{t:pca_project}).
The components $k$ represent successive projections that account for the largest variances in the data and $\{c_{jk}\}$ represent a rotational transformation that projects the dataset onto a new axis.
{Note that the $c_{jk}$ are not subscripted by the object index $i$, but are dependent on the subsample chosen to compute the PCs.  In the following, we drop the subscript $i$ for the sake of brevity unless its absence is ambiguous.}
The distribution of the projections of the different data points defined by these components are then sifted into separate classes, with the boundaries defined as one-sided Gaussians.
A typical assignment, described here for illustrative purposes {(see, e.g., the middle panel of Figure~\ref{f:riz2pc})}, is one where
the Foreground class is $\propto 1$ for $v^{(k)} < v_\mathrm{F}$, and decreases as the Gaussian $N(v_\mathrm{F},\sigma_\mathrm{F{\rightarrow}M}^2)$ for $v^(k) \ge v_\mathrm{F}$;
the Members class increases as $N(v_\mathrm{M},\sigma_\mathrm{M{\rightarrow}F}^2)$ over the range $[v_\mathrm{F},v_\mathrm{M}]$ and decreases as $N(v_\mathrm{M},\sigma_\mathrm{M{\rightarrow}B}^2)$ over the range $[v_\mathrm{M},v_\mathrm{B}]$;
and the Background class increases as $N(v_\mathrm{B},\sigma_\mathrm{B{\rightarrow}M}^2)$ for $v^{(k)} \le v_\mathrm{B}$ and is $\propto 1$ for $v^{(k)} > v_\mathrm{B}$.
The intervals and the widths are chosen separately for the $k^{\rm th}$ PC based on training set data as described below, and in some cases the directions of the transitions could be reversed.
The sum of the components is normalized to $1$ at each projected value $v^{(k)}$.

We then construct likelihoods empirically as a mixture of these three smooth components representing the foreground, association member, and background classes.
To compute the likelihood for a given object, the data defining it in the vector space of interest are projected onto the component of interest, and the probability of observing those data are defined by the relative values of the likelihood curves for the different classes.
This exercise is repeated for different data streams, generating a series of independent likelihoods of obtaining the data given the class.
This enables us to expand the likelihood factor in Equation~\ref{e:Bayes} as the product of the likelihoods obtained for each of these independent vectors.
The final probability for each class is then the product of these likelihoods for a given class, multiplied by the corresponding prior, and further normalized such that the sum across the classes is $1$.
{The likelihood that an object has the observed data values for a given membership class can then be expanded as}
\begin{eqnarray}
p(D&|&{\rm membership~class}) \nonumber \\
     &\propto&		p(v^{(2)}(\rr,\ii,\zz)|{\rm membership~class}) \nonumber \\
     &\times&	p(v^{(3)}(\rr,\ii,\zz)|{\rm membership~class}) \nonumber \\
     &\times&	p(v^{(2)}(\rp,\ip,\Ha)|{\rm membership~class}) \nonumber \\
     &\times&	p(v^{(2)}(\hh,\kk,\jj)|{\rm membership~class}) \nonumber \\
     &\times&	p(v^{(1)}(\jj,\kk)|{\rm membership~class}) \nonumber \\
     &\times&	p(v^{(2)}(\jj,\kk)|{\rm membership~class}) \nonumber \\
     &\times&	p(v^{(1)}(\qs,\qm,\qh)|{\rm membership~class}) \nonumber \\
     &\times&	p(\hrat|{\rm membership~class}) \nonumber \\
     &\times&	p(\Av|{\rm membership~class}) \,,
\end{eqnarray}
{where a component is used if and only if data are available for that source, and each component is defined as the normalized conjoined Gaussians as described above.  The probability of membership in a given class, given the data, are then computed by multiplying by the prior and normalizing the sum to 1,}
\begin{equation}
p({\rm class}|D) = \frac{ p(D|{\rm class}){\cdot}p({\rm class})}{ \sum_{c=\{F,M,B\}} p({\rm class}=c|D){\cdot}p({\rm class}=c) \,.}
\end{equation}
{As a consequence of this process, each object is normalized separately, and if some objects are missing some part of the data, those missing parts have no effect on the assigned probabilities.}

{The ranges and profiles of the Gaussians are modified using expert domain knowledge to segment the training sample into regions where one classification dominates, and thus generate acceptable classifications for each vector separately.}
We define the locations and widths of the one-sided Gaussians by using subsets of well-measured points, i.e., observations with small error bars, treated as a training sample, from each vector.
Note that this approach tends to pick out the brighter objects, which could result in biases in the likelihoods applied to fainter objects, if the faint population is qualitatively different from the bright population.
{However, we define class boundaries for the training samples by comparing how the selected classes project back into physically meaningful color-magnitude spaces (see, e.g., the right panels of Figures~\ref{f:riz2pc} and \ref{f:riz3pc}).  Since these boundaries are generally insensitive to the magnitudes of the errors in the dataset, the main effect of increased uncertainties is to decrease the contrast between the class boundaries.  Thus, all changes made to the class boundaries in principal components spaces are rooted to the color-magnitude spaces, the influence of using small error subsamples is minimized.}
In the one case where we observe the class boundaries shifting with fainter sources (for X-ray hardness ratios; see Section~\ref{s:xshape}), we use likelihoods that are designed to be less informative.
{Our training sample is also highly diversified, with each subset typically having an overlap of $\approx\frac{1}{3}$ with the remaining set (except for the $(\jj,\kk)$ set being a proper subset of $(\hh,\jj,\kk)$ set).  The number of objects chosen for the training set are in the range of a few$\times$10$^2$, compared to $1620$ unique objects in the union of the training samples.  This variety in the choice of training set population prevents potential biases in any one stream from affecting the overall calculations.}
{Note that we set the bounds independently for each subset, by evaluating the projected distributions in color-magnitude spaces over large scales.  We ignore deviations that may be present at smaller scales in color-magnitude diagrams, and thus classifications derived from a single stream are necessarily crude (this point is illustrated in Sec~\ref{s:whydothis} and Figure~\ref{f:riz2Vriz3}).  We rely on the combination of several streams of data to compute a final classification probability.  This is further supplemented by manual inspection and correction to account for special cases that the broad-scale classification misses (see Sec~\ref{s:reclassify}).}

\begin{deluxetable*}{lcc}
\tablecaption{Projections onto Principal Component axes
\label{t:pca_project}}
\tablehead{
\colhead{Attributes} & \colhead{Component used} & \colhead{Projections$^a$} \\
{\hfil} & \colhead{in Classification} & {\hfil}
}
\startdata
$(\rr,\ii,\zz)$ & 2 & {($+0.1970, +0.0035, -0.2008$)} \\
$(\rr,\ii,\zz)$ & 3 & {($-0.0403, +0.0772, -0.0382$)} \\
$(\rp,\ip,\Ha)$ & 3 & {($-0.0483, -0.0114, +0.0588$)} \\
$(\rp-\ip,\rp-\Ha)$ & 2 & {($-0.3640, +0.3640$)} \\
$(\hh,\kk,\jj)$ & 2 & {($+0.0494, +0.1931, -0.2463$)} \\
$(\jj,\kk)$ & 2 & {($+0.2363, -0.2363$)} \\
$(\jj,\kk)$ & 1$^b$ & {($+0.9717, +0.9717$)} \\
$(\qs,\qm,\qh)$ & 1 & {($+0.9433, +0.9964, +0.9294$)} \\
\hline
\multicolumn{3}{l}{{$a$: The corresponding eigenvalues are the summed squares of the projections.}} \\
\multicolumn{3}{l}{$b$: Used only if likelihood for foreground object $>$ likelihood derived from PC2$[\jj,\kk]$}
\enddata
\end{deluxetable*}

\subsection{{\rm a priori} Expectations}
\label{s:priors}

Bayesian analysis requires that priors be defined in order to convert the likelihoods, which are probabilities defined as functions of the data, to posterior probabilities, which are probabilities defined for the parameters or classes of interest.
In the absence of any prior information, a flat distribution is the best option, which in our case corresponds to $p({\rm foreground})=p({\rm member})=p({\rm background})=\frac{1}{3}$.
{This is the choice used in order to demonstrate the effect of the adopted likelihood function for each of the component streams (Figures~\ref{f:riz2pc}-\ref{f:riz2Vriz3} and \ref{f:JK1pc}-\ref{f:q1231pc}).  While this choice is appropriate to illustrate how the likelihood maps to variables of interest, for the combined analysis, we can make a more informed choice.  We define informative priors}
by estimating the number of X-ray sources that may be obtained from background quasars and from foreground Galactic sources.

We estimate the number of AGNs expected in the X-ray sample by assuming that their numbers are distributed as a broken power-law with indices $\beta_1=1.34$ for $f_{\rm X}\leq8.1\times10^{-15}$~ergs~cm$^{-2}$~s$^{-1}$, and $\beta_2=2.23$ at higher $f_{\rm X}$ (see Equation 5 of 
\citealt{Lehmer+12}).
Further assuming a nominal power-law spectrum with index $\Gamma=1.8$ to determine a flux-to-counts conversion factor, we find that for the specific exposure map of the \chandra\ \cygob\ survey, {${\approx}$1200-1800 sources would be present with net counts $>5$ for $N_{\rm H}=1-4\times10^{22}$~cm$^{-2}$.}
The sensitivity  of the survey varies across the field, as does the {diffuse background \citep{CygOB2DiffuseBkg.Facundo+2018} and the} absorbing column density, with lower detection thresholds or lower $N_{\rm H}$ potentially yielding 
more
background X-ray sources.
{The variation in $N_{\rm H}$ over the field can be evaluated by considering the range of $\Av$ estimated for the various objects in our dataset (see Sec~\ref{s:extinction}).  The $\Av$ ranges from $\approx$0.1 to $>10$, suggesting that $N_{\rm H}$ varies from $\approx 10^{20}$~cm$^{-2}$ to a few$\times 10^{22}$~cm$^{-2}$ \citep[cf.][]{Predehl.Schmitt.1995}.}

We estimate the approximate number of foreground stars using a dynamical model of the Galaxy \citep[TRILEGAL v1.6;][]{TRILEGAL2012},\footnote{\url{http://stev.oapd.inaf.it/cgi-bin/trilegal}} which produces a representation of the stellar population along the line of sight towards \cygob\ assuming an exponential thin disk, a squared hyperbolic secant thick disk, an oblate spheroid halo, and a triaxial bulge.
The number of expected X-ray detections varies by over an order of magnitude for different assumptions about the X-ray luminosity function of the field stars and the local limiting sensitivity.
For assumed fixed $L_{\rm X}=10^{27,28,29}$~ergs~s$^{-1}$, for a limiting sensitivity of $10^{-15}$~ergs~s$^{-1}$~cm$^{-2}$, we expect ${\approx}30,800,10000$ X-ray sources respectively, ranging from there being almost no foreground sources, to accounting for essentially all the detected sources.
It is plausible, however, for realistic luminosity functions with median $L_{\rm X}{\sim}10^{28}$~ergs~s$^{-1}$, to produce 500-1500 X-ray detections.

Thus, it is reasonable to assume that of the $\approx$8000
{sources considered,} 
about 2500 should be some combination of foreground and background sources.
We therefore adopt prior probabilities on the classes $$p({\rm foreground})=p({\rm background})=0.15\,,$$ that is, if no information were available, the probability that an X-ray source in the \cygob\ field of view is a member of the association is $0.7$.
The specific values of the prior are important only when the data are not informative, and when they are, the posterior estimates are driven by the likelihoods.
We also point out that it is theoretically unjustifiable to change the priors after the analysis, e.g., by iteratively adjusting them until the posterior counts match the adopted fractions.\footnote{As it happens, we find that (see Table~\ref{t:posthoc}) the relative fractions of the foreground, member, and background objects to be $\approx 6\%, 77\%, 17\%$ respectively.}
Nevertheless, we have explored the sensitivity of the class assignments to the prior by considering how many foreground objects are ``lost'' when $p({\rm foreground})=0.05$ and how many background objects are ``gained'' when $=p({\rm background})=0.2$, commensurate with the fractions that are found upon carrying out the classification procedure (see Section~\ref{s:classify} below).
We find that the changes are small: in the former case the number of foreground objects decreases by $\approx20$ ($3$\%) and in the latter case, the number of background objects increase by $\approx100$ ($5$\%), similar to the magnitude of the systematic uncertainties present in the process (see Section~\ref{s:reclassify}).

The X-ray properties of sources matched with the OIR catalog differs substantially from those with no matches 
\citep{Guarcello+15a}.
In particular, the sources with no matches exhibit a bimodal distribution in $\qm$, with nearly half of the population exhibiting $\qm>3$~keV.
It is thus reasonable to consider whether different values of $=p({\rm background})$ should be used for the samples of matched vs.\ unmatched sources.
However, as noted above, the precise values of the priors are not a significant factor in the classification, so for simplicity of calculation we maintain the same prior for the whole sample.
Nevertheless, because sources with no OIR counterparts are classified entirely through their X-ray properties alone, we flag them in the final catalog (see Section~\ref{s:catalog}) as such, and also indicate how robust that classification is.

For each object, the likelihood that it belongs to a given class is computed independently for each data stream (the PCs in Table~\ref{t:pca_project}, and the mixture components of $\Av$ and $\hrat$), while accounting for measurement errors (see Appendix~\ref{s:errorbars}).
The product of these likelihoods and the priors are then computed for each class, and renormalized such that the sum adds up to $1$.
The probability values span a continuum between $[0,1]$, but for the sake of specificity, we assign a specific class to each object as that which has the highest of the three probabilities.
These assigned classes are then reviewed and those that are clearly misclassified are reassigned (see \S\ref{s:reclassify}).

\subsection{$\rr,\ii,\zz$ \label{s:riz}}

\begin{figure*}[htb!]
\includegraphics[width=2.5in]{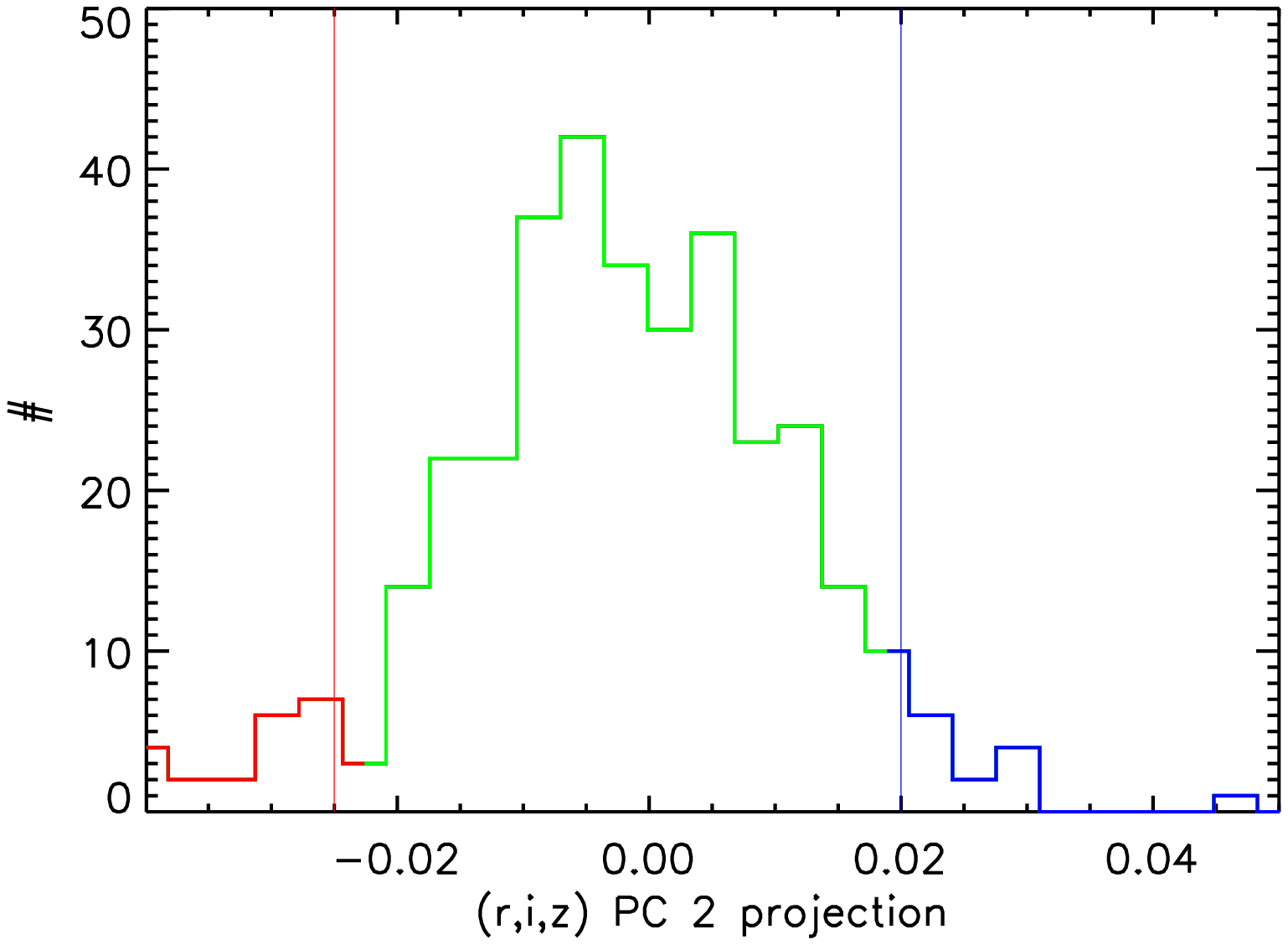}
\includegraphics[width=2.5in]{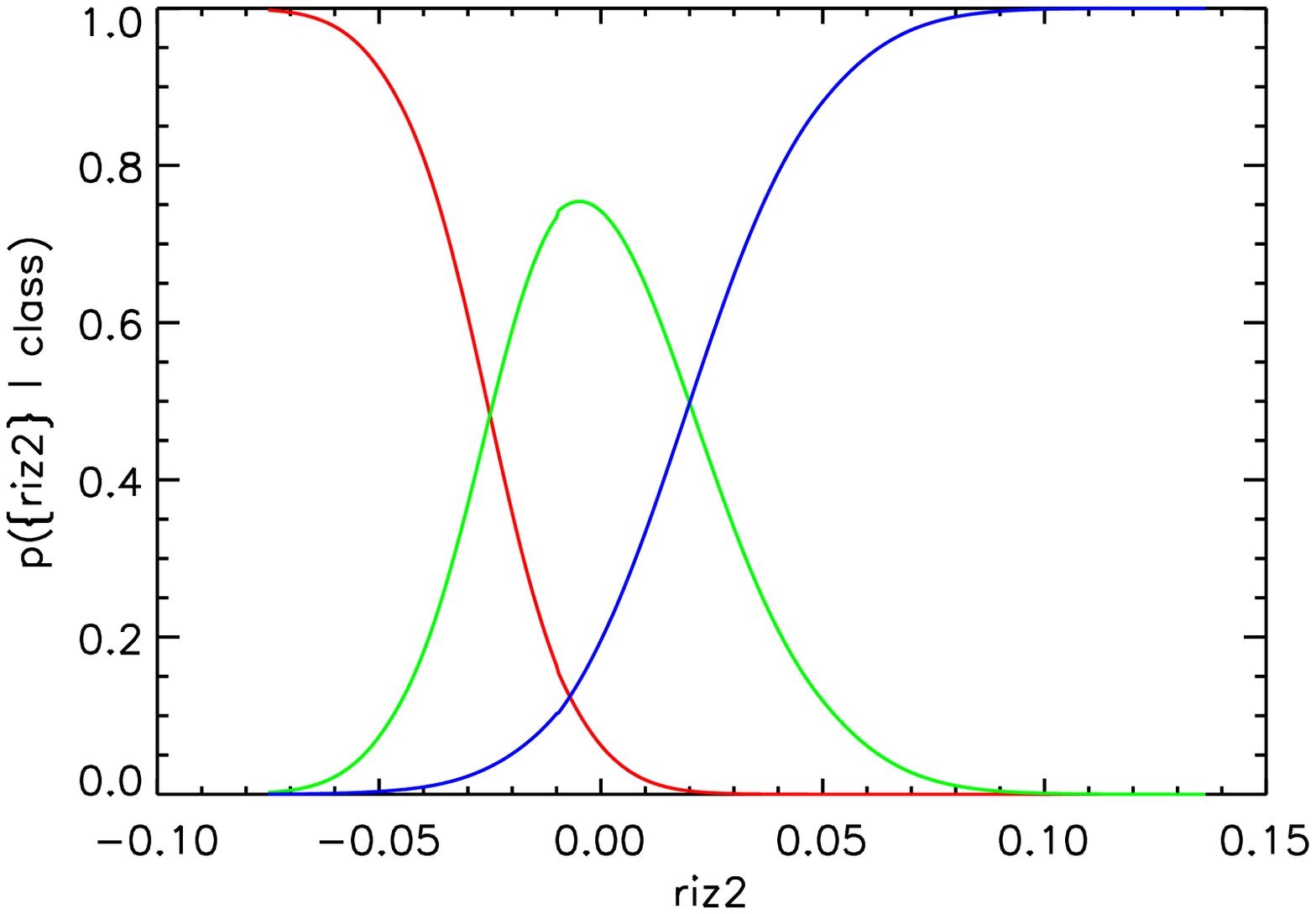}
\includegraphics[width=2.5in]{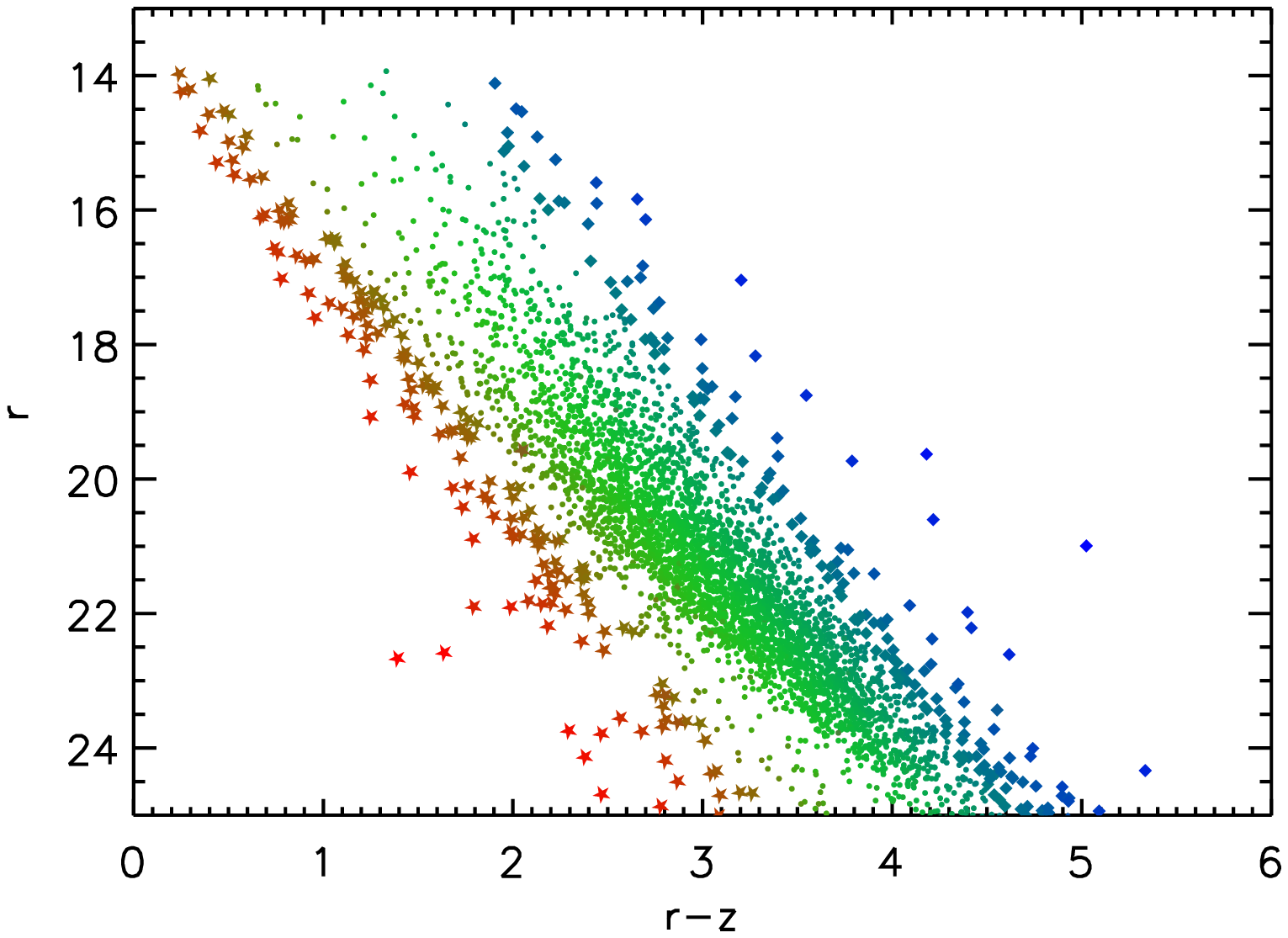}
\caption{
Demonstrating the process of generating the likelihoods of foreground, association member, and background classification using PCs for $\{\rr,\ii,\zz\}$.
A subset with well-measured magnitudes are used as a training set, and the histogram of the projections onto the $2^{nd}$ PC is shown at {\sl left} (see Table~\ref{t:pca_project}), with vertical lines denoting the approximate separation between the different classes, along with the likelihoods constructed from the normalized profiles ({\sl middle}).
{A nominal classification to illustrate these likelihoods, carried out using non-informative priors are also shown as extrapolated to the full dataset ({\sl right}). Each source is color-coded as an rgb-tuple with the relative probability of it belonging to the foreground (red asterisks), cluster (green circles), or background (blue diamonds) class.}
}
\label{f:riz2pc}
\vskip 0.5in
\end{figure*}

\begin{figure*}[htb!]
\includegraphics[width=2.5in]{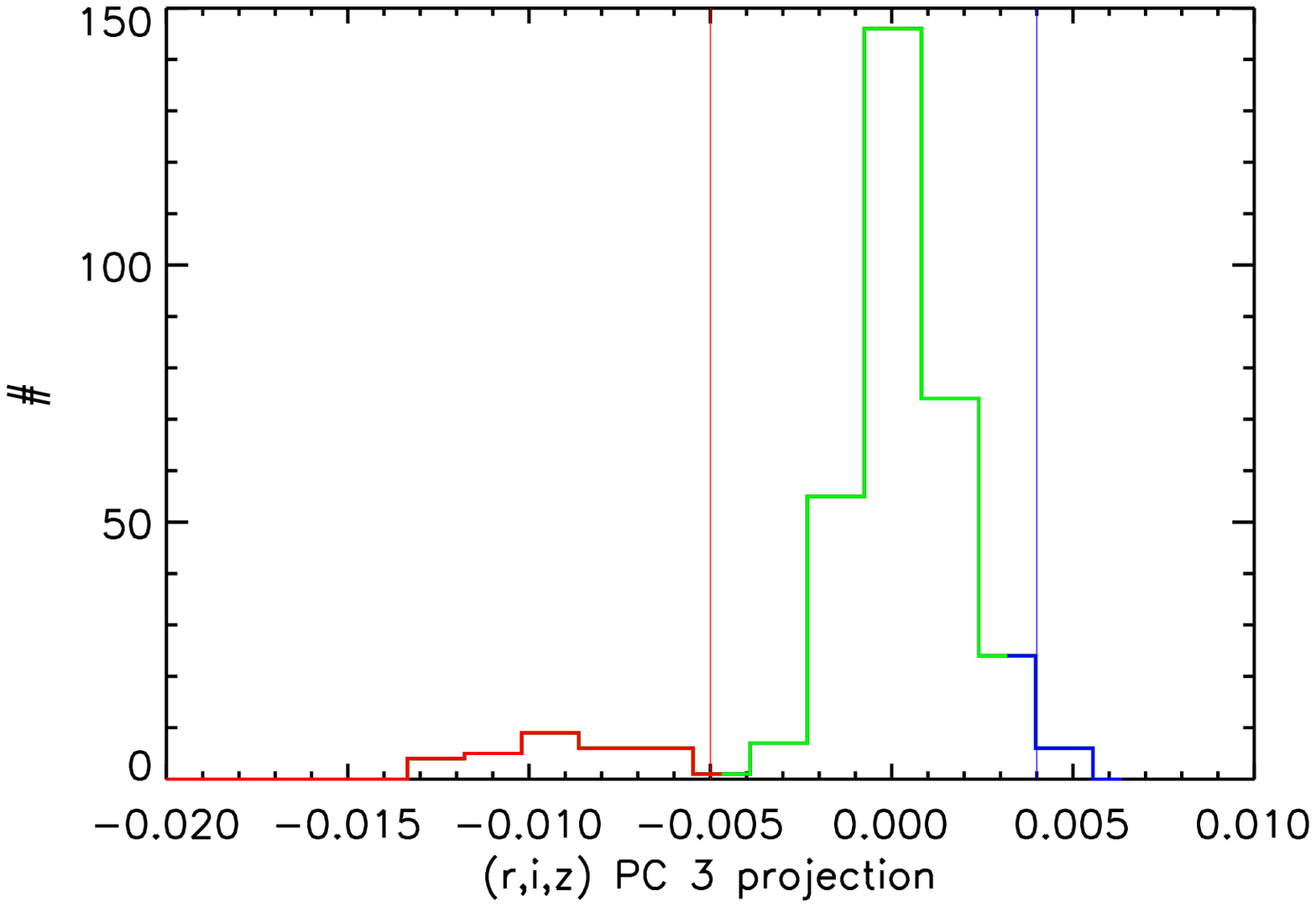}
\includegraphics[width=2.5in]{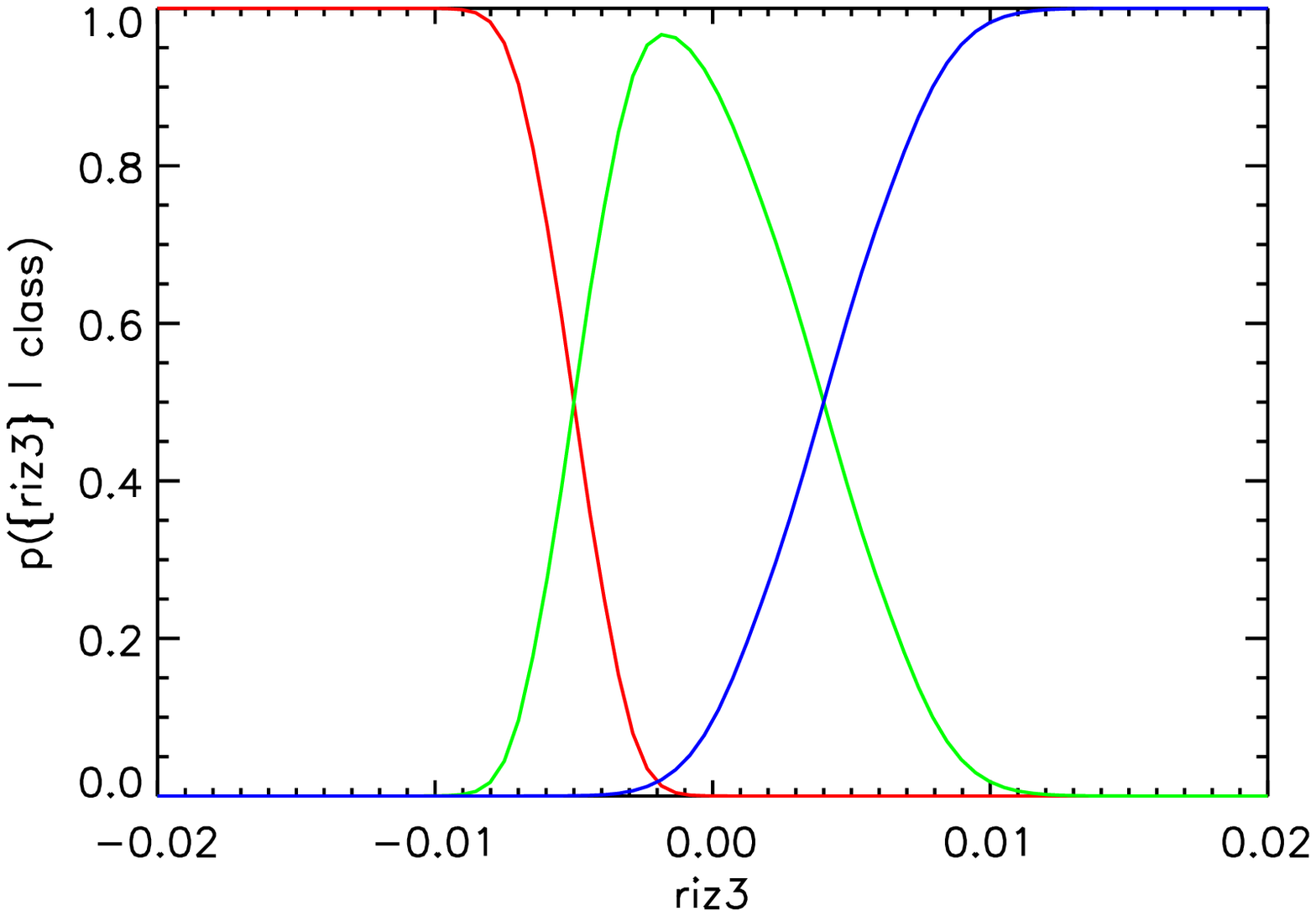}
\includegraphics[width=2.5in]{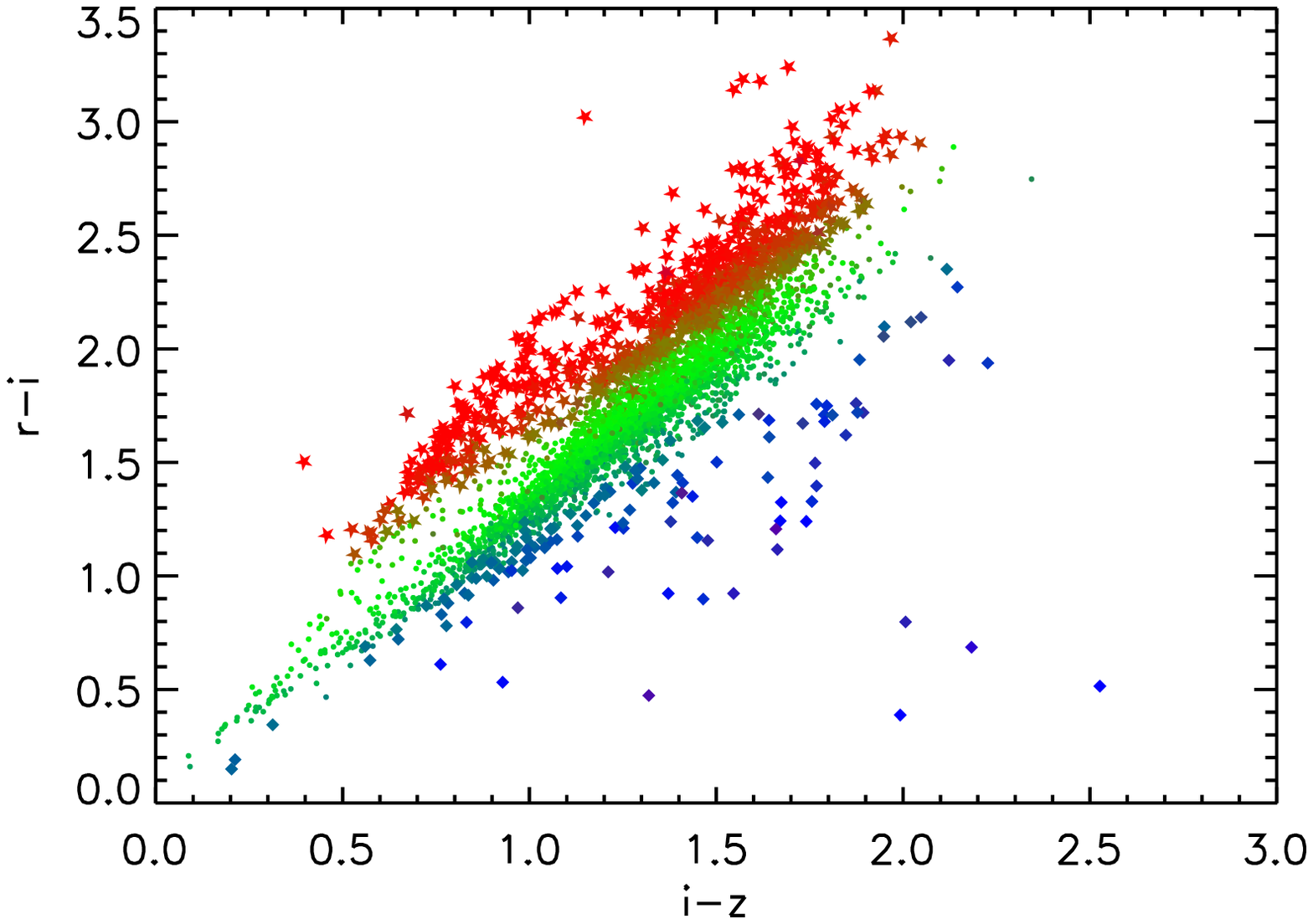}
\caption{
As Figure ~\ref{f:riz2pc}, for the $3^{rd}$ Principal Component of $\{\rr,\ii,\zz\}$.
}
\label{f:riz3pc}
\vskip 0.5in
\end{figure*}

The usual way to sift sources into different classes is to display them on color-color diagrams and identify regions where there is a higher propensity for members of one class to appear.
When the $\{\rr,\ii,\zz\}$ triplet is available, for example, foreground objects stand out along the leftward edge in the $\rr$~vs.~$(\rr-\zz)$ diagram, and along the upper edge of the $(\rr-\ii)$~vs.~$(\ii-\zz)$ diagram.
We extract this information using the $2^{nd}$ and $3^{rd}$ PCs of a subset of $\{\rr,\ii,\zz\}$ data points.
Notice that we do not use the $1^{st}$ component, even though it accounts for the largest fraction of the variance in the dataset, as it is not informative for the purpose of separating the different classes.
We apply similar judgements for other subgroups, and exclude all cases where the projections of the classes overlap significantly or are not easily modeled (e.g., $\frac{f_{\rm X}}{f_{\rm opt}}$).
For the training sample, we compile a list of 200 sources that have the smallest error bars in each band, which results in a total of 348 unique sources\footnote{We employ the same method to compile training samples in Sections~\ref{s:JHK},\ref{s:riHa}, and \ref{s:xshape} where PCA is used for likelihood generation.}.
The projected components are shown as a histogram in the left panels of Figures~\ref{f:riz2pc} and \ref{f:riz3pc}, colored according to which range is preferentially dominated by members of which class (red denotes foreground, green denotes association members, and blue denotes background).
Note that at this stage the sources are sifted into different classifications by construction, that is, according to our expectation of how they are likely to be distributed amongst the different classes.
The boundary between the classes is not sharp, and the transition from one dominant class to another is assumed to occur smoothly.
The distributions of the number of objects, projected along the PC, for each class, is assumed to be approximated as Gaussians and smoothly varying.
They are then normalized such that each component has a maximum of $1$, and are further renormalized at each point along the PC such that their sum adds up to $1$.
The likelihoods thus obtained are shown in the middle panels with the same color scheme.
These likelihoods are then applied to the full $\{\rr,\ii,\zz\}$ dataset (that is, not just the training sample), incorporating uncertainties (see Appendix~\ref{s:errorbars}).
The resulting classifications (using flat priors set to $\frac{1}{3}$ for each class), constructed using {\sl only this data stream and none others}, are shown in the right panels, with color hues ranging from red (denoting foreground) to green (denoting members) to blue (denoting background).
We discuss the necessity of using multiple components in \S\ref{s:whydothis}.

\subsection{$\jj,\hh,\kk$}
\label{s:JHK}

For the combination of $\{\jj,\hh,\kk\}$, we first consider the color-magnitude diagram $\jj$~vs.~$(\jj-\kk)$.
For this, we carry out PCA on a sample of 287 objects with the best-measured $\{\jj,\kk\}$.
Ultimately, we primarily use the $2^{nd}$ component, but use the higher of the computed foreground likelihoods for foreground objects, since a bright $\jj$ strongly suggests that the object is in the foreground.

We also carry out a PCA on a subset of 332 of the best-measured triples $\{\jj,\hh,\kk\}$, using the $2^{nd}$ component that separates foreground and background objects in $(\jj-\hh)$~vs.~$(\hh-\kk)$ diagrams (Figure~\ref{f:HJK2pc}).

\subsection{$\rp,\ip,\Ha$}
\label{s:riHa}

We consider the triple $\{\rp,\ip,\Ha\}$ as a group since these are measured with IPHAS.
We compute PCs for both the triple and for the paired colors $(\rp-\Ha)$~vs.~$(\rp-\ip)$ using a training sample of 212 well-measured objects.
The $3^{rd}$ component of the former (Figure~\ref{f:riHa3pc}) and the $2^{nd}$ component of the latter (Figure~\ref{f:riHa2pc}) sift the region into similar groupings, but with slight differences that indicate a complex projection of components.
We use both components.
We note that $(\rp,\rr)$ and ($\ip,\ii)$ are strongly correlated (Pearson's $\rho\approx1$), but display complex behavior in their errors (e.g., the ratio of the errors is correlated with $\rr-\ii$ with $\rho\approx0.5$), and thus are able to provide additional discriminatory power to the PCs used in Section~\ref{s:riz}.

\subsection{X-ray Spectral Shape}
\label{s:xshape}

ACIS spectra typically have $>800$ usable pulse-height channels, but their shapes are effectively characterized by counts in a small number, $\sim$5 of passbands {\citep[similar to the number of free parameters in thermal spectra usually used while fitting the X-ray spectra of weak sources, e.g.,][]{Flaccomio+18}}.
We use three measures of quantiles (defining the energies that include $25\%$, $50\%$, and $75\%$ of all the observed counts within the source regions) and two measures of hardness ratios (extremes of {fractional hardness} $HR=\frac{H-S}{H+S}$ to motivate a model that is applied to {color} $C=\ln{S/H}$) as proxies to characterize the dispersion that describes X-ray spectral shapes.

We use the $1^{st}$ PC of the analysis of the best-measured sample of 357 objects of the triple $\{\qs,\qm,\qh\}$, as it groups typically unabsorbed thermally emitting foreground objects as having a soft spectrum, and background objects which are likely to be power-law sources and heavily absobed as having a hard spectrum {(Figure~\ref{f:q1231pc})}.
For detailed discussions of the spectra, see \citet{Flaccomio+18}.

\begin{figure*}[htb!]
\includegraphics[width=2.5in]{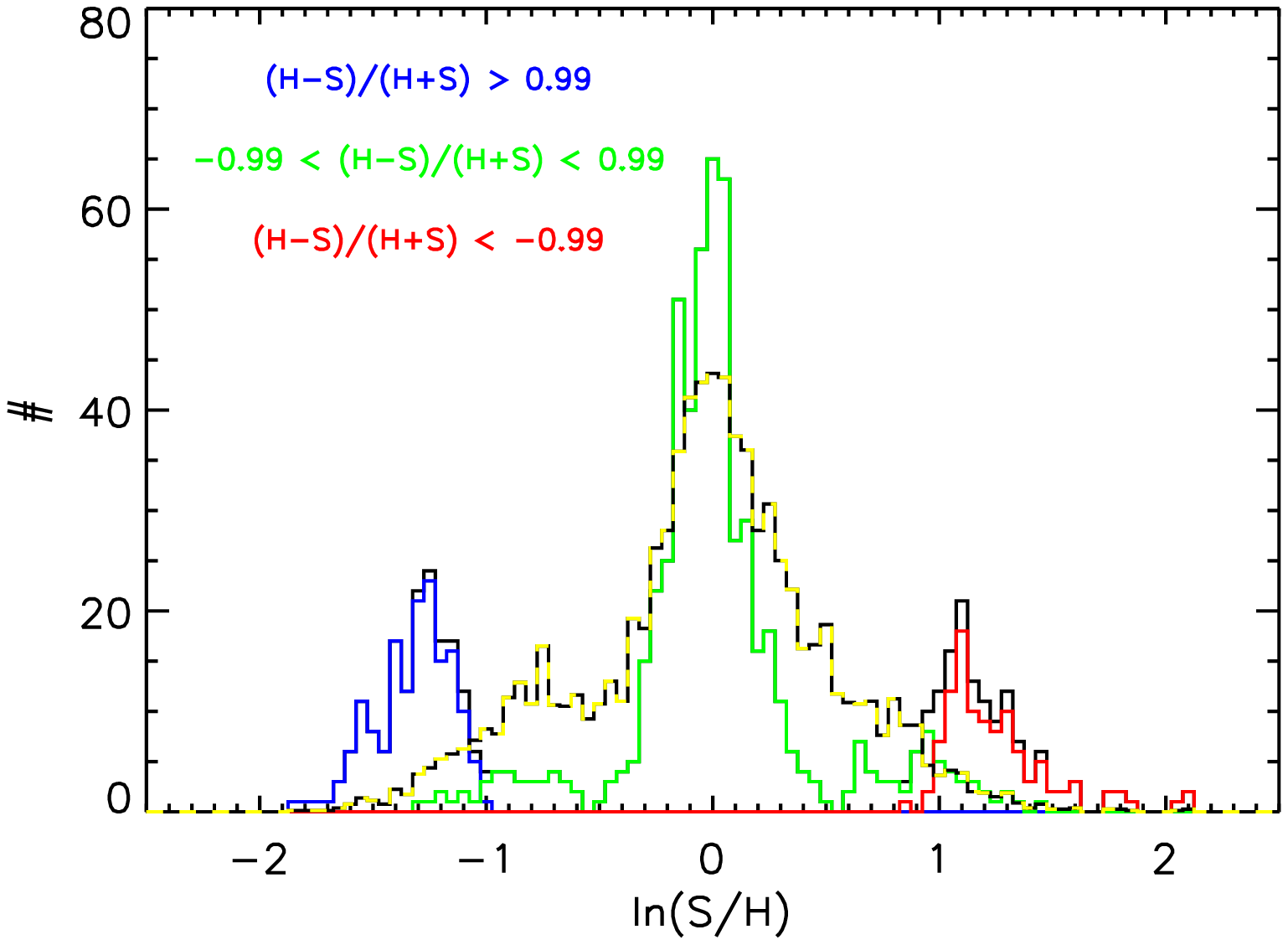}
\includegraphics[width=2.5in]{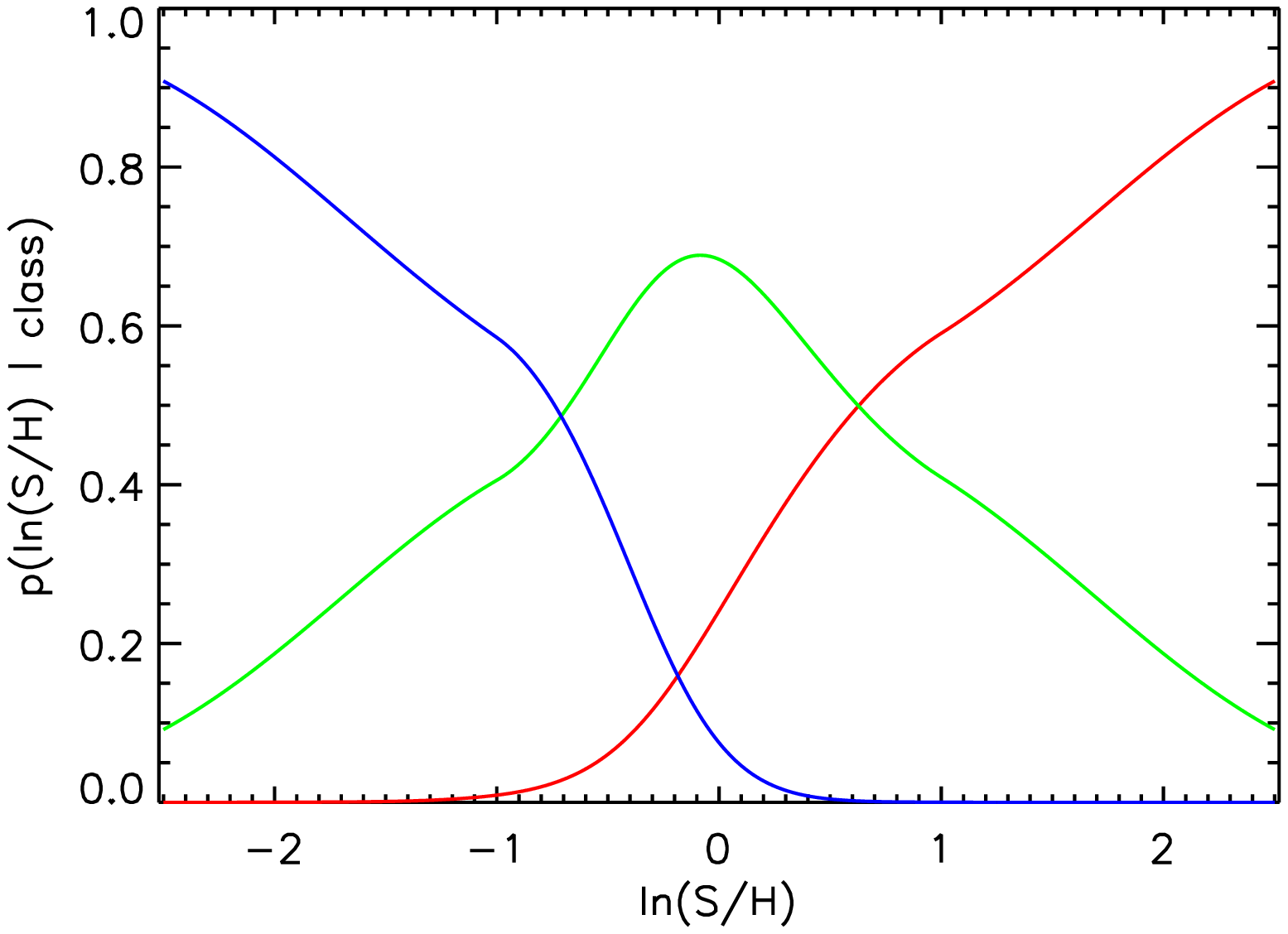}
\includegraphics[width=2.5in]{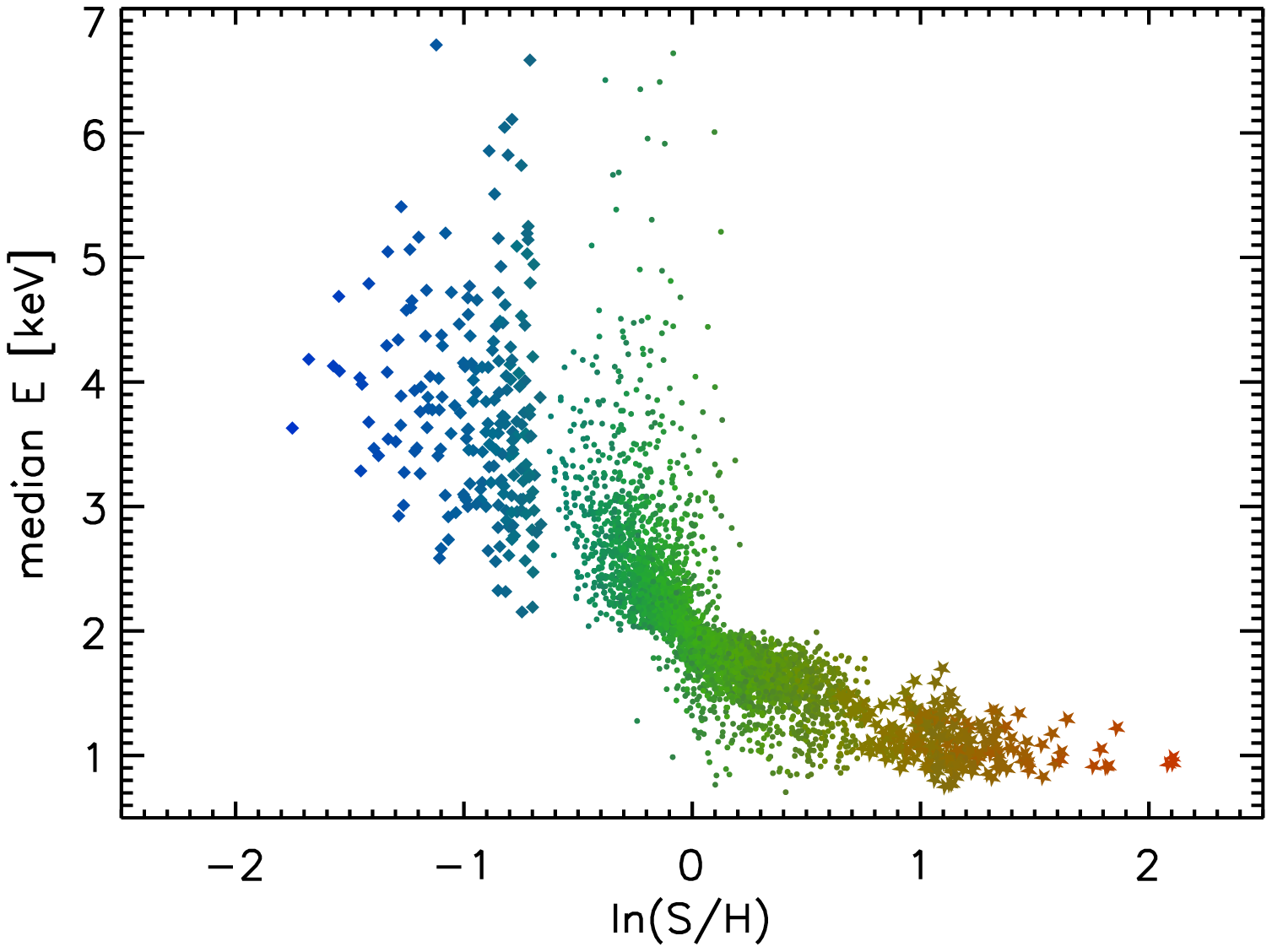}
\caption{
Likelihood generation for X-ray colors.
{\sl Left:} Histogram of X-ray colors filtered on hardness ratios for a well-measured subsample, with the histogram for the full sample shown as the dashed black-yellow line normalized to the same number of objects and overplotted.
{\sl Middle:} The likelihoods generated by square-root transformation of the Gaussian components, with the tails modified to be linearly decreasing.
{\sl Right:} Class assignments with points color coded as rgb-tuples as in Figure~\ref{f:riz2pc}.
}
\label{f:xcol}
\vskip 0.5in
\end{figure*}

Though there is overlap in the information codified by the quantiles and hardness ratios {(and $\Av$; see Sec~\ref{s:extinction} below)}, comparisons of typical model grids show that they encode the spectral shape information differently.
Furthermore, there is a dearth of background objects in the training sample made using the quantile data (see, e.g., Figure~\ref{f:q1231pc}), which makes it necessary to include a broader measure.
We thus complement the separation obtained from the quantiles with color $C=\ln{S/H}$, computed using BEHR \citep[Bayesian Estimation of Hardness Ratios][]{Park+06}.
Because $C$ is one-dimensional, and the counts in individual bands are sensitive to normalization, we cannot use PCA to determine the optimal axes as above.
{Instead, we model it as a mixture of Gaussians.}
We build a weakly informative likelihood model by considering the behavior of the distribution of $C$ in extreme cases and extrapolating the model to more ambiguous cases.
Considering only the extremely soft ($HR\leq-0.99$; likely foreground) and extremely hard ($HR\geq+0.99$; likely background) sources, we see that the distribution of the posterior modes of the colors $C$ of such sources splits into two distinct and well-separated components (see left panel of Figure~\ref{f:xcol}; red and blue curves respectively).
In contrast, a sample of 823 of the best-measured subset of sources with $-0.99<HR<+0.99$ (likely members) occupy a third component in between the extremes (green curve).
The presence of such distinct components suggests a simple parameterization of the likelihood function centered on each of the components.
However, {this measure is subject to imperfect domain adaptation:} as larger samples are examined, the outer peaks move inward, eventually smoothing out the trimodal structure, suggesting that the extreme values do not form a high-fidelity training set (dashed black histogram).
We therefore {seek to construct a measure which accounts for the gross separation without addressing the detailed shape or the changes in the distribution of colors as samples with larger uncertainties are included; i.e., we seek to avoid overfitting to the distributions by choosing a deliberately imprecise scheme.  We thus} choose Gaussians centered at $C=(-1,0,+1)$, with widths corresponding to the standard deviation of the subsets, to describe the different classes{,} further dilute the sharp divisions between the components by using a square-root transformation on the Gaussians to reduce the contrast, and enforce a linear decline in the tails of the central component to avoid numerical instability at large deviations.
Since we expect there to be a a great deal of mixing between the different classes due to the large dynamic ranges present in X-ray luminosities, these corrections 
avoid the X-ray colors being the dominant contributor to the class likelihood assignments.
The result of these modifications is shown in the middle panel of Figure~\ref{f:xcol} {(which shows that the transitions between classes are not sharp, but the extremes are indeed unambiguously assigned)}, and the corresponding class assignments to the full dataset in the right panel.

\subsection{Extinction, $\Av$}
\label{s:extinction}

\begin{figure*}[htb!]
\includegraphics[width=2.5in]{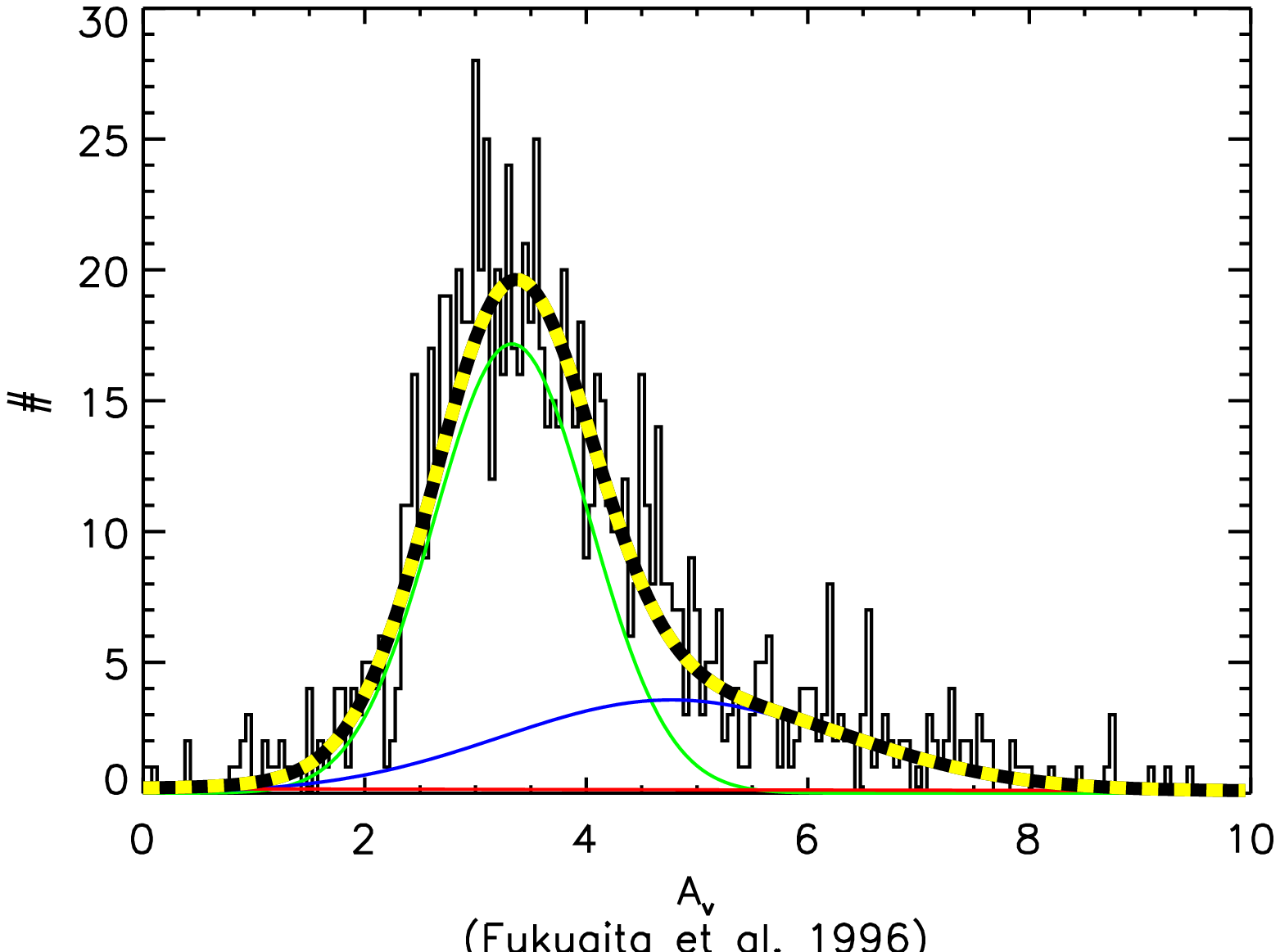}
\includegraphics[width=2.5in]{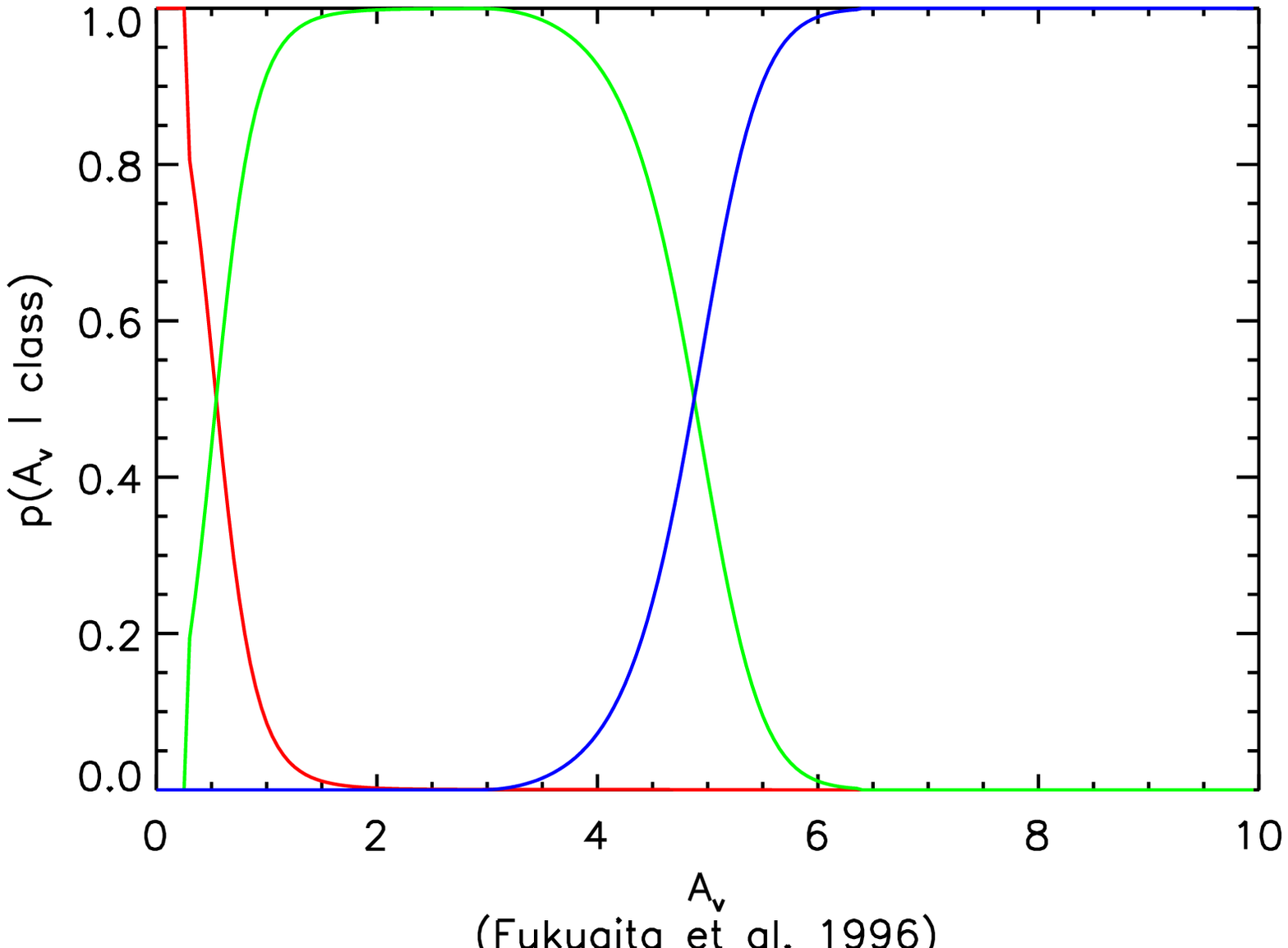}
\includegraphics[width=2.5in]{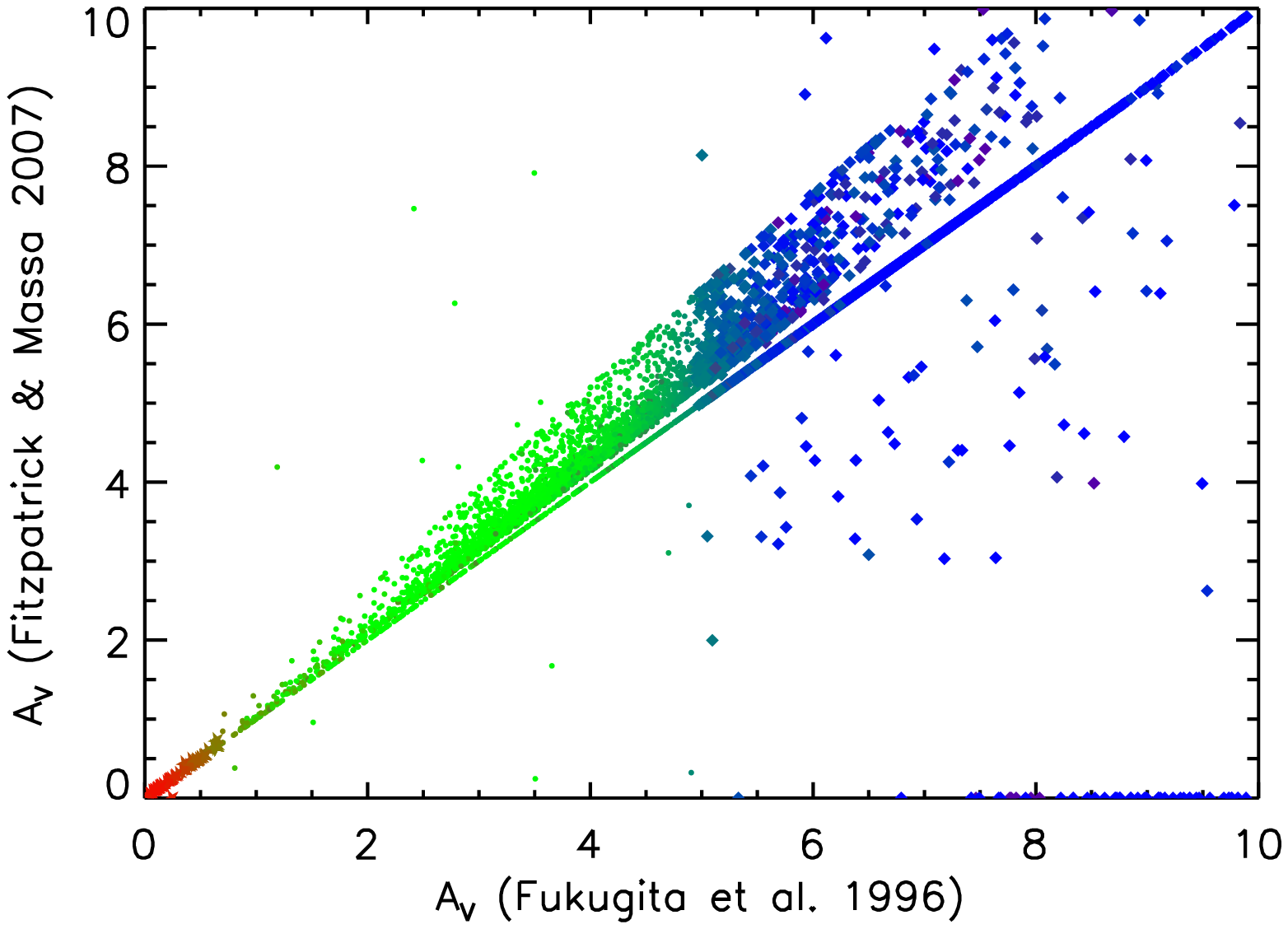}
\caption{
Separating \cygob\ sources using extinction, $\Av$.
The observed histogram ({\sl left}) is modeled as a mixture of three Gaussians, with the components that predominate at low, medium, and high $\Av$ assumed to represent the foreground (red), member (green), and background (blue) objects.
These Gaussian components, after removal of the population size effect and renormalization are identified with the corresponding likelihoods ({\sl middle}).
A scatter plot showing the difference in estimated $\Av$ when using the extinction laws of 
F96 
and 
FM 
is shown at {\sl right}, with the points color coded as rgb-tuples as in Figure~\ref{f:riz2pc}. 
}
\label{f:bestAv}
\vskip 0.5in
\end{figure*}

Low extinction is a strong diagnostic of whether a source is in the foreground of the cluster or not, and conversely, high extinction suggests that an object is in the background.
{Because of this high sensitivity of class assignment to extinction, we compute $\Av$ from optical and IR data, and incorporate it as additional data stream.
Note that while this information is partially included in the spectral shapes data extracted from the X-ray colors (see Sec~\ref{s:xshape} above), extinction affects spectra at different temperatures differently, and an independently generated data stream can provide additional information to define the classification.  However, because}
extinction is not {spatially} uniform across the cluster, a simple cut across $\Av$ is not a good method to assign classes.
{As in the case of X-ray color $C$ (see Sec~\ref{s:xshape}), $\Av$ is also an 1-dimensional data stream, and we model a subset of 1003 well-estimated $\Av$ as a mixture of three Gaussians, each directly representing the three classes of interest} (see Figure~\ref{f:bestAv}).

Our $\Av$ values are estimated based on $\jj, \hh, \rr, \ii, \zz$, as well as 2MASS and UKIDSS magnitudes.
(Note that due to the non-linear interplay between the optical and IR magnitudes, $\Av$ forms another complementary combination of these attributes that were used in linear combinations with PCA above.)
Individual extinctions are calculated using one of two methods.
The main method consists of calculating the displacement of stars in the $\rr-\ii$ vs.\ $\ii-\zz$ diagram along the extinction vector from the 
$3.5\,$Myrs isochrone.
For those stars with unreliable or absent optical photometry 
\citep[see][]{Guarcello+12},
individual extinctions are calculated from 2MASS and UKIDSS photometry using the Near-Infrared Color Excess Revisited (NICER) algorithm
\citep{Lombardi.Alves:2001}.
Errors are calculated propagating the photometric uncertainties in these two colors.
We have considered both the extinction vector in the SDSS bands from 
\citet[hereafter F96]{Fukugita+96}
and the more recent from 
\citet[hereafter FM]{FitzpatrickMassa07}.
Individual extinctions obtained with these two laws typically differ by
$\approx{7\pm10\%}$
with the $\Av$ distribution obtained with the
F96 
law peaking at about 5.5$^m$ and that with the 
FM 
law at about 4$^m$.
Our classification scheme primarily relies on the 
F96 
extinction law.
We discuss the difference using the different extinction laws makes on our classifications in Appendix~\ref{s:FF}.

\section{Discussion}
\label{s:discuss}

\subsection{Validation}
\label{s:whydothis}

\begin{figure*}[htb!]
\includegraphics[width=3.1in]{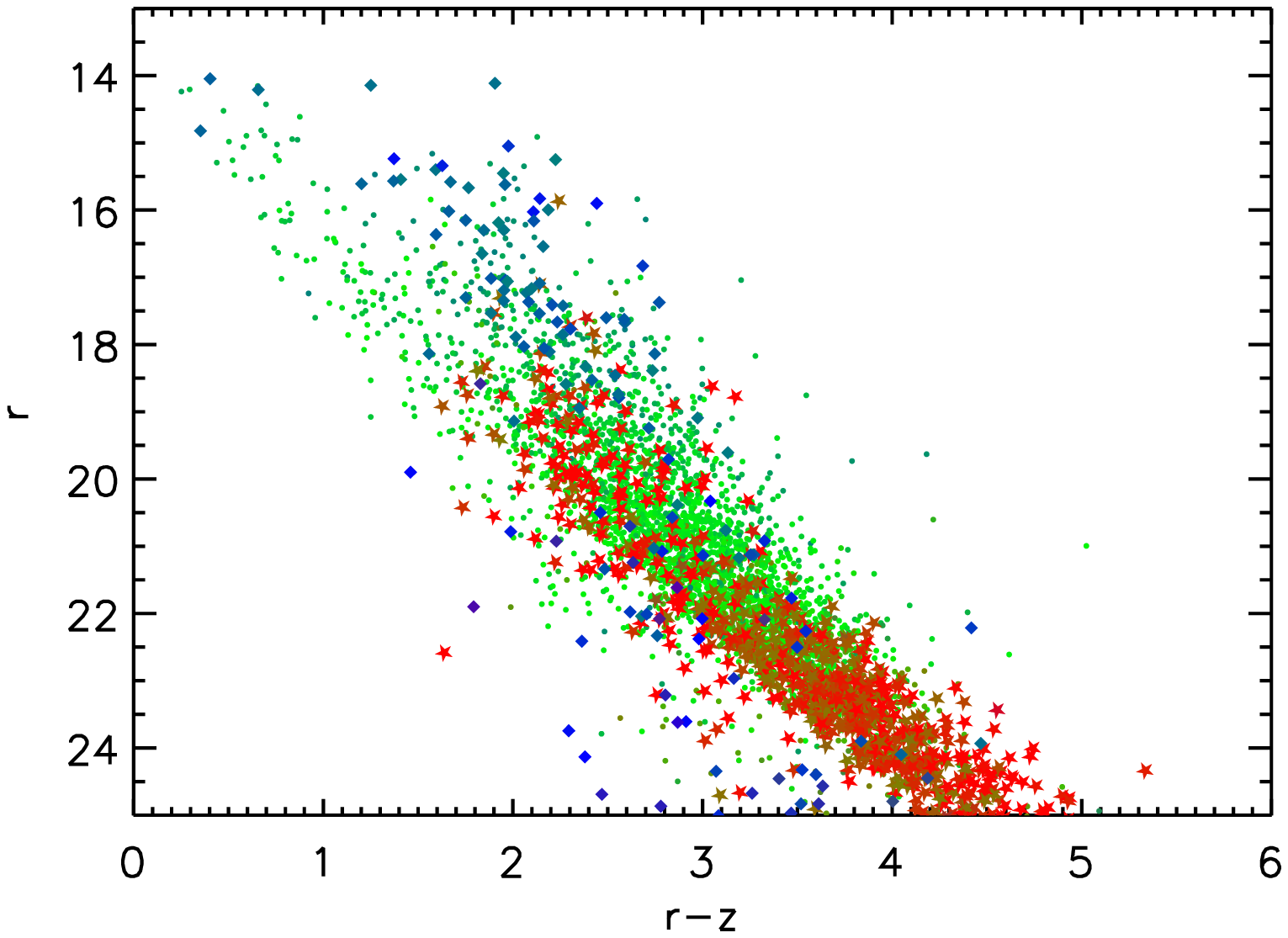}
\includegraphics[width=3.1in]{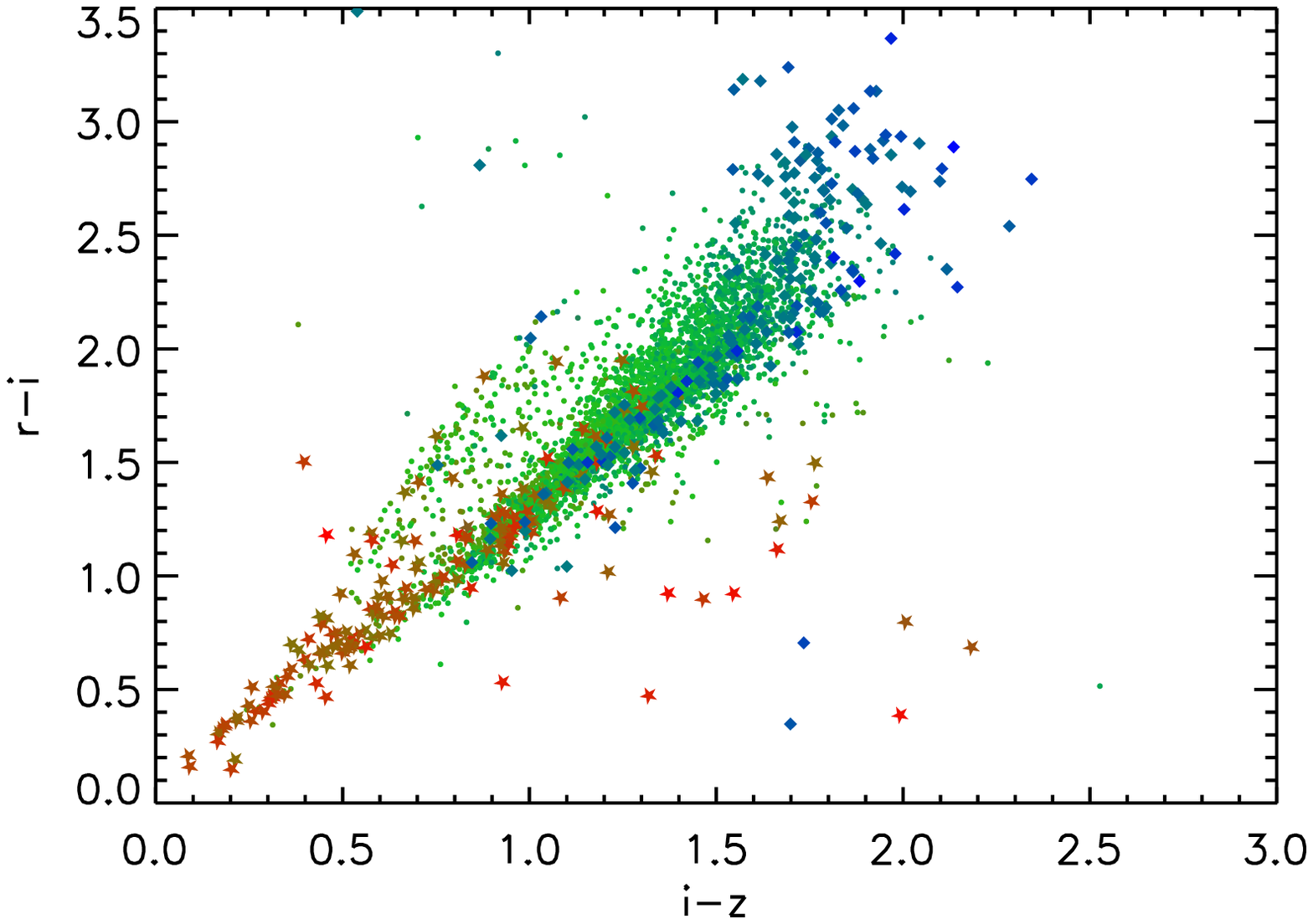}
\caption{Demonstrating the necessity of using multiple channels of data to derive class memberships.
The left figure {shows $\rr$~vs.~$\rr-\zz$ as} in Figure~\ref{f:riz2pc}-{\sl right}, but using the $3^{rd}$ PC as in Figure~\ref{f:riz3pc}.
The right figure shows the reverse case, {i.e., $\rr-\ii$~vs.~$\ii-\zz$ as} in Figure~\ref{f:riz3pc} but using class probabilities derived using the $2^{nd}$ PC as in Figure~\ref{f:riz2pc}.
}
\label{f:riz2Vriz3}
\vskip 0.5in
\end{figure*}

A question that arises is whether it is necessary to go through all these steps to obtain a classification, instead of placing cuts in a color-color diagram, say as {{for $\rr-\ii$~vs.~$(\ii-\zz)$,} as in Figure~\ref{f:riz3pc}.}
The necessity of a comprehensive analysis is best demonstrated by example, as in Figure~\ref{f:riz2Vriz3}.
Here, the probability assignments made using each of the two $\{\rr,\ii,\zz\}$ PCs (see \S\ref{s:riz}) are applied to the other: that is, the classification from the $2^{nd}$ {component, trained using $\rr$~vs.~$(\rr-\zz)$ is shown plotted with $(\rr-\ii)$~vs.~$(\ii-\zz)$,} and the classification from the $3^{rd}$ {component, trained using $(\rr-\ii)$~vs.~$(\ii-\zz)$, is shown plotted with $\rr$~vs.~$(\rr-\zz)$.}
These plots show that there is considerable mixing evident between the different classes when viewed with different projections.
It is therefore crucial that a combination of data streams be used, encompassing different pieces of information, in order to obtain reliable class assignments.

Standard methods of validation like cross-validation using $k$-fold training or bootstrapping on test samples are impractical for this analysis, since our training sample is small, and due to data incompleteness the overlap between components in the training sample is also small.
Furthermore, since any single data stream can predict classes that are inconsistent with the information present in the rest of the data, cross-validation of the process by leaving out one component is not a workable method of testing the results.
Instead, we manually screened the classifications in various color-magnitude spaces, including some not used during the classification process such as IRAC colors like $[4.5]-[5.8]$ vs.\ $[5.8]-[8.0]$, and other color projections like $\gs$ and $\gs-\rr$ vs.\ $\rr-\ii$, 
to expose outliers, which we reassigned where necessary as explained in more detail below {(see Sec~\ref{s:reclassify}).}
Based on this manual reclassification (see Table~\ref{t:posthoc} below), as well as comparing the effects of different extinction laws on the classification (see Appendix~\ref{s:FF}), we estimate that the error rate in our classification is $\approx5\%$.

\subsection{Classification}
\label{s:classify}

\subsubsection{Reclassification}
\label{s:reclassify}

Our classification can be affected by some stellar properties.
For instance, during X-ray flares stellar X-ray spectrum becomes harder, which mimics the expected X-ray energy quantiles typical of background sources.
Futhermore, compared to naked photospheres, stars with circumstellar disks have red infrared colors that can be confused with background sources at large extinction, or they may have blue optical colors typical of foreground stars because of the accretion process or the presence of scattered light 
\citep{Guarcello+10}.
Besides, the adopted NBC classification scheme is expected to have issues for very bright stars, whose photometry is typically saturated.
For these reasons, the results of the automated classification described above (\S\ref{s:likeli}) has been retested using OIR color-color and color-magnitude diagrams and the X-ray energy quantiles, which provides further leverage to separate the three classes.
We also forced the classification of known stars with disks as selected by 
\citet{Guarcello+13,CygOB2PhotoEvaporation.Guarcello+2016}
to be association members.
Overall, a total of {381} objects changed classification (see Table~\ref{t:posthoc}), 
{including 49 (out of 833) of those that had originally been included in the training sets with optical and IR colors and magnitudes (and an additional 208 of 1796 of those used only with X-ray quantiles, color, and $\Av$ analyses).  This yields an overall error rate in the automated classification of $\approx$6\% from}
among those X-ray sources with optical matches, and $\approx5\%$ for all objects (note that this includes cases with multiple matches).
{The nominal accuracy of classification of cluster membership is $\approx$96\%.  Because of the large size of this subset (6169 objects), the majority of the reclassification involves cluster members.  This class gains 199 and loses 71 to the foreground class, and gains 109 and loses 2 to the background class.  Because of the much smaller sizes of the foreground (491 objects) and background (1360 objects) classes, the nominal accuracy of their assignment is dominated by the cluster membership accuracy, at $\approx$74\% and $\approx$92\% respectively.
}

\begin{deluxetable*}{lccc|c}
\tablecaption{Post-hoc Reclassifications: Number of sources that each class with Naive Bayes approach and after reclassification, showing how many in each class change classifications
\label{t:posthoc}}
\tablehead{ \colhead{Original} & \colhead{\hfil} & \colhead{Reclassified} & \colhead{\hfil} & \colhead{subtotal Original}}
\startdata
\hfil & Foreground & Member & Background & \hfil \\
Foreground & 420 & 199 & 0 & 619 \\
Member & 71 & 5861 & 2 & 5934 \\
Background & 0 & 109 & 1358 & 1467 \\
\hline
subtotal Reclassified & 491 & 6169 & 1360 & 8020 \\
\enddata
\end{deluxetable*}

\begin{figure*}[htb!]
\centering
\includegraphics[width=3in]{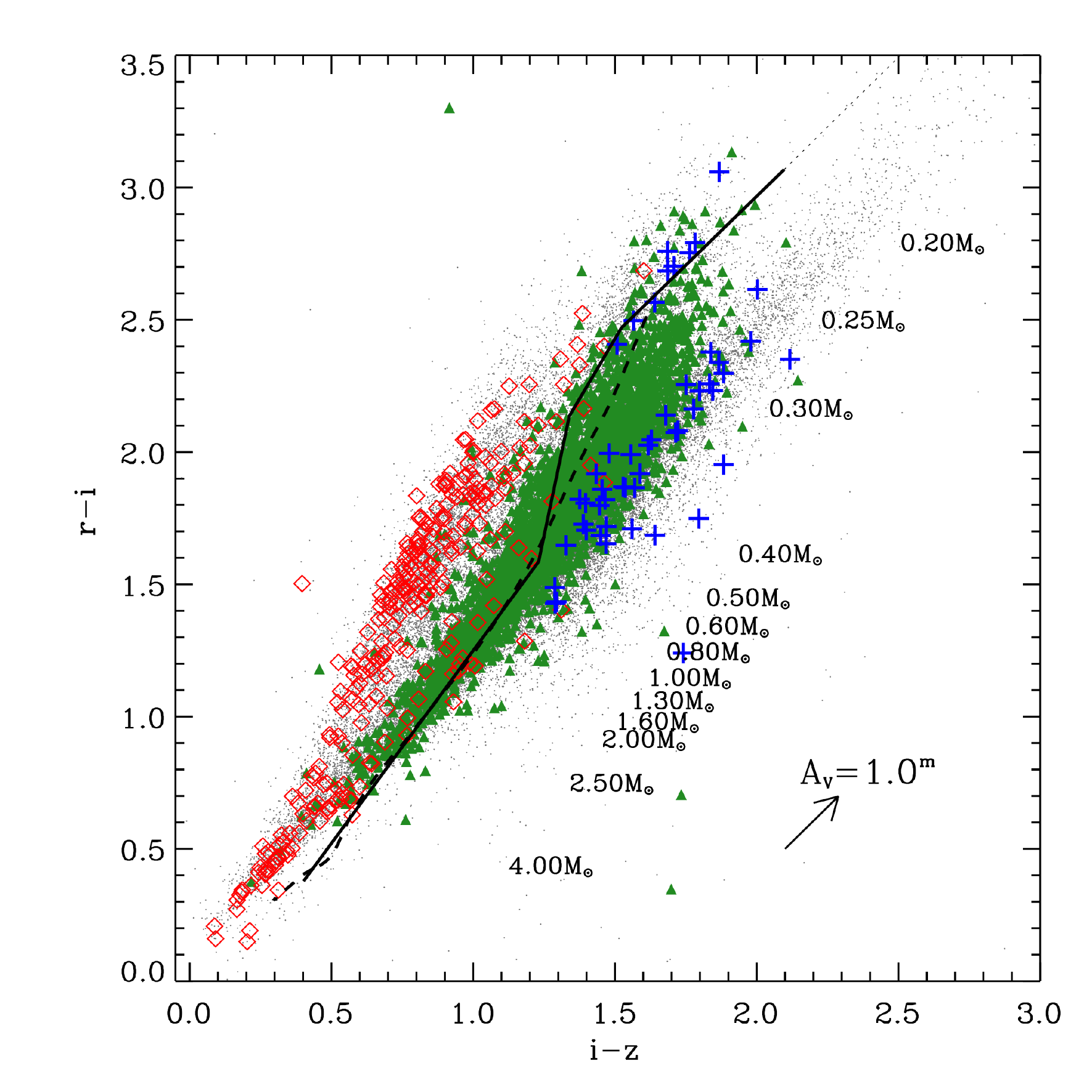}
\includegraphics[width=3in]{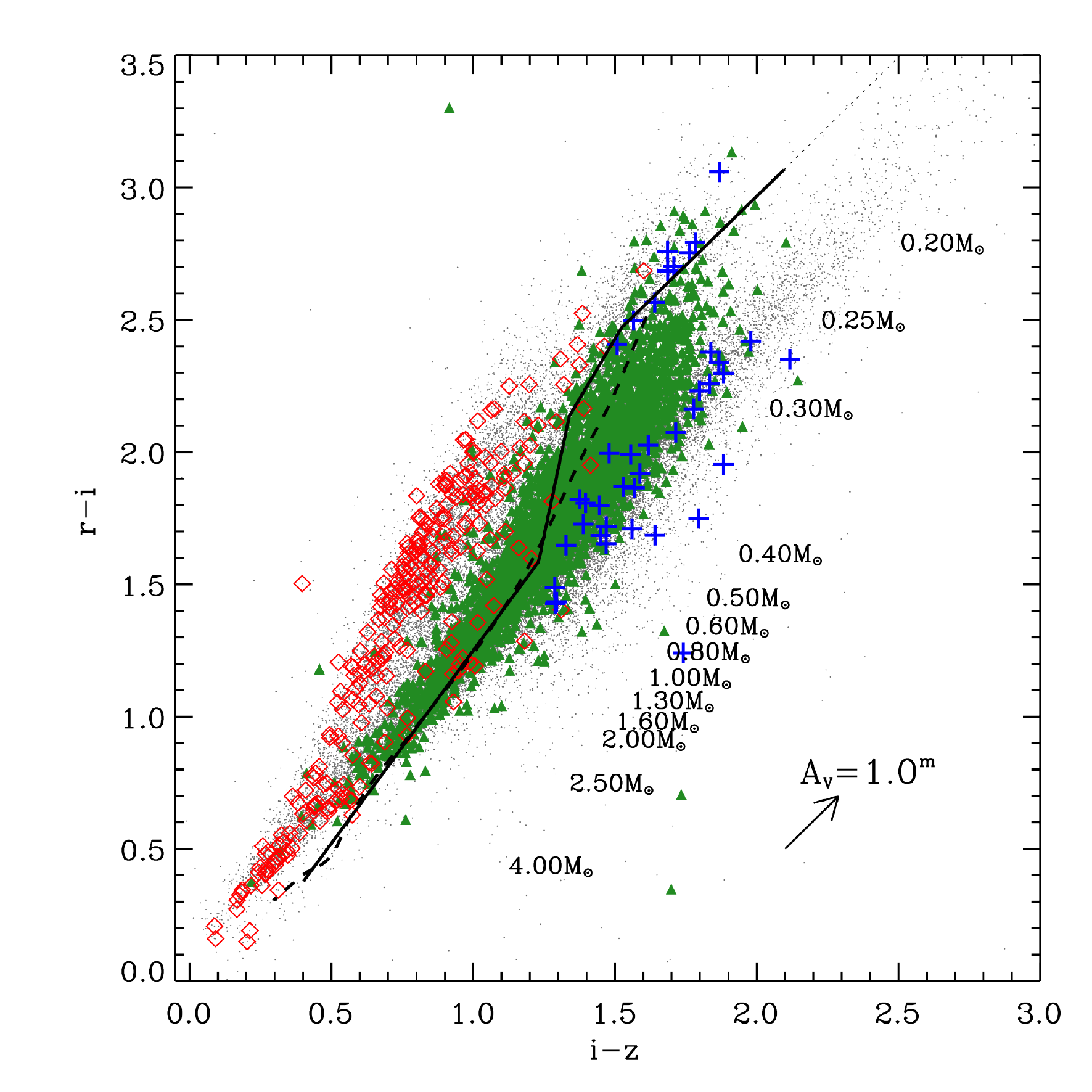}
\caption{$\rr-\ii$ vs. $\ii-\zz$ diagrams of all the sources with good OSIRIS or SDSS photometry.
The solid line is the 2.5 Myrs isochrone from 
\citet{Siess+00}
after applying the 
F96 
transformation between the UBVRI and the $\up\,\gp\,\rp\,\ip\,\zp$ photometric systems, and the dashed line is from the MIST database.
The extinction vector is obtained from 
\citet{ODonnell+94}.
The corresponding masses are indicated with a horizontal displacement from the isochrone.
The classifications are according to the automated NBC method ({\sl left}) and after manual reclassification ({\sl right}), and show foreground (red diamonds), members (green dots), and background (blue crosses) objects.
}
\label{b:1}
\vskip 0.5in
\end{figure*}

We compare the differences between the automated classification and manual reclassification in Figure~\ref{b:1} and Figures~\ref{b:2}-\ref{b:6}.
In each panel of the figures, the small grey points mark the colors or magnitudes of the sources in the survey area with good quality photometry and errors smaller than 0.15 for colors and 0.1 for magnitudes.
In each figure, the X-ray sources classified according to the automated NBC scheme are in the left panel, and the revised version based on manual inspection in the right panel.
The diagrams shown in Figures~\ref{b:1}, \ref{b:2}, and \ref{b:3} allow us to separate the sources with low (foreground), intermediate (members) and high (background) extinction and thus select stars that are likely wrongly classified by the NBC method, such as the foreground and background stars which populate the same locus as the candidate members in these diagrams.
Even after the revision, a small number of stars appear to lie on the ``wrong'' part of these diagrams, such as the background sources between the E$_{B-V}=0$ and E$_{B-V}=1$ main sequences in the left panel in Fig. \ref{b:2}.
These are cases where no clear indications come from the diagrams and the X-ray photon energy quantiles, for instance because of mismatches between counterparts of different catalogs.
In these cases, we have kept the classification from the NBC method.
It is also important to note that the position of candidate members with disks in these diagrams can be affected by accretion and/or scattered light, for instance increasing the $\rp - \Ha$ color or decreasing the $\gs-\rr$ color, thus pushing the sources above the E$_{B-V}=0$ main sequence in Figure~\ref{b:2} or below the $\Av=0$ isochrone in Figure~\ref{b:3}.

The $\hh-\kk$~vs.~$\jj-\hh$ diagrams shown in Figure~\ref{b:4} allow us to separate sources with different extinction, and thus stars in the foreground, background and those within the association.
In this diagram it is also clear that the NBC method fails to classify as members very bright stars which are clearly massive members of \cygob\ according to their photometric properties, soft X-ray spectra, and in some case existing spectral classification.
Furthermore, these stars are clearly clustered in the various subclusters of \cygob.

Given that extinction does not seriously affect the Spitzer/IRAC colors, the diagrams of IRAC colors in Figure~\ref{b:5} are not useful for separating stars affected by different extinction.
However, together with the set of diagrams used in 
\citet{Guarcello+13},
they are useful for selecting candidate stars with disks, typically populating the regions corresponding to colors larger than $0.5^m$, and background galaxies populating loci that are empirically defined by various authors.
Recall that we enforced the classification as ``members'' of all the candidate stars with disks selected by 
\citet{Guarcello+13}.

The spatial distributions of the X-ray sources classified before and after the revision are shown in Figure~\ref{b:6}.
Comparing the two panels, it is evident that several stars whose classification has been turned into ``members'' are indeed clustered in the center of the region or in other subclusters.
In both panels regions with dense nebulosity can be identified by the contours of the IRAC $[8.0]\mu$m continuum emission levels.
Several X-ray sources classified as ``background'' objects fall within high-extinction regions such as within DR~18, at approximately $\alpha=308.7$, $\delta=41.2$.
These sources are likely to be embedded young stars detected in X-rays, but their classification has not been changed due to the lack of a good quality OIR counterpart. 

\subsubsection{Impact of \gaia}
\label{s:gaia}

Unlike our method, which is based on a probabilistic weighing of proxy information, precise measurements of distances to matched OIR sources can fix the classification exactly.
The 
\gaia\ DR2 catalog has parallaxes for over a billion stars \citep{Gaia+18}, and could be a source of such distance measurements.
Of the X-ray and matched OIR sources, 1128 (60 foreground, 749 members, 319 background; $\approx14\%$ of the catalog) have probable counterparts in \gaia\ DR2 based on overlaps of position error circles.
Of these, choosing the nearest \gaia\ counterpart, 329 (52 foreground, 252 members, 25 background) have non-zero parallaxes, and only 140 (47 foreground, 91 members, 2 background) have well-measured distances (parallaxes measured at better than $3\sigma$).
The change in the mix of classification (from $\approx6\%$ foreground to $\approx35\%$ foreground) is consistent with nearer objects having better distance measurements.
It is reasonable to expect that foreground stars will be the ones best characterized {by \gaia,} and any misclassifications in the probabilistic scheme will be dominated by these sources.
Indeed, we find that 41 sources are apparently misclassified, of which 31 are classified as members but are at distances $<1.2$~kpc, and are plausibly foreground stars.
In addition, one \gaia-matched source classified as background is likely a member (distance$\approx1.3$~kpc), and four sources classified as foreground stars are at distances $>1.2$~kpc and could be considered association members or background sources.
This is consistent with the error rate of the classification scheme (see Tables~\ref{t:posthoc}, \ref{t:cmpFuFM}, and \ref{t:posthocFM}).
Thus, the \gaia\ DR2 release has a negligible effect on both our method and our results.
Here, we report the potential changes in classification due to \gaia\ parallax measurements as part of the catalog (see Sec~\ref{s:catalog} below), but otherwise do not incorporate it in our procedure, and defer a more detailed analysis that looks at the individual matches and an assessment of the systematic uncertainties in the \gaia\ catalog at large distances and large $\Av$ to a later work.

\subsubsection{Catalog}
\label{s:catalog}

\begin{figure*}[htb!]
\centering
\includegraphics[width=2.8in]{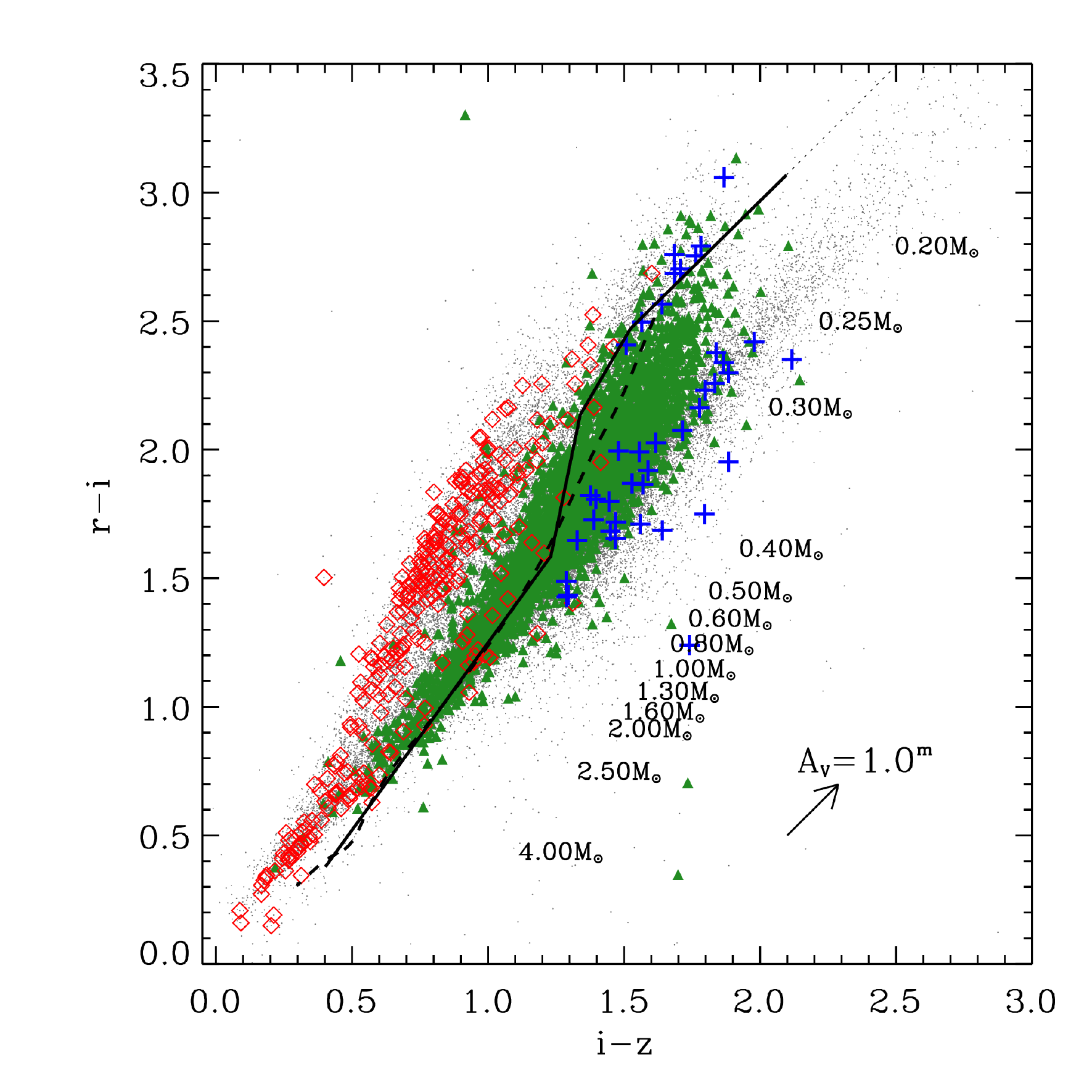}
\includegraphics[width=2.8in]{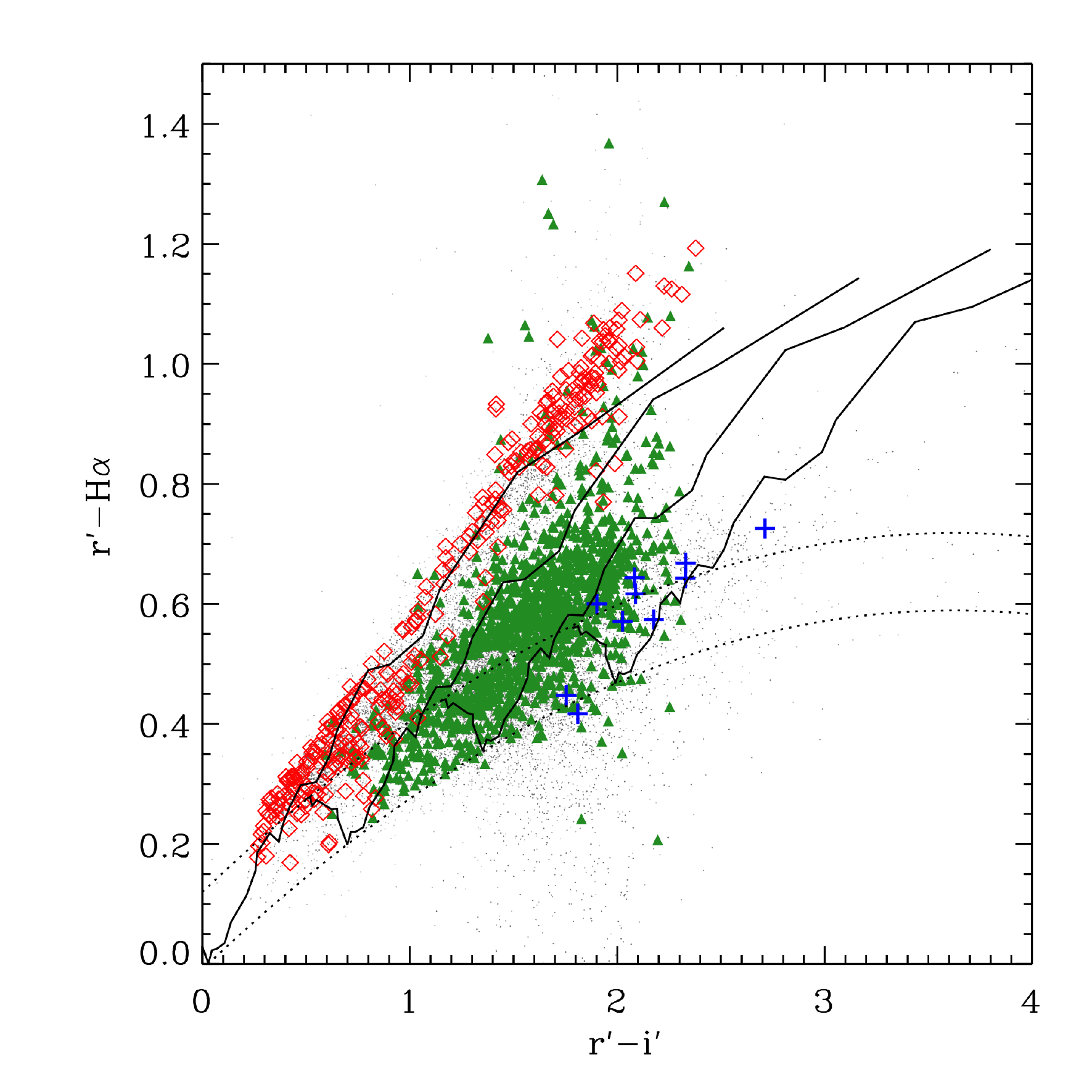}
\includegraphics[width=2.8in]{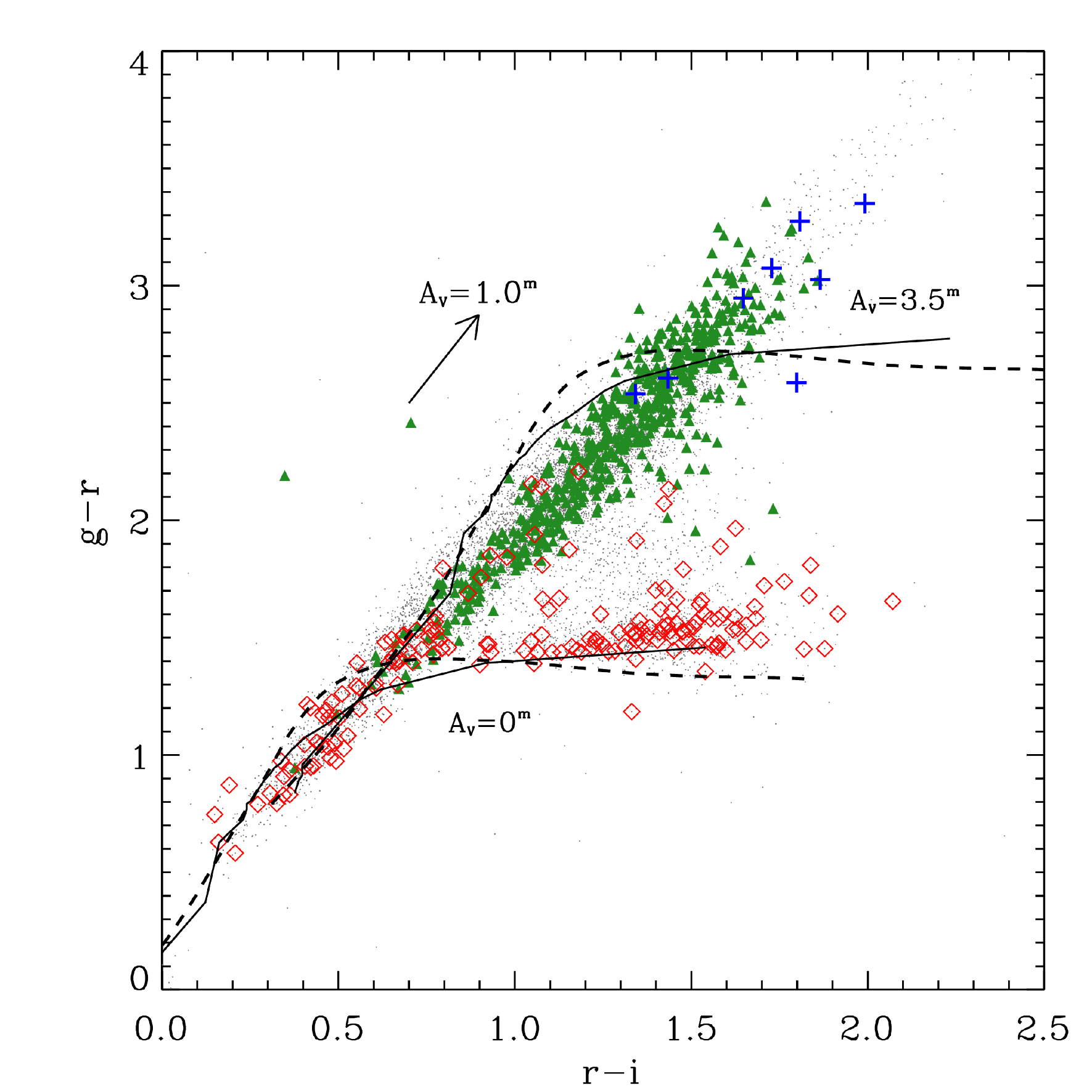}
\includegraphics[width=2.8in]{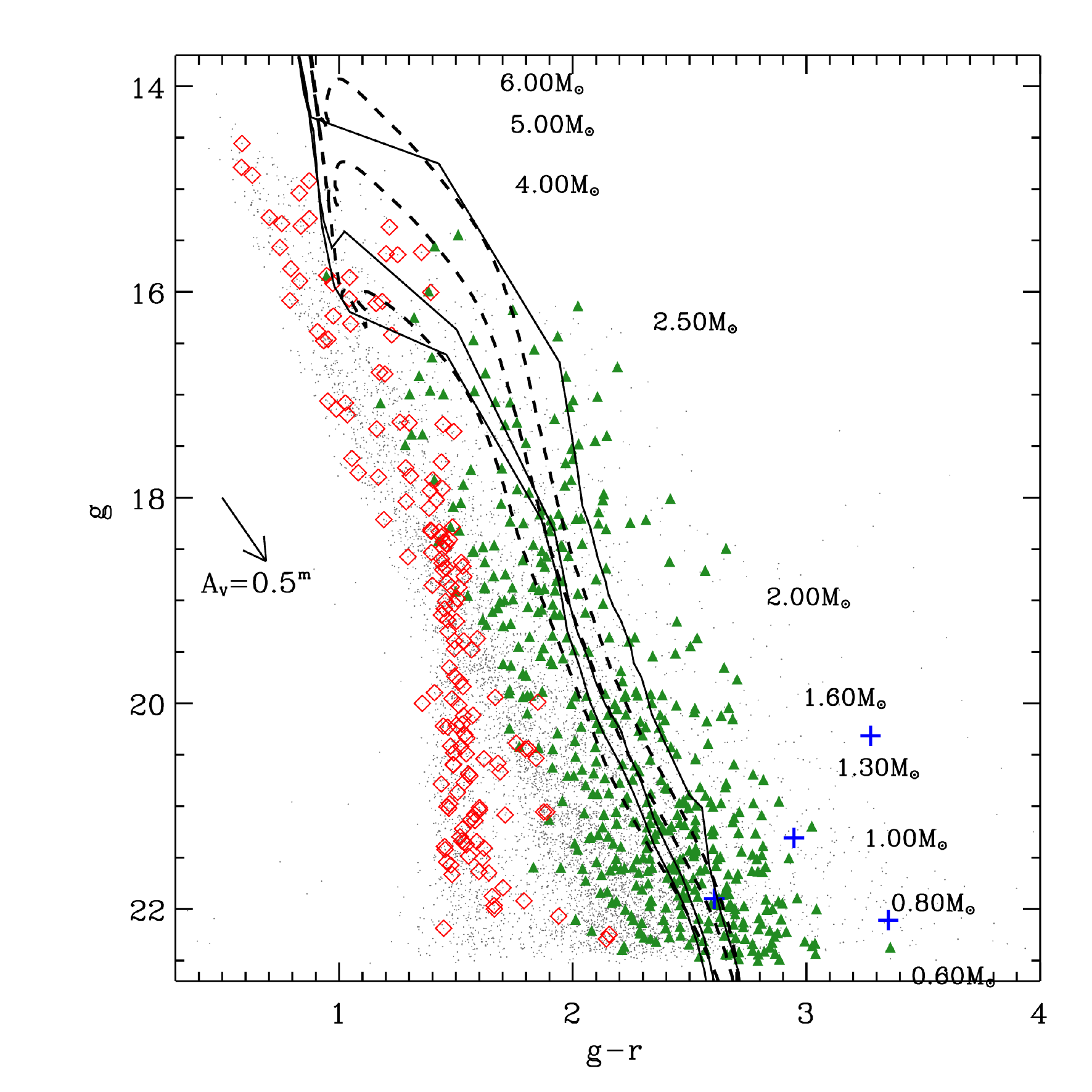}
\includegraphics[width=2.8in]{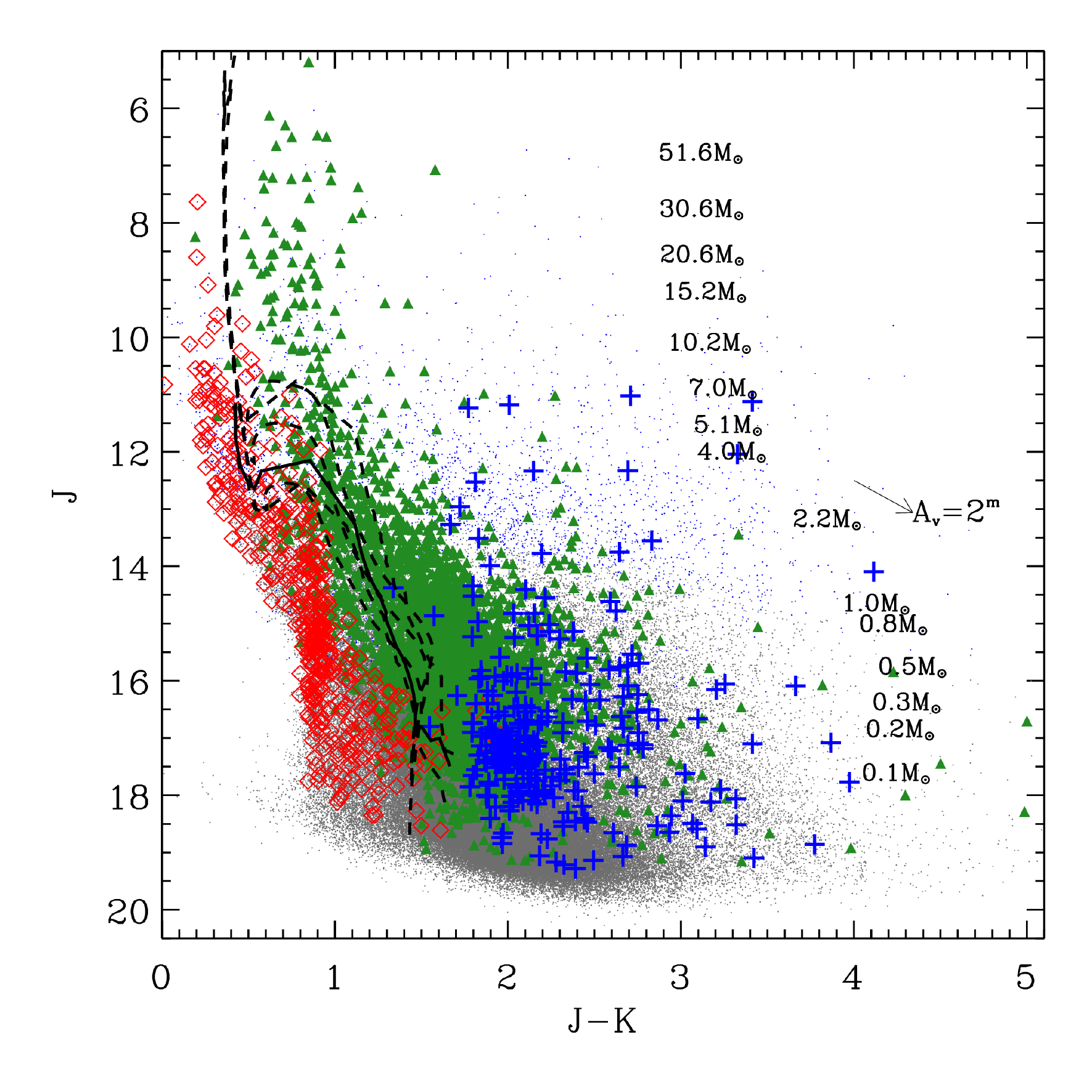}
\includegraphics[width=2.8in]{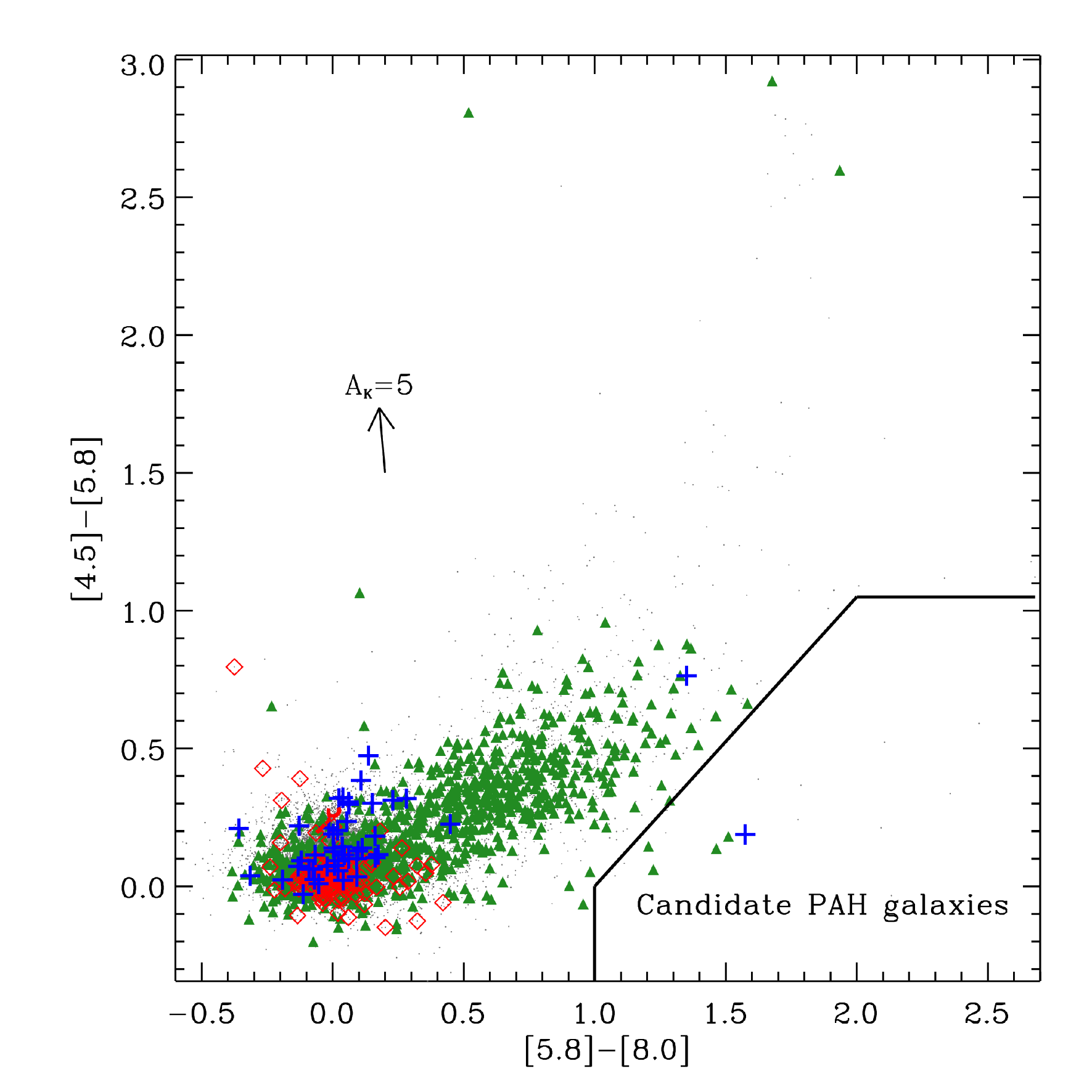}
\caption{Set of diagrams showing colors and magnitudes of: all the sources of our OIR catalog with good photometry in the involved bands (gray dots), candidate OIR+X-ray foreground sources (red diamonds), \cygob\ members (green dots), background sources (blue crosses).
We also show the extinction vectors and the 2.5 Myrs isochrone from
\citet{Siess+00}
(solid lines) and the MIST database (dashed line), assuming a distance of 1400$\,$pc and A$_V=3.5^m$ with corresponding mass values labeled with a horizontal displacement.
In the upper-right panel the solid lines are instead ZAMS at increasing E$_{B-V}$ (0, 1, 2, 3).}
\label{f:mario_diagrams}
\vskip 0.5in
\end{figure*}

The final classifications are shown in Figures~\ref{f:mario_diagrams} using suitable color-color and color-magnitude diagrams where the foreground, background, and association sources can be distinguished.
In each panel the small gray points mark those sources in the survey area with good quality photometry and errors smaller than 0.15 in the colors and 0.1 in magnitudes.
(The criteria for ``good photometry data'' in these catalogs are described in 
\citet{Guarcello+12,Guarcello+13}).
The OIR+X-ray sources are marked with colors coding their final classification.
We also mark a possible locus of candidate members of \cygob\ using the 2.5 Myrs isochrone from 
\citet{Siess+00}
(converted into the $\up\,\gp\,\rp\,\ip\,\zp$ photometric system), the MIST isochrones 
\citep{Dotter+16,Choi+16},
and in the IPHAS color-color diagram the ZAMS defined by 
\citet{Drew+05},
assuming the distance of 1400$\,$pc found by 
\citet{Rygl+12}.

In all the diagrams, with the exception of the IRAC color-color diagram, given the direction of the reddening vector, the foreground stellar population can be separated from the stars at the distance of the association or those further away.
This is particularly evident in the IPHAS color-color diagram and in the SDSS diagrams, the latter mainly for low-mass stars.
The vast majority of the sources sorted as ``foreground stars'' populate these low-extinction loci.
The background population is more evident in the $\jj\hh\kk$ diagram, mainly in a high extinction locus (faint and red stars).
In the IRAC color-color diagram some of the background objects lie in the locus where stars with disks are typically found.
These stars have been excluded from the list of stars with disks and classified as ``background contaminants'' by 
\citet{Guarcello+13}.
The cluster locus lies between the foreground and the background population in most of the diagrams.
Mainly in the IPHAS color-color diagram it lies within the 1$\leq$E$_{B-V} \leq 3$ range and in the $\jj\hh\kk$ diagram within the 3$\leq$A$_V \leq 10$ range. 
Of those objects classified as likely members based on their IR excess by \citet{Guarcello+13}, {439 (of 1843) are detected in X-rays.}
{Among the corresponding 510 optical matches, there are 365 Class~II, 16 Class~I, 58 Flat Spectrum, 43 Transition and pre-transition disks, 22 accretors according to their H$_\alpha$ line, and 6 have blue excesses}

\begin{figure*}[htb!]
\includegraphics[width=2.5in]{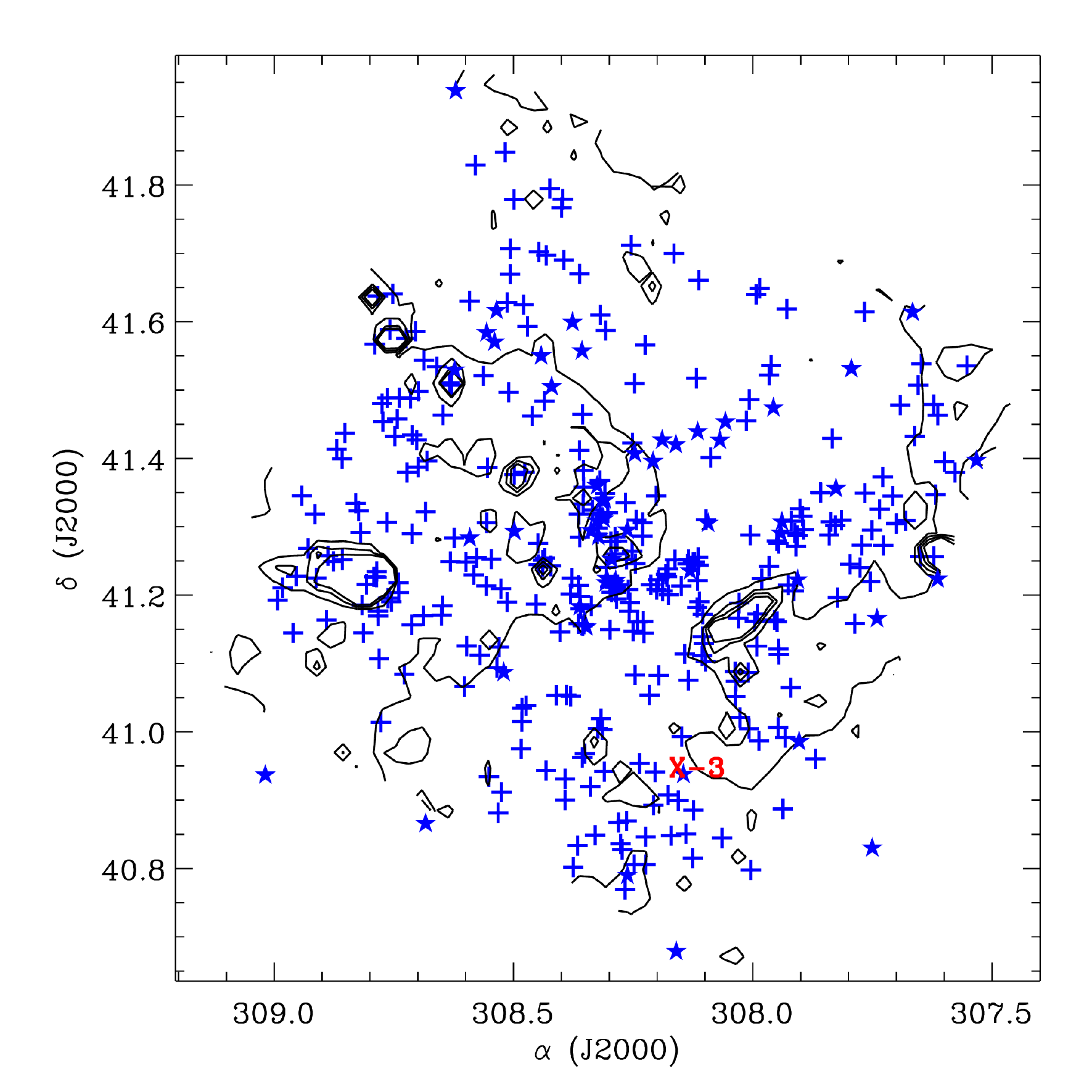}
\includegraphics[width=2.5in]{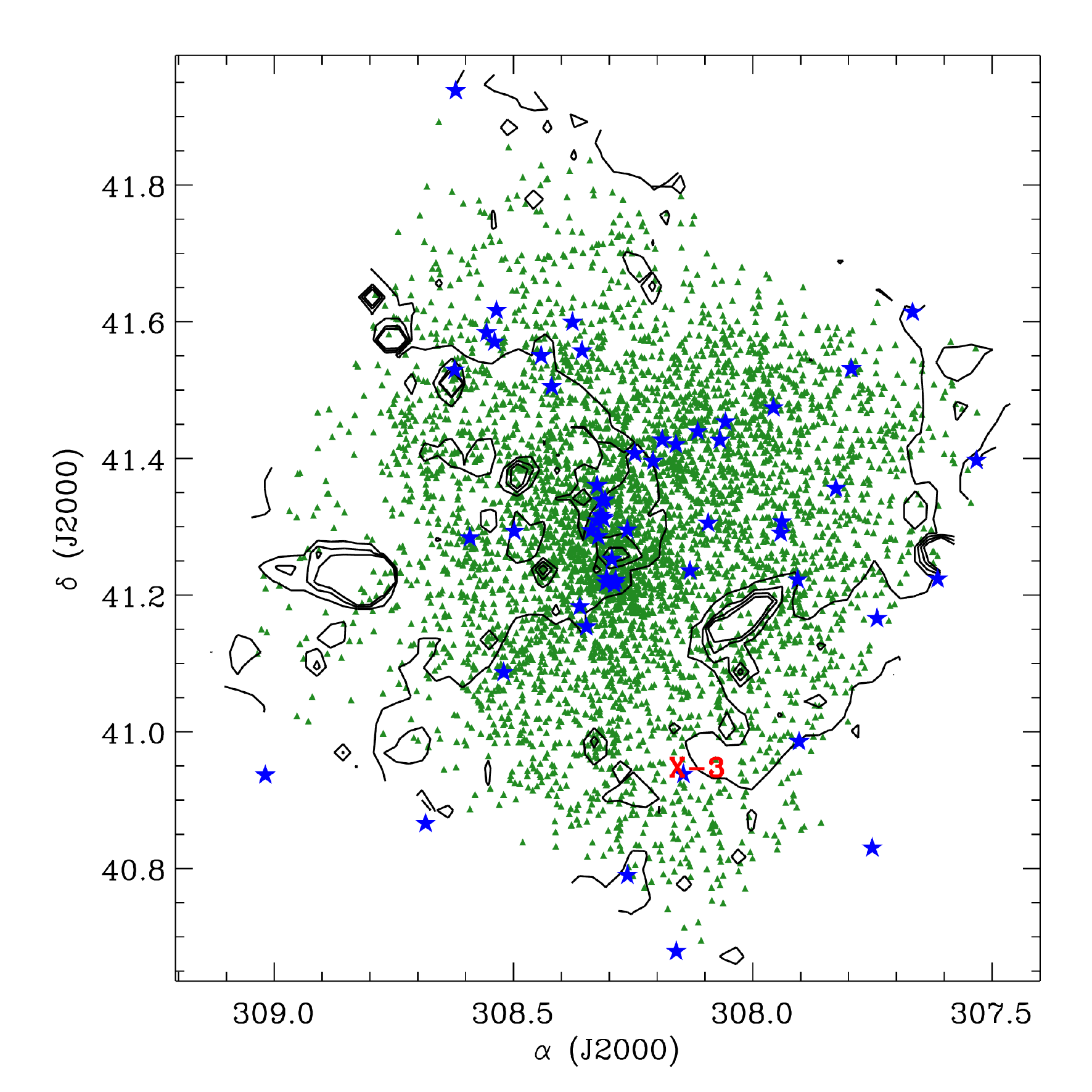}
\includegraphics[width=2.5in]{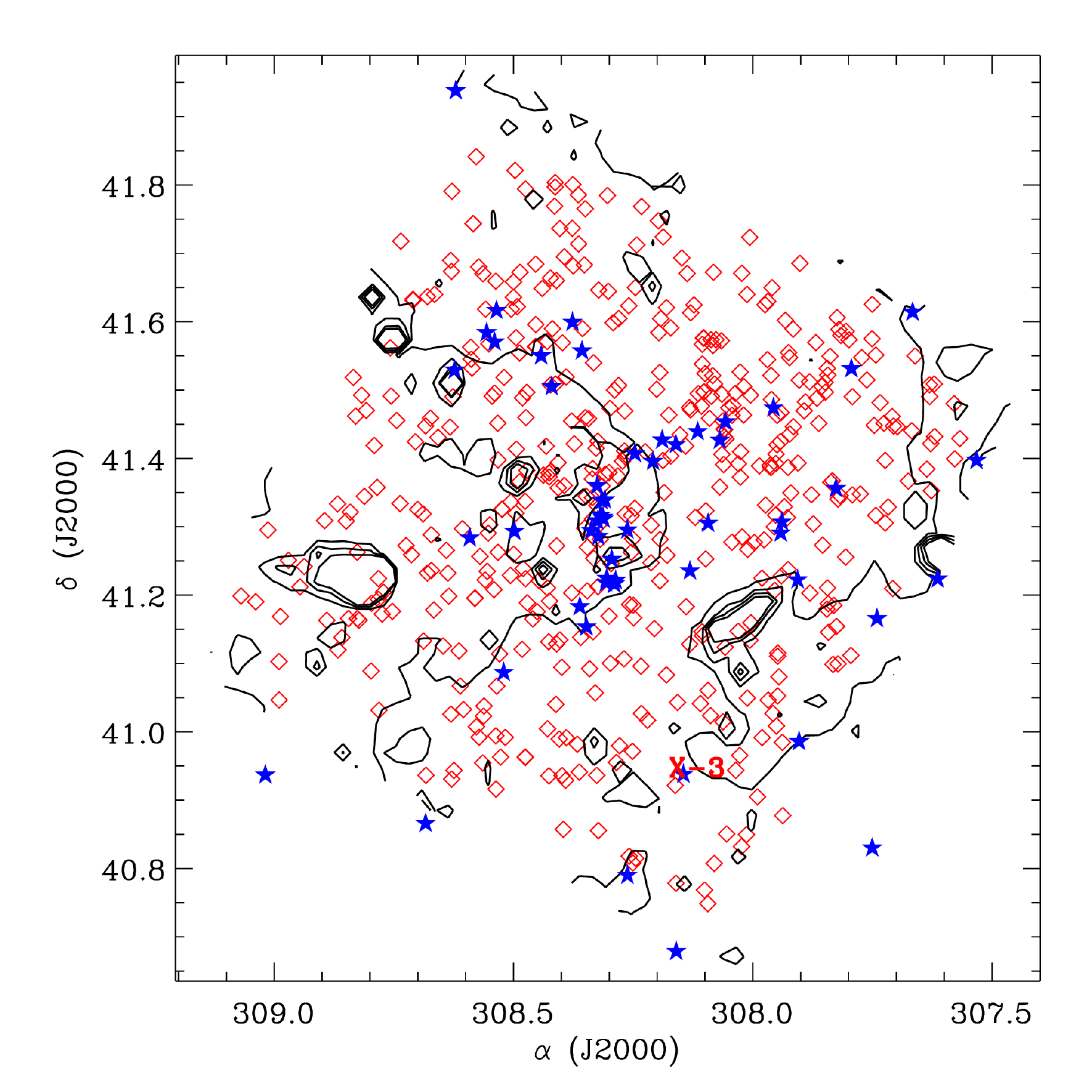}
\caption{Spatial distribution of {background} ({\sl left}), association member ({\sl middle}), and foreground ({\sl right}) sources {with OIR matches.  
The contours mark continuum emission levels at $[8.0]\mu$m Spitzer band, denoting the presence of dust and high $\Av$.  The position of Cyg~X-3 is marked in red.}  {The filled blue asterisks denote O stars.}}
\label{f:mario_spadis}
\vskip 0.5in
\end{figure*}

The  spatial distribution of the foreground, member, and background objects are shown in Figure~\ref{f:mario_spadis}.
Objects classified as members show a clear concentration that corresponds to the association, while the distribution of foreground objects is more isotropic.
Objects classified as background show a small enhancement over regions of small $\Av$ {\citep[see also][]{CygOB2DiffuseBkg.Facundo+2018}}.

A short exemplar version of the catalog of classification is in Table~\ref{t:shortcat} (the full catalog is available online at Zenodo at \url{https://doi.org/10.5281/zenodo.8025756}),
and lists the CXO ID, {whether it has a} corresponding OIR match, 
the computed probabilities of classification from the NBC method, and the final classification.

\smallskip
\begin{deluxetable*}{ll|ccc|r}
\tablecaption{Catalog of Classifications$^\S$.  
(Full catalog is available at Zenodo at \url{https://doi.org/10.5281/zenodo.8025756}) 
\label{t:shortcat}}
\tablehead{ \colhead{CXO ID} & \colhead{OIR Match} & \colhead{$p_{\rm foreground}$} & \colhead{$p_{\rm member}$} & \colhead{$p_{\rm background}$} &  \colhead{Classification} }
\startdata
1 &  Yes  & 0.00 & 1.00 & 0.00 & member \\
4 &  No  & 0.00 & 0.03 & 0.97 & background \\
5 &  Yes  & 0.00 & 0.99 & 0.01 & member$^\Box$ \\
6 &  Yes  & 0.00 & 0.00 & 1.00 & background$^\Box$ \\
9 &  Yes  & 0.00 & 1.00 & 0.00 & foreground$^\ddag$ \\
11 &  Yes  & 0.13 & 0.87 & 0.00 & member$^\star$ \\
15 &  Yes  & 0.09 & 0.91 & 0.00 & foreground$^\ddag$$^\star$ \\
17 &  Yes  & 0.00 & 0.60 & 0.40 & member$^\star$$^\Box$ \\
37 &  Multiple  & 1.00 & 0.00 & 0.00 & foreground \\
\hfil &  Multiple  & 0.00 & 1.00 & 0.00 & member$^\Box$ \\
81 &  Yes  & 1.00 & 0.00 & 0.00 & member$^\ddag$$^\blacksquare$ \\
227 &  Yes  & 1.00 & 0.00 & 0.00 & foreground$^\blacksquare$ \\
\hline
\multicolumn{6}{l}{$\S$: Representative subset, for demonstration.} \\ 
\multicolumn{6}{l}{$\ddag$: Classification changed by inspection} \\
\multicolumn{6}{l}{$\star$: Used as part of training set} \\
\multicolumn{6}{l}{$\Box$: Matched to a {\sl Gaia} DR2 source within position error} \\
\multicolumn{6}{l}{$\blacksquare$: {Final classification} inconsistent with {\sl Gaia} distance} \\
\enddata
\end{deluxetable*}

\clearpage
\section{Summary}
\label{s:summary}

We have classified the X-ray sources observed towards \cygob\ as being foreground objects, members of the association, or background objects using a variety of associated data, including optical and infra-red magnitudes, X-ray quantiles and hardness ratios, and extinction estimates.
We adopt a Naive Bayes method to obtain automated classifications.
We use domain knowledge of expected distributions of well-measured stars observed in different passbands to construct likelihoods that are then applied to the full dataset.
Likelihoods are constructed by using a semi-supervised training method that uses objects with well-measured magnitudes, transformed using a Principal Component Analysis {or modeled with mixtures of Gaussians} to perform efficient separations for each channel.
{The probability that each source belongs to a given class is then computed, and sources are sifted into the appropriate class.  This is then augmented with visual inspection of several color-color and IRAC magnitude diagrams and correlated against known properties like the presence of disks that can cause systematic mis-assignments in the automated classification.}
We consider the effects of measurement as well as systematic uncertainties due to extinction, and estimate that the residual error in our classification is $\approx5$\%.

We construct a catalog that includes a probabilistic assessment of the class that each source belongs to.
Adopting a hard threshold that states that the highest of the triad of $\{p({\rm foreground}), p({\rm member}), p({\rm background})\}$ determines the class, we find that $\approx75$\% of the catalog sources are members of the \cygob\ association, $\approx5$\% are foreground stars, and the remainder are background objects.

\facility{CXO}

\software{PINTofALE \citep{2000BASI...28..475K},
TRILEGAL \citep{TRILEGAL2012},
IDL (v8.4)}

\acknowledgments
We thank the anonymous referee for a careful reading of the paper and for comments which significantly improved its clarity.
This work was supported by \chandra\ grant GO0-11040X.
VLK, JJD, and TLA were supported by NASA contract NAS8-03060 to the {\it Chandra X-ray Center} and thank the director, B.Wilkes, and the CXC science team for continuing support and advice.
MGG and NJW were supported by Chandra Grant GO0-11040X during the course of this work. 
MGG also acknowledges the grant PRIN-INAF 2012 (P.I. E. Flaccomio).
NJW acknowledges a Royal Astronomical Society Research Fellowship. 
JFAC is a researcher of CONICET and acknowledges their support. 
This work has made use of data from the European Space Agency (ESA) mission {\it Gaia} (\url{https://www.cosmos.esa.int/gaia}), processed by the {\it Gaia} Data Processing and Analysis Consortium (DPAC, \url{https://www.cosmos.esa.int/web/gaia/dpac/consortium}); funding for the DPAC has been provided by national institutions, in particular the institutions participating in the {\it Gaia} Multilateral Agreement.

\bibliographystyle{apj}
\bibliography{nbc.bib}
\appendix

\section{Accounting for uncertainties in evaluating likelihoods}
\label{s:errorbars}

Naive Bayes analyses allow for computing the probability of a class by evaluating the likelihood at an observed value.
However, observations often have large errors, and often the size of the error bars is comparable to the scale at which the likelihoods vary.
Point evaluations do not account for such uncertainties and can introduce a bias in the classification.
We have developed a method to incorporate measurement uncertainties into the classification probabilities.

We model likelihoods locally as 4$^{th}$-degree polynomials around the estimate of a given observed measure, and weight them with an assumed Normal error distribution to obtain an uncertainty weighted likelihood.
The likelihood $p(D|\theta)$, where $D$ represent the data, and $\theta$ is a parameter of interest, is typically a slowly varying function, especially in the context of Naive Bayes applications.
Around a specific value $\theta_0$, it can be expanded as a Taylor series,
\begin{equation}
p(D|\theta) \approx p(D|\theta_0) + \sum_{k=1}(\theta-\theta_0)^k \frac{\partial^kp(D|\theta)}{\partial\theta^k}\Bigg|_{\theta_0}
\end{equation}
It can thus be locally approximated as a $K^{th}$-degree polynomial,
\begin{equation}
\lk(x) = \sum_{k=0}^{K} a_k x^k
\end{equation}
where $x=\theta-\theta_0$ is the variable of interest.

\subsection{Quartic Polynomials}



Consider evaluations of $\lk(x)$ at various values surrounding $x=0$, $x={0, \pm\dx, \pm2\dx}$.
Suppose that it has the values $$\lk(-2\dx)=\lkm,~~ \lk(-\dx)=\lm,~~ \lk(0)=\lz,~~ \lk(+\dx)=\lp,~~ \lk(+2\dx)=\lkp.$$
A polynomial that goes through these five points has the form
\begin{eqnarray}
\lk(x) =&& a + b\,x + c\,x^2 + d\,x^3 + e\,x^4 \nonumber \\
\equiv  && A + \nonumber \\
        && B\,(x+2\dx) + \nonumber \\
        && C\,(x+2\dx)\,(x+\dx) + \nonumber \\
        && D\,(x+2\dx)\,(x+\dx)\,(x) + \nonumber \\
        && E\,(x+2\dx)\,(x+\dx)\,(x)\,(x-\dx) \,.
\label{e:ppt}
\end{eqnarray}
After some algebra, we obtain the coefficients as expressions of $\lk(\cdot)$,
\begin{eqnarray}
A &=& \lkm \nonumber \\
B &=& \frac{1}{\dx}\left( -\lkm + \lm \right) \nonumber \\
C &=& \frac{1}{2\dx^2}\left( \lkm - 2\,\lm + \lz \right) \nonumber \\
D &=& \frac{1}{6\dx^3}\left( -\lkm + 3\,\lm - 3\,\lz + \lp \right) \nonumber \\
E &=& \frac{1}{24\dx^4}\left( \lkm - 4\,\lm + 6\,\lz - 4\,\lp + \lkp \right)
\end{eqnarray}
and
\begin{eqnarray}
a &=& \lz \nonumber \\
b &=& \frac{1}{12\dx}\left[ \lkm - 8\lm + 8\lp - \lkp \right] \nonumber \\
c &=& \frac{1}{24\dx^2}\left[ -\lkm +16\lm -30\lz +16\lp -\lkp \right] \nonumber \\
d &=& \frac{1}{12\dx^3} \left[ -\lkm +2\lm -2\lp +\lkp \right] \nonumber \\
e &=& \frac{1}{24\dx^4} \left[ \lkm - 4\,\lm + 6\,\lz - 4\,\lp + \lkp \right] \,.
\label{e:a_k}
\end{eqnarray}

\subsection{Weighting by Normal}

We assume that the error distributions on the data points are Normal, of the form
\begin{equation}
\ee(x) = \frac{1}{\sig\sqrt{2\pi}} e^{-\frac{x^2}{2\sig^2}} \,.
\end{equation}
The weighted likelihood estimate is then
\begin{equation}
p = \int_{-\infty}^{+\infty}~dx~\lk(x)~\ee(x) \,,
\end{equation}
since the process can be thought of as averaging over an ensemble of observations, with $\ee(x)$ providing an importance weight for independent draws.

Noting that
$\int_{-\infty}^{+\infty}~dx~x^{2n}~f(x) = 2 \int_0^{\infty}~dx~x^{2n}~f(x)$
and
$\int_{-\infty}^{+\infty}~dx~x^{2n+1}~f(x) = 0$
for symmetric $f(x)$ and integer $n$, and using known calculations of the integral (see, e.g., Gradshteyn \& Ryzhik, Equations 3.461.2-3),
we obtain the uncertainty-weighted likelihood,
\begin{eqnarray}
\lk|_{\ee} &=& a + c\sig^2 + 3e\sig^4 \nonumber \\
&=& \lz + \frac{\sig^2}{24\dx^2}(-\lkm+16\lm-30\lz+16\lp-\lkp) \nonumber \\
&& + \frac{3\sig^4}{24\dx^4}(\lkm-4\lm+6\lz-4\lp+\lkp)
\end{eqnarray}

{Choosing the natural scale in the problem, $\dx = \sig/z$,
\begin{eqnarray}
\lk|_{\ee} &=& \frac{1}{24} [ \lkm(3z^4-z^2) + \lm(-12z^4+16z^2) + \lz(18z^4-30z^2+24) \nonumber \\
&& + \lp(-12z^4+16z^2) + \lkp(3z^4-z^2) ]
\end{eqnarray}
}

For $z=1$, the above reduces to
\begin{equation}
\lk|_{\ee} = \frac{1}{24}\left[ 2\lkm + 4\lm + 12\lz + 4\lp + 2\lkp \right] \,.
\end{equation}


Notice that these expressions have some desirable mathematical properties:
the error-weighted likelihood is positive definite;
the coefficients of each $x^k$ are symmetric in how $\lk(.)$ at $\pm\dx$ and $\pm2\dx$ are included;
if the likelihood function is flat, all coefficients of $x^k$ for $k>0$ vanish;
and finally, if the likelihood function is flat, $\lk|_{\ee} \equiv \lz$.

\section{Extinction}
\label{s:FF}

\subsection{Fukugita+ vs.\ Fitzpatrick \& Massa}

As discussed in \S\ref{s:extinction}, we primarily use the extinction law of 
\citet[F96]{Fukugita+96}
to compute $\Av$.
However, the extinction law based on the newer study of 
\citet[FM]{FitzpatrickMassa07,FitzpatrickMassa09}
is a viable alternative.
We do not use the latter as our primary reference because the peak of the distribution of $\Av$ is shifted lower by $\approx1.5^m$.
Here we consider the effect of changing the extinction law on the classification.
In Table~\ref{t:cmpFuFM}, we show how many sources change their Naive Bayes based classification based on this change.
We then carry out the same analysis as in \S\ref{s:reclassify}, reclassifying sources manually, and show how many sources are reclassified in Table~\ref{t:posthocFM}.
In both cases, we find a similar fraction of changes, suggesting that our statistical classification is effectively at the limit defined by potential systematic errors in the datasets.

\begin{deluxetable}{lccc|c}
\tablecaption{Changes in Naive Bayes classification upon changing extinction law from 
F96 
to 
FM 
\label{t:cmpFuFM}
}
\tablehead{ \colhead{F96} & \colhead{\hfil} & \colhead{FM} & \colhead{\hfil} & \colhead{subtotal F96}}
\startdata
\hfil & Foreground & Member & Background & \hfil \\
Foreground & 610 & 7 & 2 & 619 \\
Member & 70 & 5740 & 124 & 5934 \\
Background & 3 & 83 & 1381 & 1467 \\
\hline
subtotal FM & 683 & 5830 & 1507 & 8020 \\
\enddata
\end{deluxetable}

\begin{deluxetable}{lccc|c}
\tablecaption{Post-hoc Reclassifications (as in Table~\ref{t:posthoc}) but based on 
FM 
extinction law
\label{t:posthocFM}
}
\tablehead{ \colhead{Original} & \colhead{\hfil} & \colhead{Reclassified} & \colhead{\hfil} & \colhead{subtotal original}}
\startdata
\hfil & Foreground & Member & Background & \hfil\\
Foreground & 491 & 192 & 0 & 683 \\
Member & 60 & 5768 & 2 & 5830 \\
Background & 0 & 110 & 1397 & 1507 \\
\hline
subtotal Reclassified & 551 & 6070 & 1399 & 8020 \\
\enddata
\end{deluxetable}

\begin{figure*}[htb!]
\includegraphics[width=3.1in]{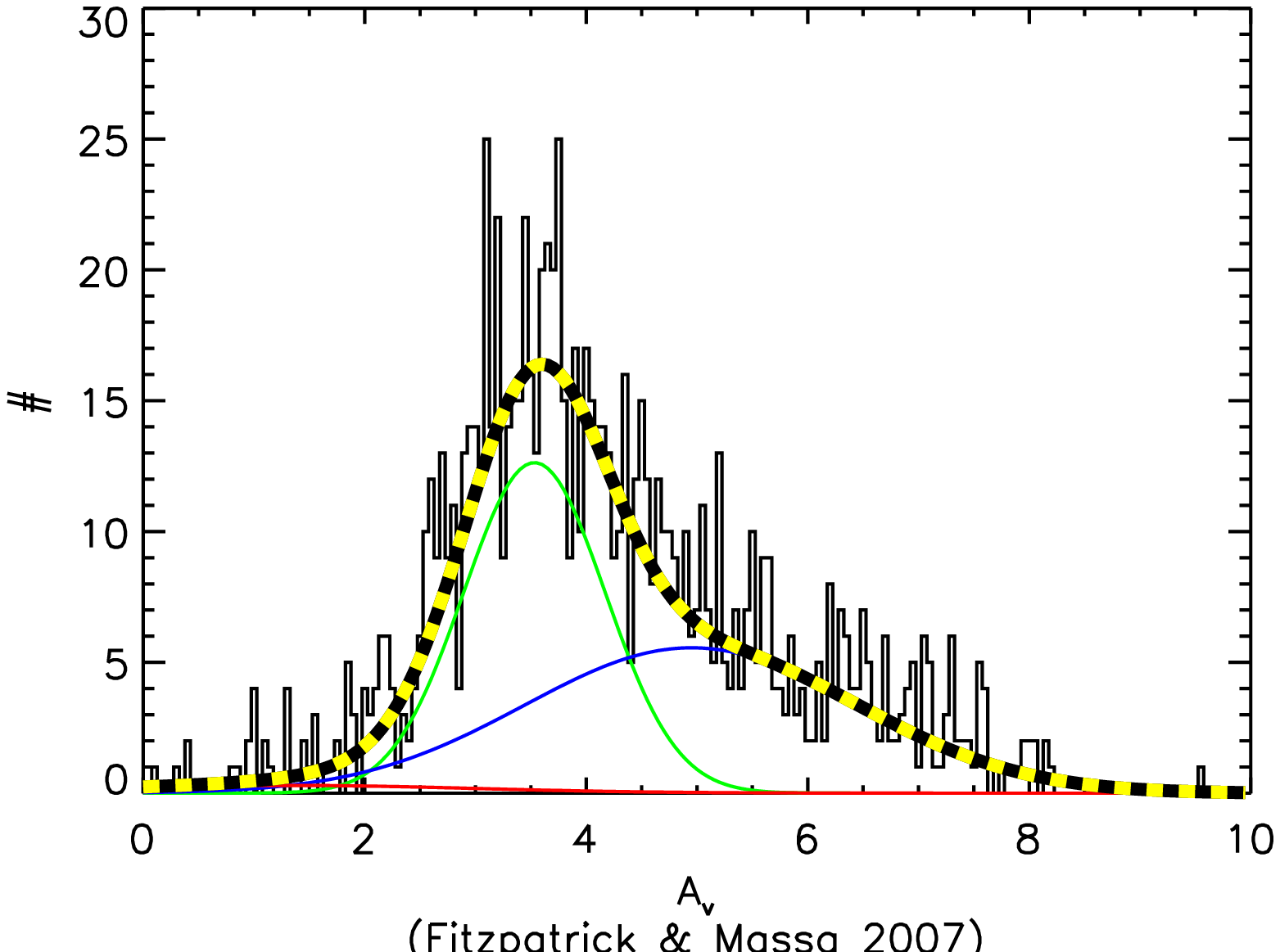}
\includegraphics[width=3.1in]{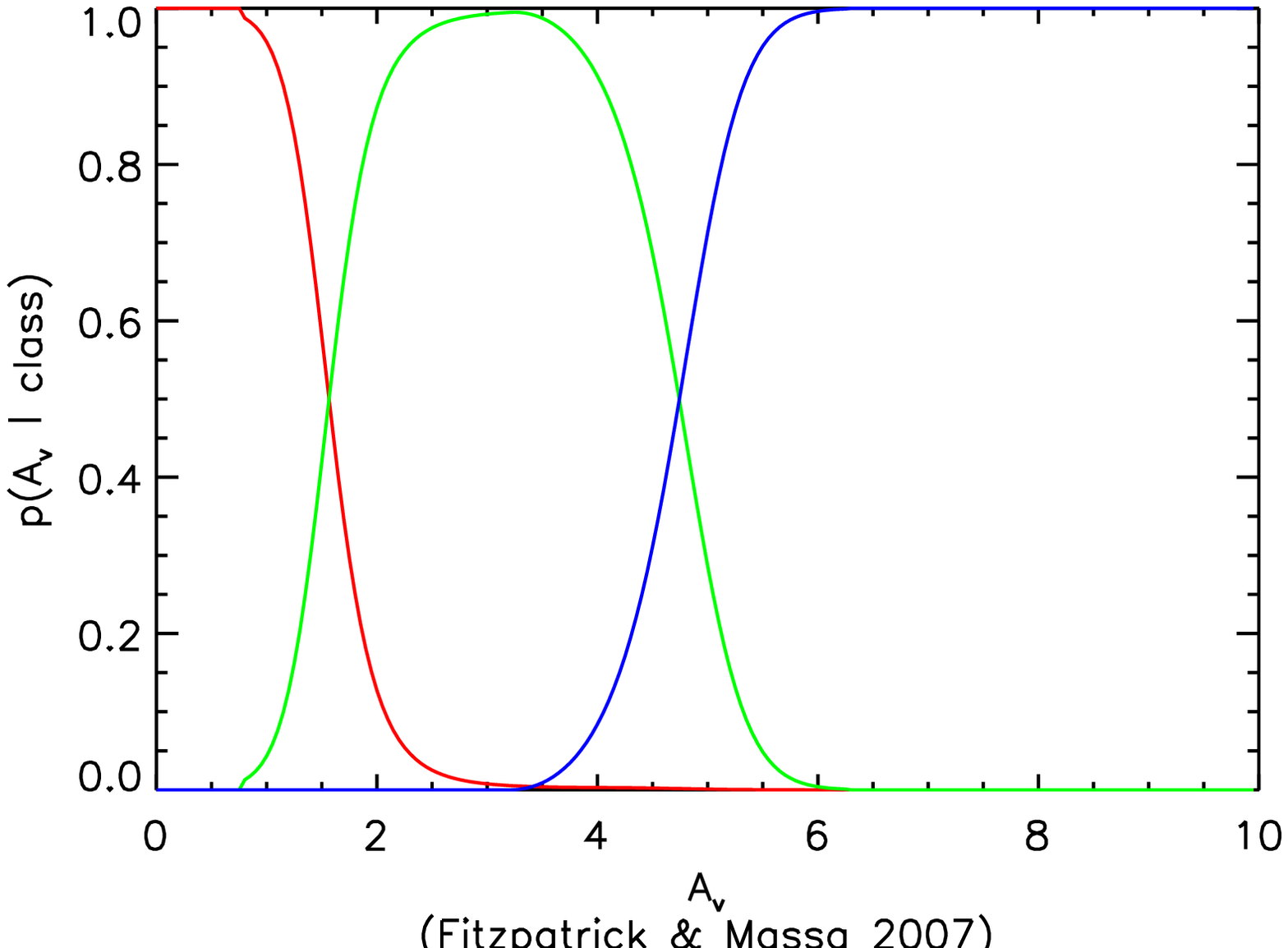}
\caption{
As in the first two figures of Figure~\ref{f:bestAv}, but for extinctions derived from 
FM 
}
\label{f:bestAvFM}
\vskip 0.5cm
\end{figure*}

\subsection{$\Av$ across the FOV}

Individual extinction for stars associated with \cygob\ is calculated with a similar approach of Guarcello et al. 2012.
In that paper, individual extinction of the X-ray sources with optical counterpart from the OSIRIS or SDSS catalogs is calculated from the displacement along the extinction vector from the A$_V$=0 3.5$\,$Myrs isochrone from Siess et al. 2000 in the $r-i$ vs. $i-z$ diagram.
This method is feasible thanks to the almost monotonic shape of the isochrone in this color space, but it requires the use of suitable color transformations from the Johnson-Cousin $UBVRI$ to the $\up\,\gp\,\rp\,\ip\,\zp$ photometric system.
\citet{Guarcello+12}
adopted the transformations from 
F96.  \par 

In this paper, in order to avoid the use of any photometric transformation, we adopted the MIST isochrones which are provided in several photometric systems, one of which is $\up\,\gp\,\rp\,\ip\,\zp$\footnote{MIST isochrones can be downloaded from http://waps.cfa.harvard.edu/MIST/index.html}.
Besides, these isochrones span a wider range of stellar mass, allowing the calculation of individual extinction also for stars more massive than 7$\,$M$_\odot$, which is the upper mass limit in the \citet{Siess+00} isochrones.
Additionally, in this paper we calculate the individual extinction of those 2MASS/UKIDSS sources without a good optical counterpart using the NICER method 
\citep{Lombardi.Alves:2001},
based on the $H-K$ color. \par  

\begin{figure*}[htb!]
\centering
\includegraphics[width=3.2in]{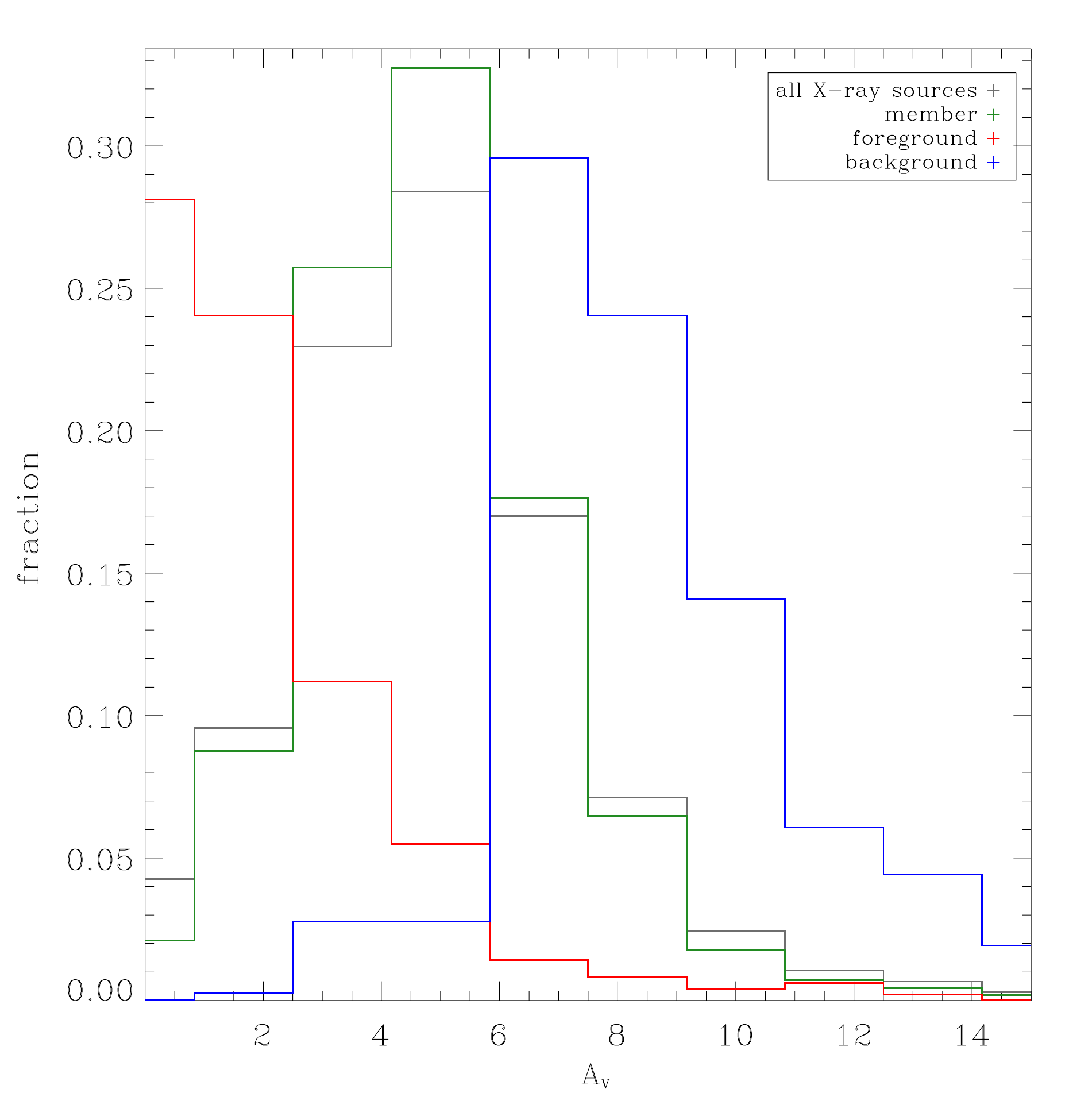}
\includegraphics[width=3in]{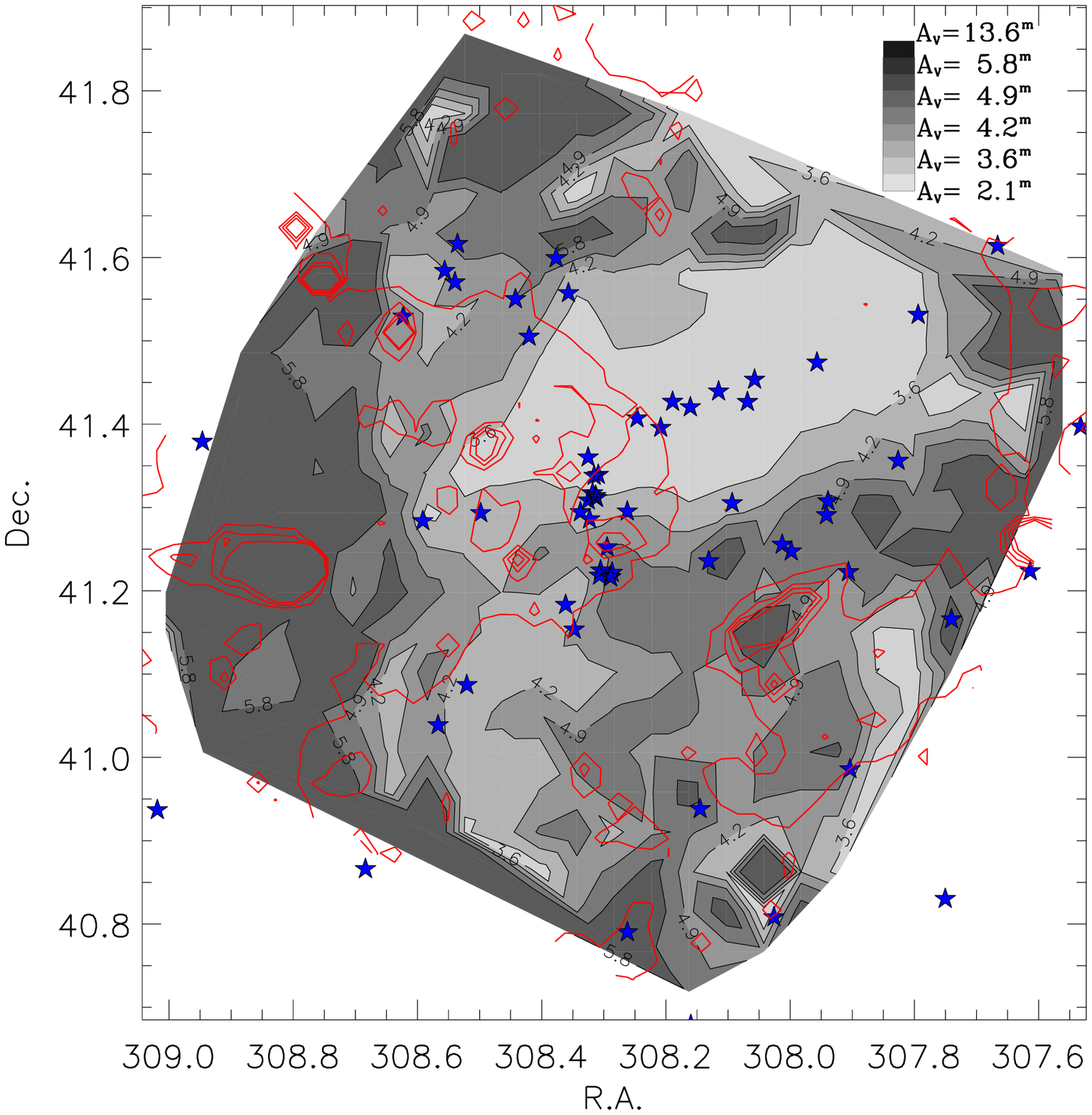}
\caption{{\sl Left:} Distribution of the individual extinctions of the sources in the catalog. 
{\sl Right:} Map of stellar extinction across the area of the survey.
The red contours mark the continuum emission levels at 8.0$\, \mu$m.
The blue stars mark the positions of known O stars.
}
\label{f:mario_av}
\vskip 0.5in
\end{figure*}

The left panel in Fig. \ref{f:mario_av} shows the distribution of the resulting individual extinctions for the X-ray sources associated with \cygob.
The median value is 4.2$^m$, with the 10\% quantile equal to 2.7$^m$ and the 90\% quantile to 7.9$^m$, quite similar to previous estimate 
\citep[e.g.,][]{Drew+08,Sale+09,Wright+10}.
The right panel shows the spatial map of extinction across the area of the survey, comparing it with the level of continuum emission at 8.0$\, \mu$m which mark the dust emission.
The well known low extinction region at north-west is evident, as well as the large extinction regions corresponding to some of the dusty structures in the cloud. 

\section{X-ray sources with no OIR matches}
\label{s:nomatch}


A detailed description of the matching procedure to determine OIR counterparts to the X-ray sources is given in \citet{Guarcello+15a}.
For the 1428 X-ray sources classified here as members but have no counterparts, \citet{Guarcello+15a} list the nearest match in a supplemental table.
Here, we consider some properties of this population.

\begin{figure*}[]
\includegraphics[width=3.5in]{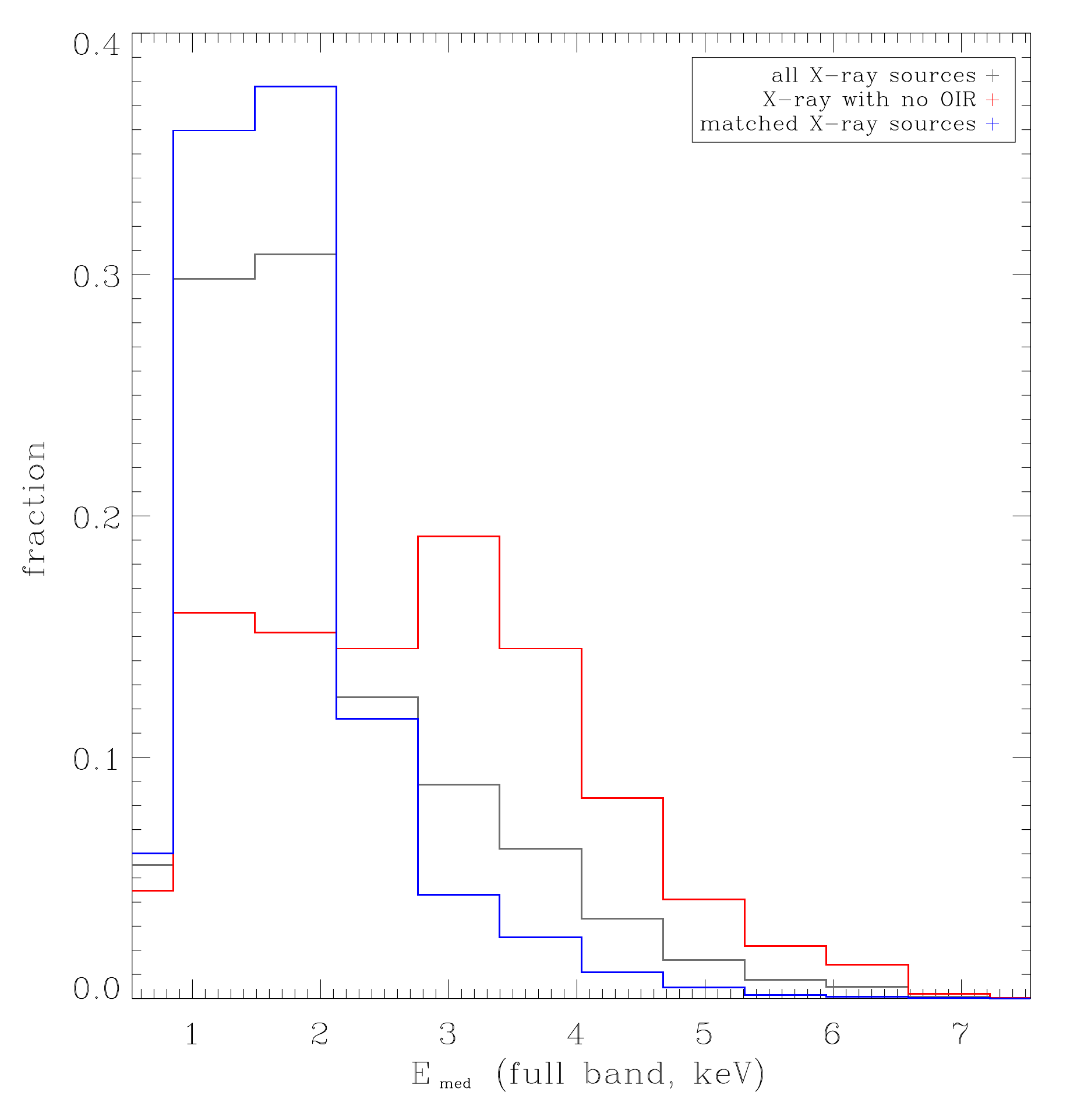}
\includegraphics[width=3.5in]{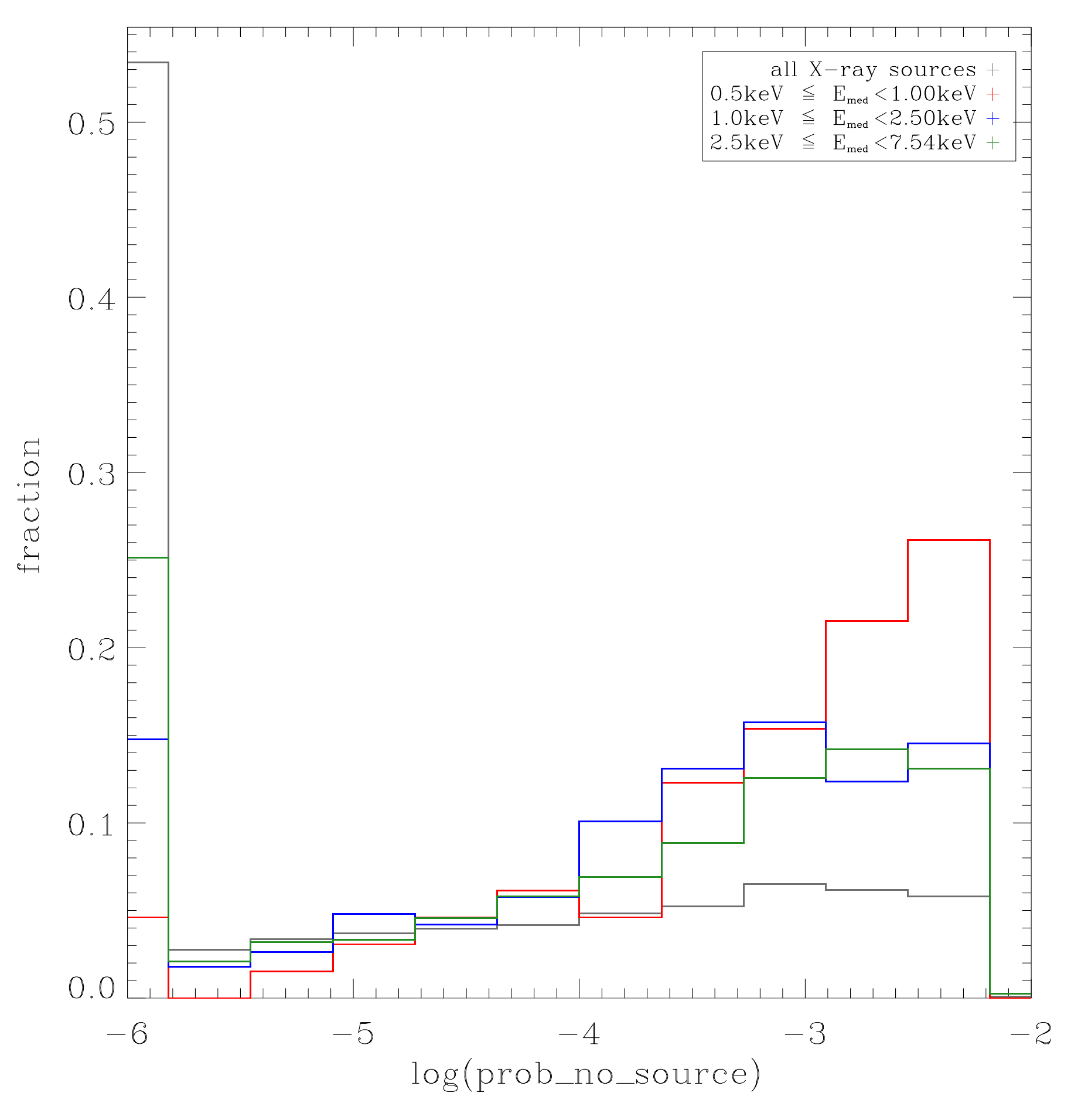}
\caption{{\sl Left:} Distributions of the median energy of the detected photons for all the X-ray sources detected in our survey (grey), for those matched with OIR counterparts (blue), and the X-ray sources without OIR counterpart (red).
{\sl Right:} Distribution of the ACIS Extract parameter {\it prob-no-source} for the X-ray sources with no OIR counterpart separated according to their median photon energies, with soft ($0.5-1$~keV; red), medium ($1-2.5$~keV; {blue}), hard ($2.5-7.5$~keV; green), and all ({black}).}
\label{f:medE}
\vskip 0.5in
\end{figure*}

The presence of different populations in the sample of the X-ray sources without OIR counterpart is evident by looking at the distribution of their photon median energy and the parameter {\it prob-no-source} evaluated by ACIS Extract 
\citep{Broos+10}.
The latter parameter indicates the reliability of the source in terms probability that it is a background fluctuation.
The distribution of the photons' median energy of the X-ray sources with no OIR counterpart is clearly different (see Figure~\ref{f:medE}) than that of the whole sample, being almost flat while the latter peaks at about 1.8$\,$keV.
The high energy tail of the X-ray sources with no OIR counterpart is due to background sources.
The nature of the soft sources in this sample can be investigated by looking at the distribution of the prob-no-source parameter.
For soft X-ray sources with no OIR counterpart (red distribution) the distribution of this parameter clearly peaks at high values, while the low-value bins are populated by sources with higher median photon energy, suggesting that the soft X-ray sources with no OIR counterpart are mainly candidate spurious sources, while those with higher photon energy are mainly genuine detections.

\begin{figure*}[]
\includegraphics[width=4.0in]{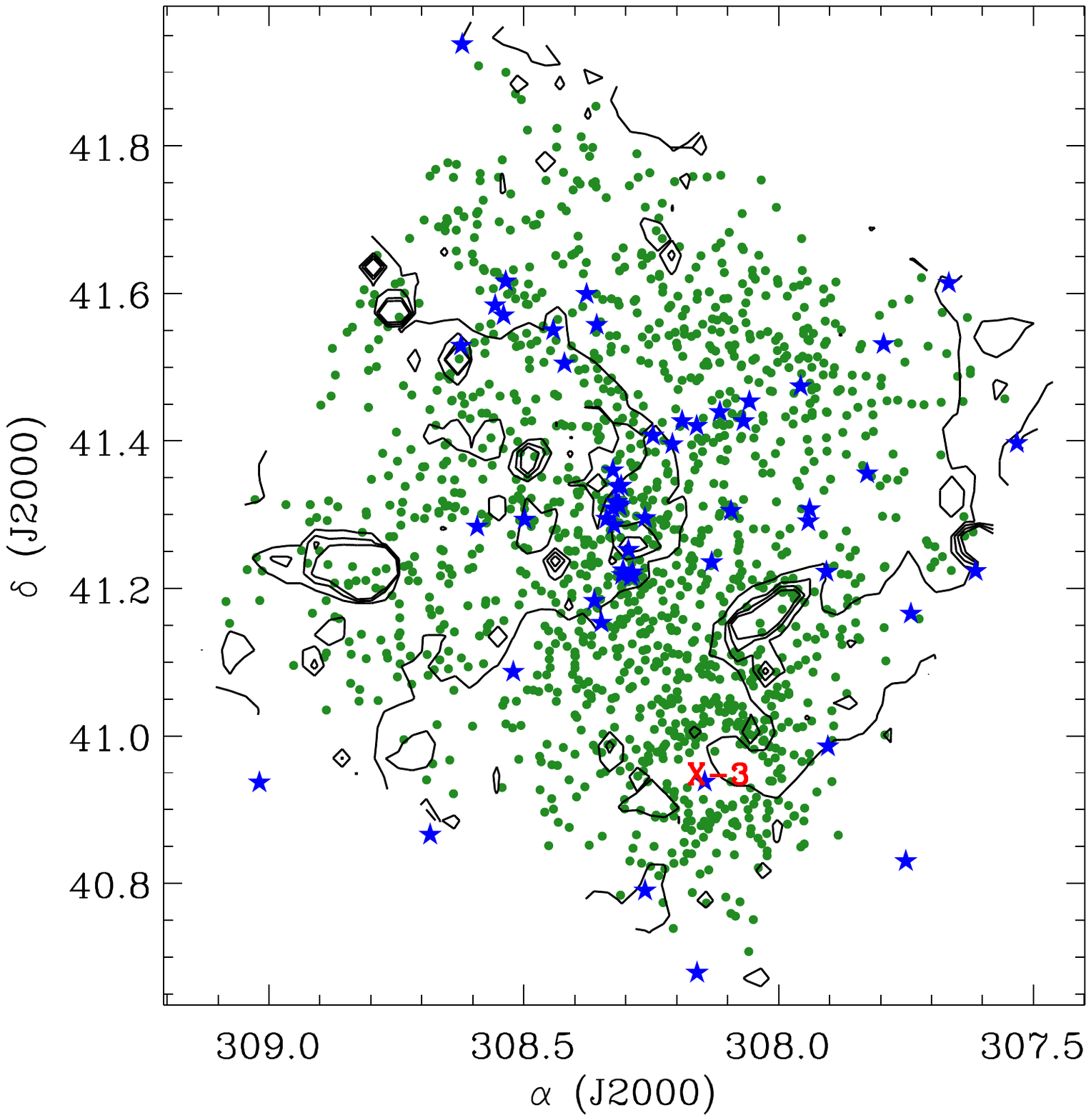}
\includegraphics[width=4.0in]{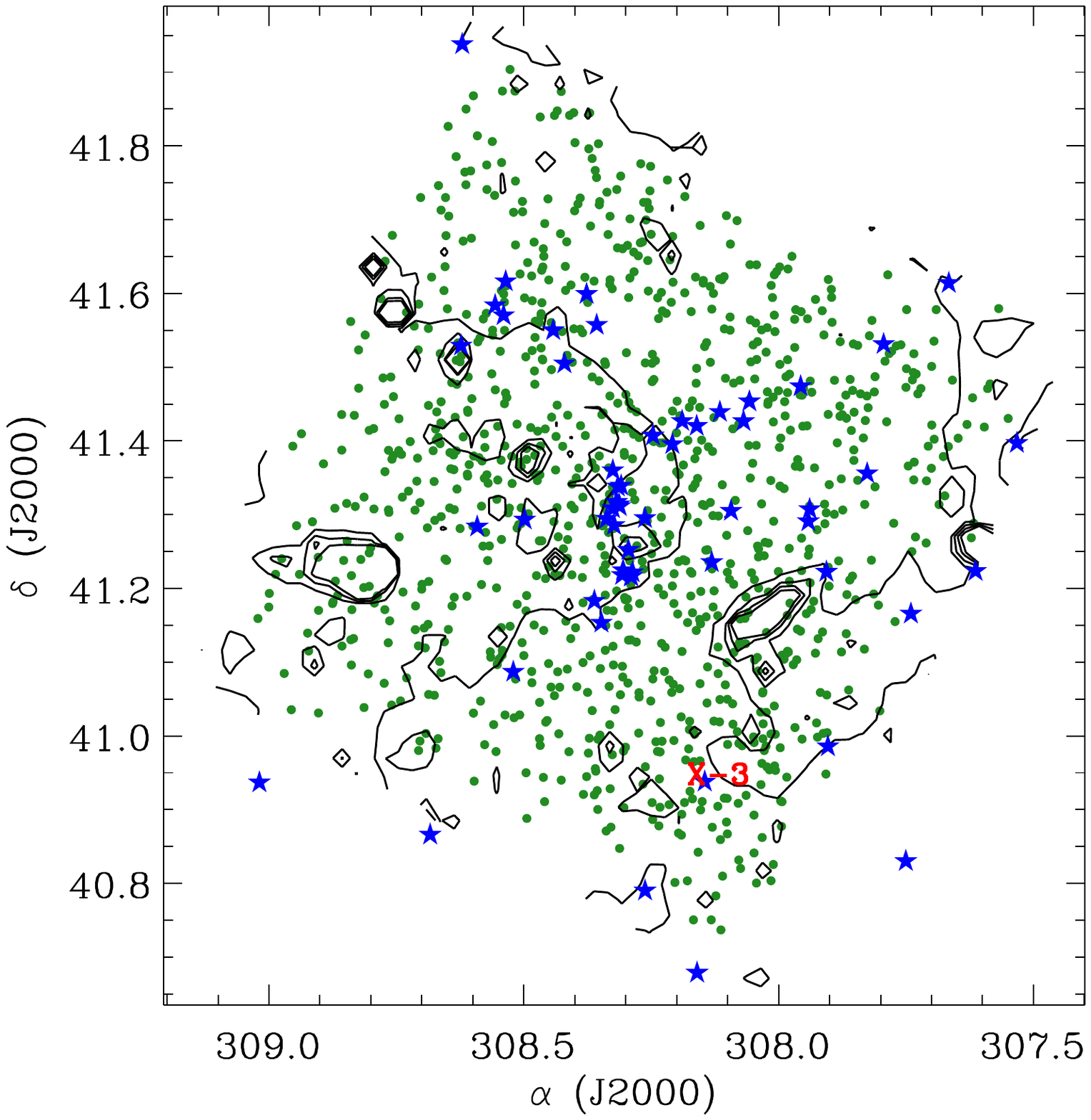}
\caption{Spatial distributions of the X-ray sources with no OIR counterpart classified as members (left panel) and background sources (right panel). In each panel, the blue stars mark the positions of the known O stars; the contours mark the background brightness at {8.0~$\mu$m}, tracing the dust emission. The position of Cygnus~X-3 is also indicated.}
\label{f:spadis}
\vskip 0.5in
\end{figure*}

Figure~\ref{f:spadis} shows the spatial distributions of the X-ray sources without OIR counterparts classified as members and those classified as sources separately. While the latter are more uniformly distributed, the candidate members show some level of clustering corresponding to various Cygnus~OB2 subclusters, as expected for real background and clusters sources. Both samples also show a clear halo of sources around the position of Cygnus X-3, which are more likely spurious X-ray sources that survived the pruning process of the X-ray catalog.

To further verify the nature of the X-ray sources with no OIR counterparts that are classified as members, we have inspected their positions in various optical and infrared diagrams and the optical images of the closest source in the OIR catalog of each of these sources, selecting 46 candidate false negatives produced in the match between the X-ray and the OIR catalogs.
Table~\ref{t:xnooir} lists their CXO-IDs and the separation in arcsec from the closest OIR source.
Note that some of these sources have an OIR star closer than 1 arcsec.
The listed sources are divided into four categories.
In four cases the closest OIR source is a known star with disk classified by 
\citet{Guarcello+13}.
In the remainder the closest OIR source falls into the loci defined by Cyg~OB2 members in the various diagrams.
Five stars have $\jj<12^m$ (candidate bright members), 19 sources have $13^m<\jj<16^m$ (candidate members), and 18 sources have $16^m<\jj<19^m$ (candidate low mass or highly extinguished members).
In particular, the OIR source 0.1 arcsec from the X-ray source 3532 is compatible with being a member in infrared but not in optical, likely being a false coincidence between a low-extinction optical source and a high-extinction IR source; the X-ray source 4675 is close to The O8.5V star MT91-8D, which has been matched with the X-ray source 4673.

\begin{deluxetable}{lc}
\tablecaption{Candidate X-ray vs.\ OIR false negatives
\label{t:xnooir}}
\tablehead{ \colhead{CXO~ID} & \colhead{Separation (")} }
\startdata
\multicolumn{2}{c}{stars with disks} \\
\hline
      56   & 1.9 \\
    2327   & 2.4 \\
    3099   & 1.6 \\
    3624   & 2.2 \\
\hline
\multicolumn{2}{c}{candidate OIR bright members} \\
\hline
     796   & 2.2 \\
    3692   & 1.9 \\
    4602   & 1.7 \\
    4675   & 1.9 \\
    7115   & 1.2 \\
\hline
\multicolumn{2}{c}{candidate members} \\
\hline
     121   & 3.2 \\
    1791   & 2.1 \\
    1822   & 1.8 \\
    2214   & 2.3 \\
    2397   & 1.2 \\
    2788   & 0.6 \\
    2910   & 5.4 \\
    3056   & 1.5 \\
    3141   & 1.8 \\
    3185   & 1.5 \\
    3759   & 1.0 \\
    3942   & 0.6 \\
    4590   & 2.8 \\
    5130   & 5.1 \\
    5430   & 1.8 \\
    5442   & 0.2 \\
    6905   & 2.0 \\
    7655   & 2.7 \\
    7699   & 3.1 \\
\hline
\multicolumn{2}{c}{candidate faint members} \\
\hline
     102   & 2.3 \\
     222   & 0.8 \\
    2558   & 2.5 \\
    2850   & 0.6 \\
    3006   & 5.1 \\
    3118   & 4.2 \\
    3346   & 6.5 \\
    3532   & 0.1 \\
    3838   & 2.4 \\
    4005   & 0.7 \\
    4835   & 1.6 \\
    5573   & 0.6 \\
    6167   & 2.5 \\
    6518   & 1.9 \\
    6593   & 1.4 \\
    6681   & 0.9 \\
    6861   & 0.5 \\
    7201   & 3.0 \\ 
\enddata
\end{deluxetable}

\section{Adopted Likelihoods}

\begin{figure*}[htb!]
\includegraphics[width=2.5in]{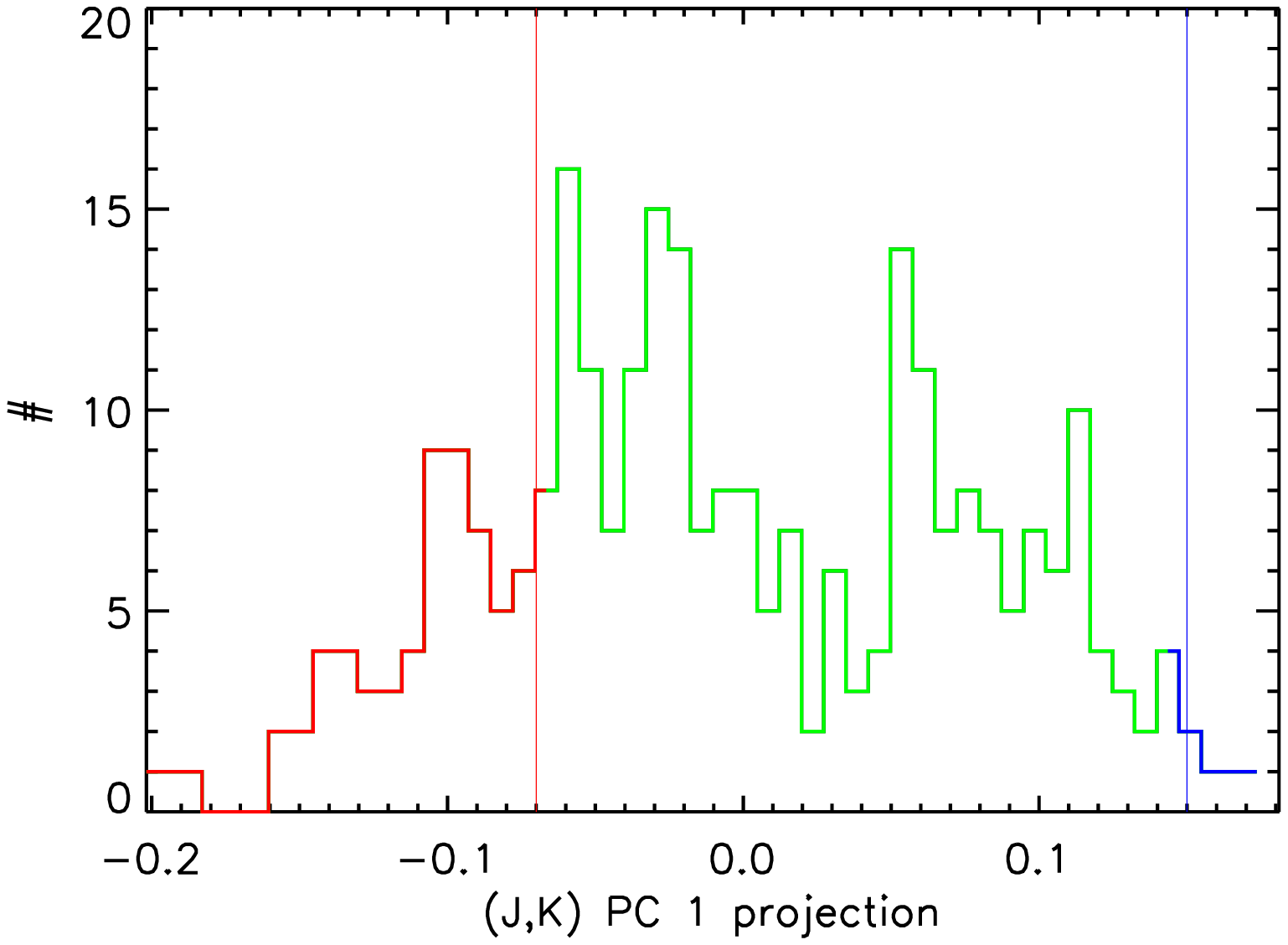}
\includegraphics[width=2.5in]{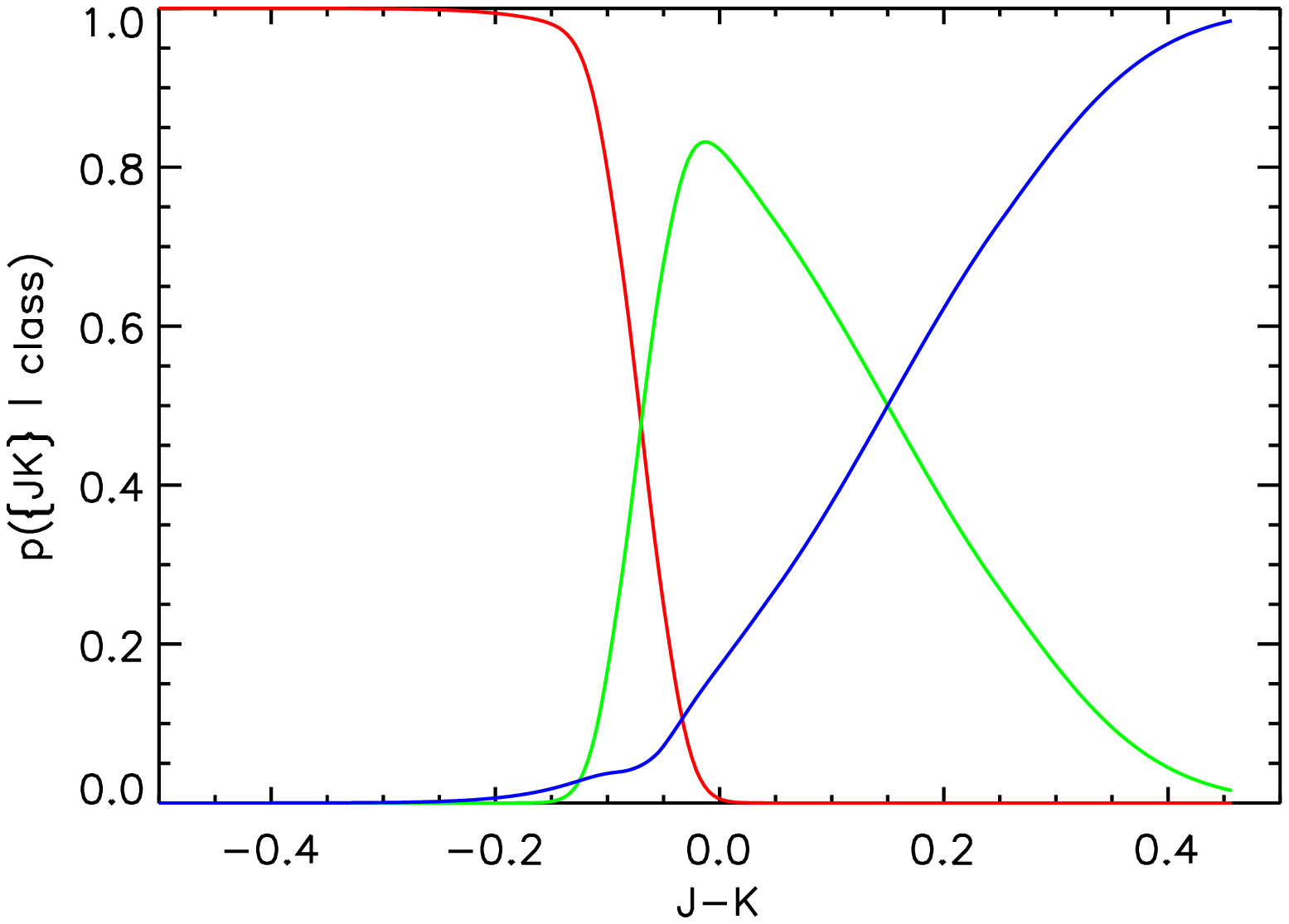}
\includegraphics[width=2.5in]{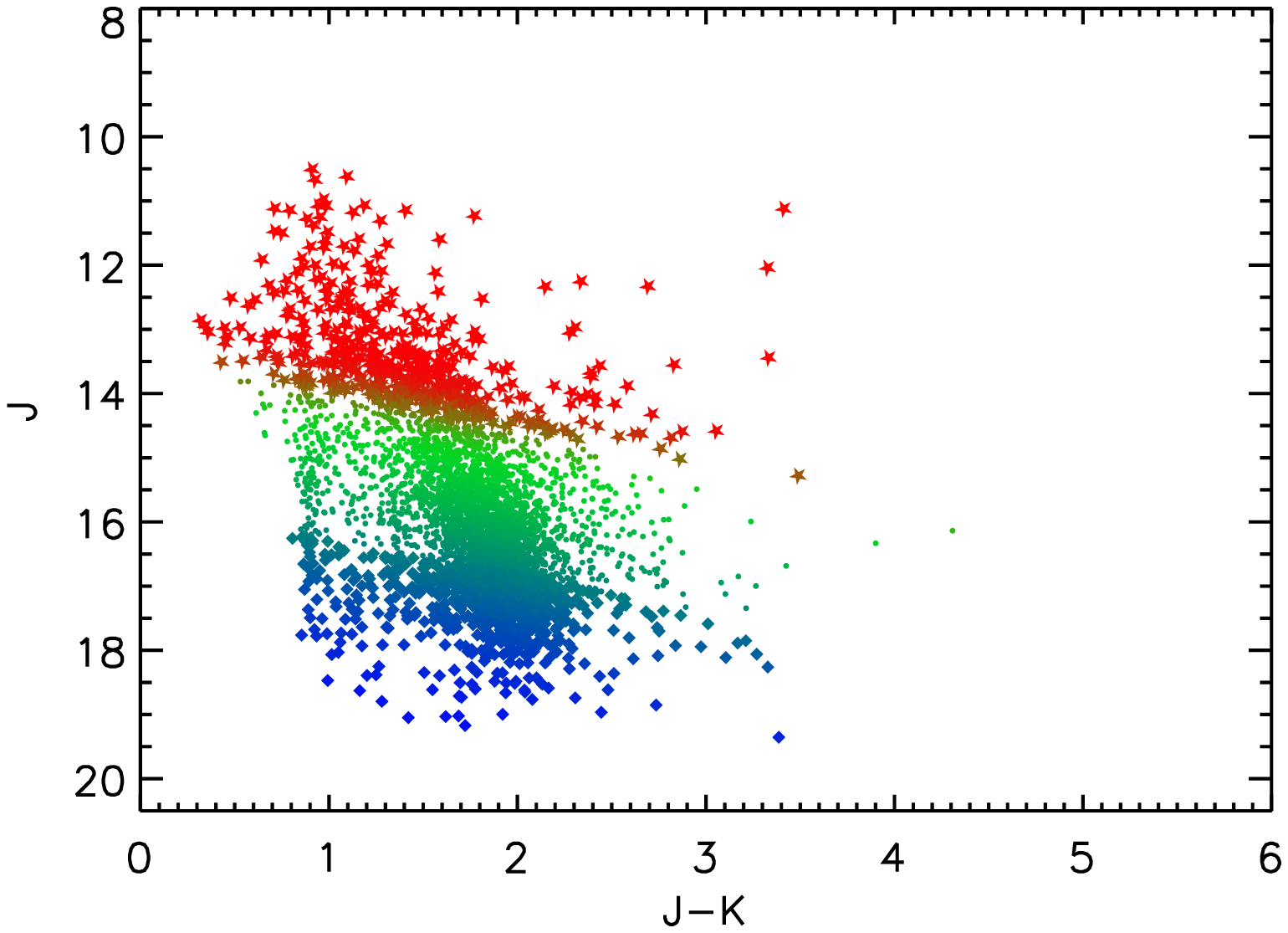}
\caption{
As Figure~\ref{f:riz2pc}, for the $1^{st}$ Principal Component of $\{\jj,\kk\}$.
}
\label{f:JK1pc}
\vskip 0.5in
\end{figure*}

\begin{figure*}[htb!]
\includegraphics[width=2.5in]{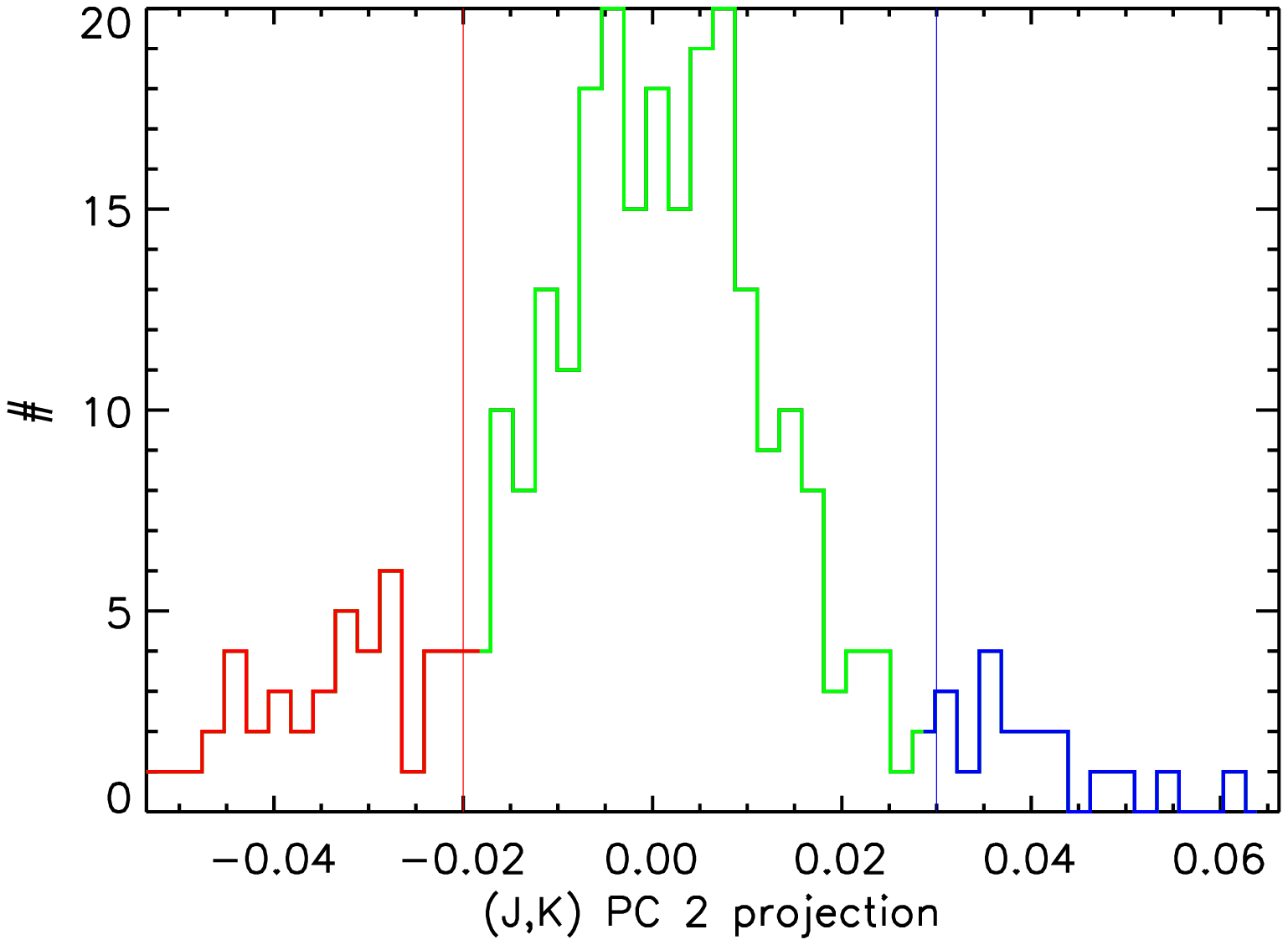}
\includegraphics[width=2.5in]{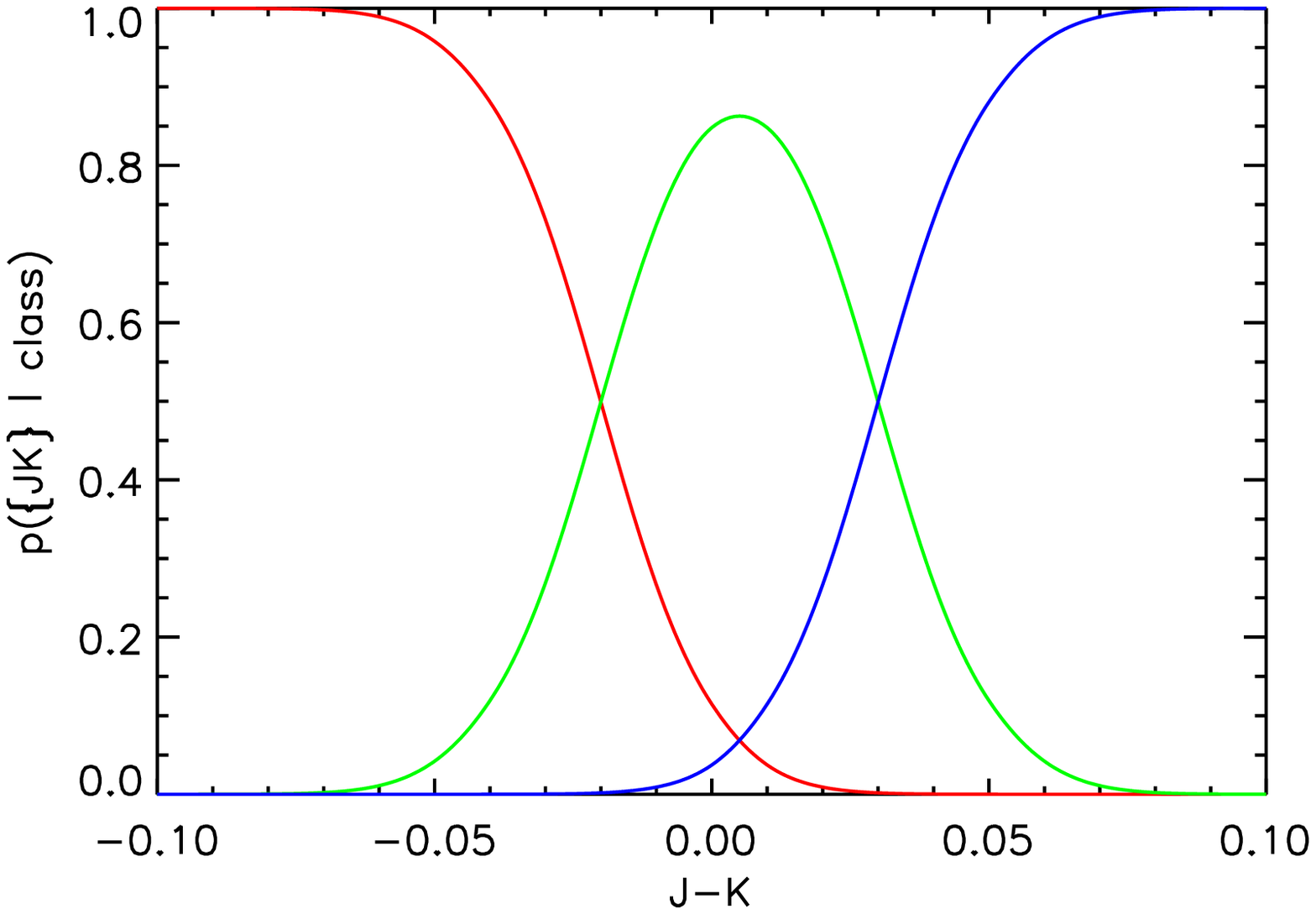}
\includegraphics[width=2.5in]{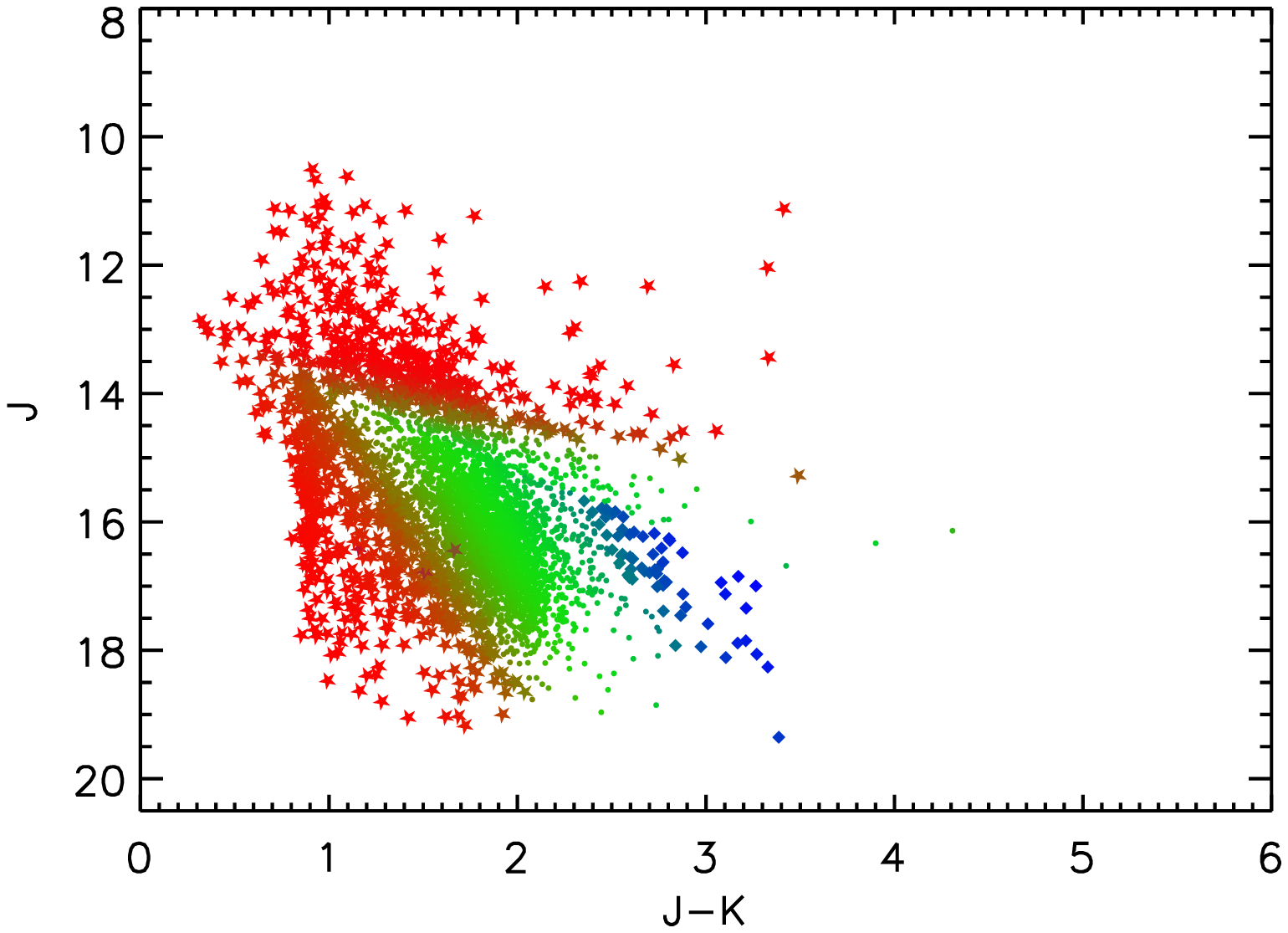}
\caption{
As Figure~\ref{f:riz2pc}, for the $2^{nd}$ Principal Component of $\{\jj,\kk\}$.
Additionally, during the classification ({\sl right panel}), the higher of the foreground likelihoods between the $1^{st}$ (Figure~\ref{f:JK1pc}) and $2^{nd}$ components is chosen.
}
\label{f:JK2pc}
\vskip 0.5in
\end{figure*}

\begin{figure*}[htb!]
\includegraphics[width=2.5in]{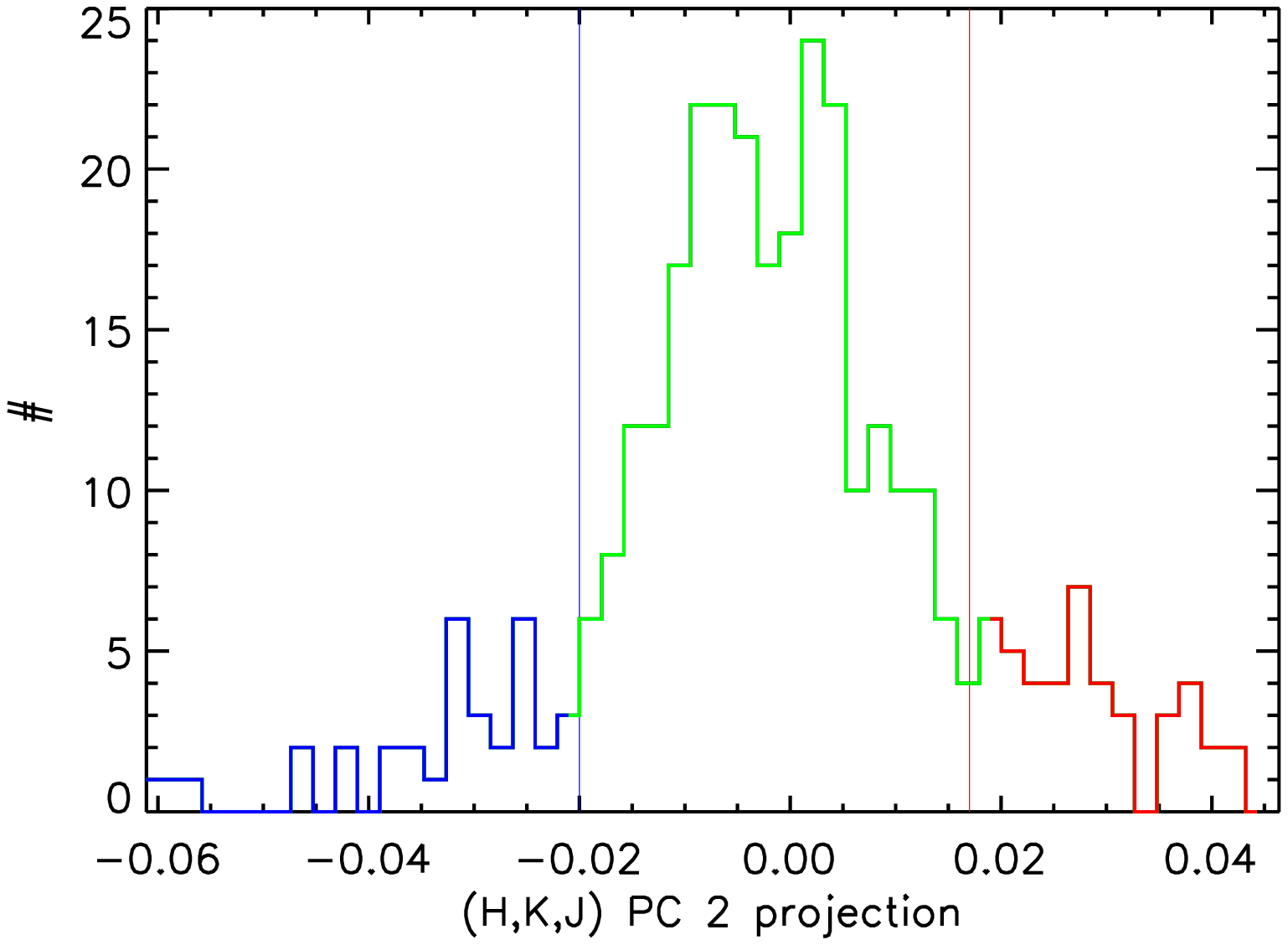}
\includegraphics[width=2.5in]{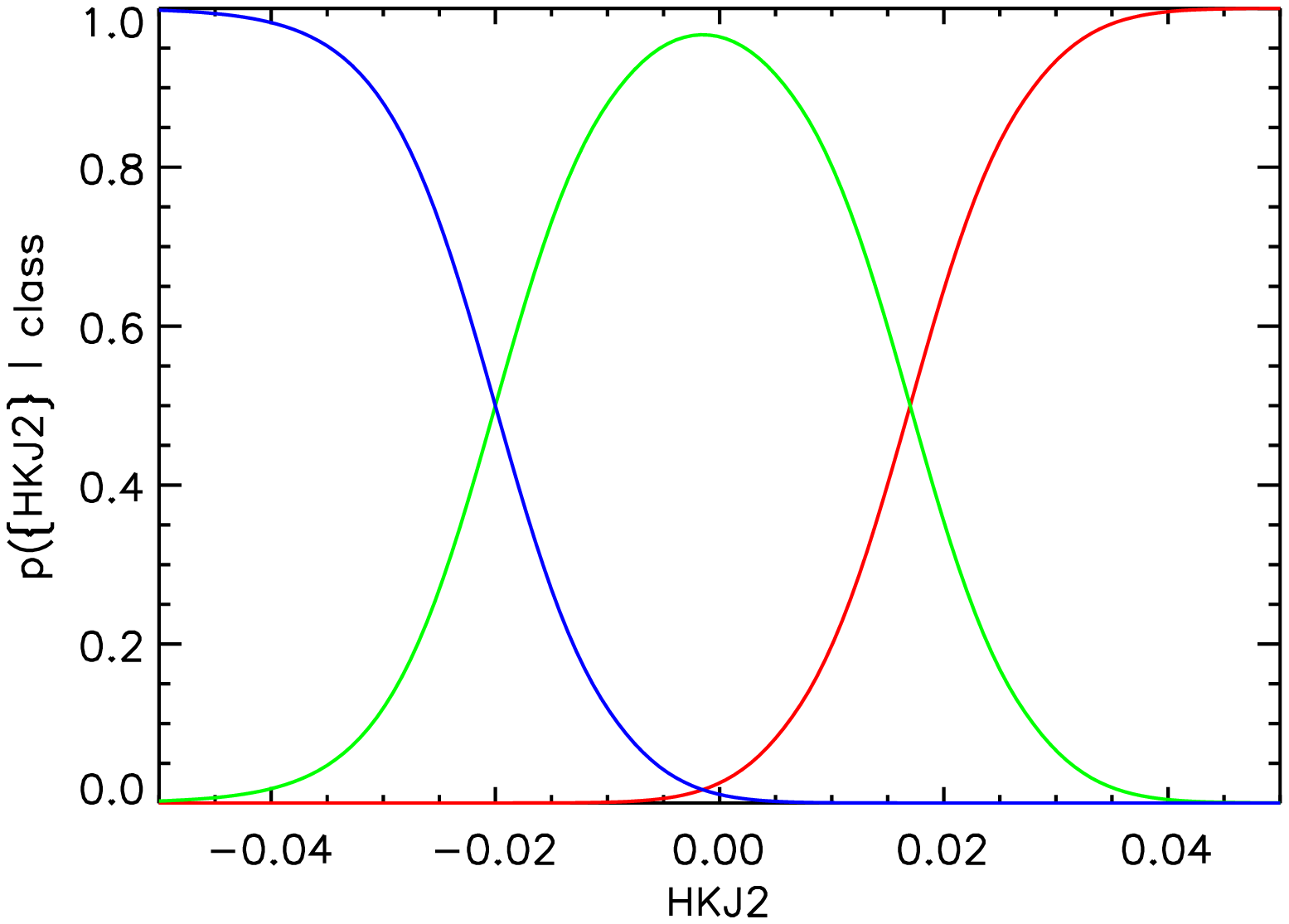}
\includegraphics[width=2.5in]{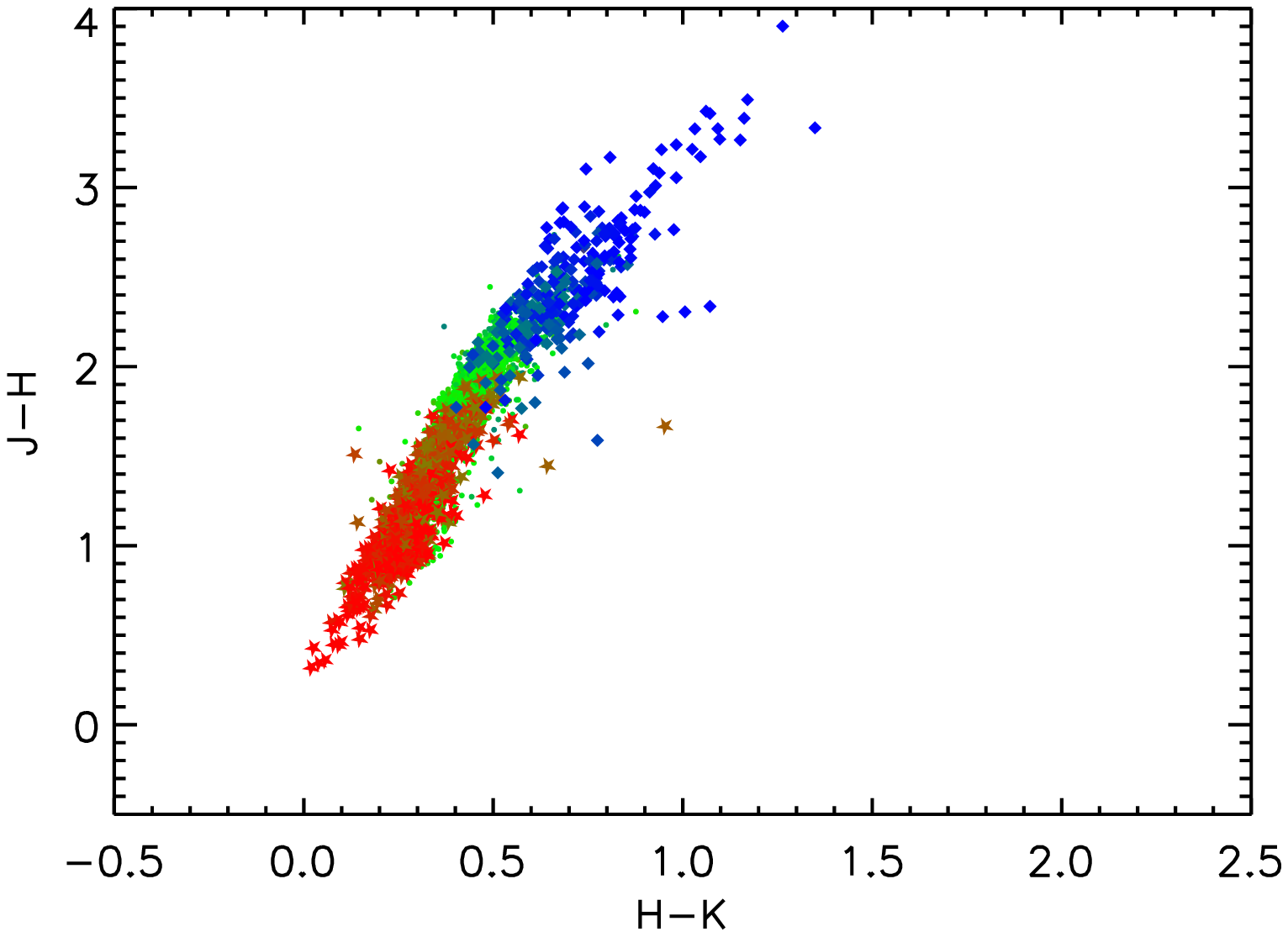}
\caption{
As Figure~\ref{f:riz2pc}, for the $2^{nd}$ Principal Component of $\{\hh,\jj,\kk\}$.
}
\label{f:HJK2pc}
\vskip 0.5in
\end{figure*}

\begin{figure*}[htb!]
\includegraphics[width=2.5in]{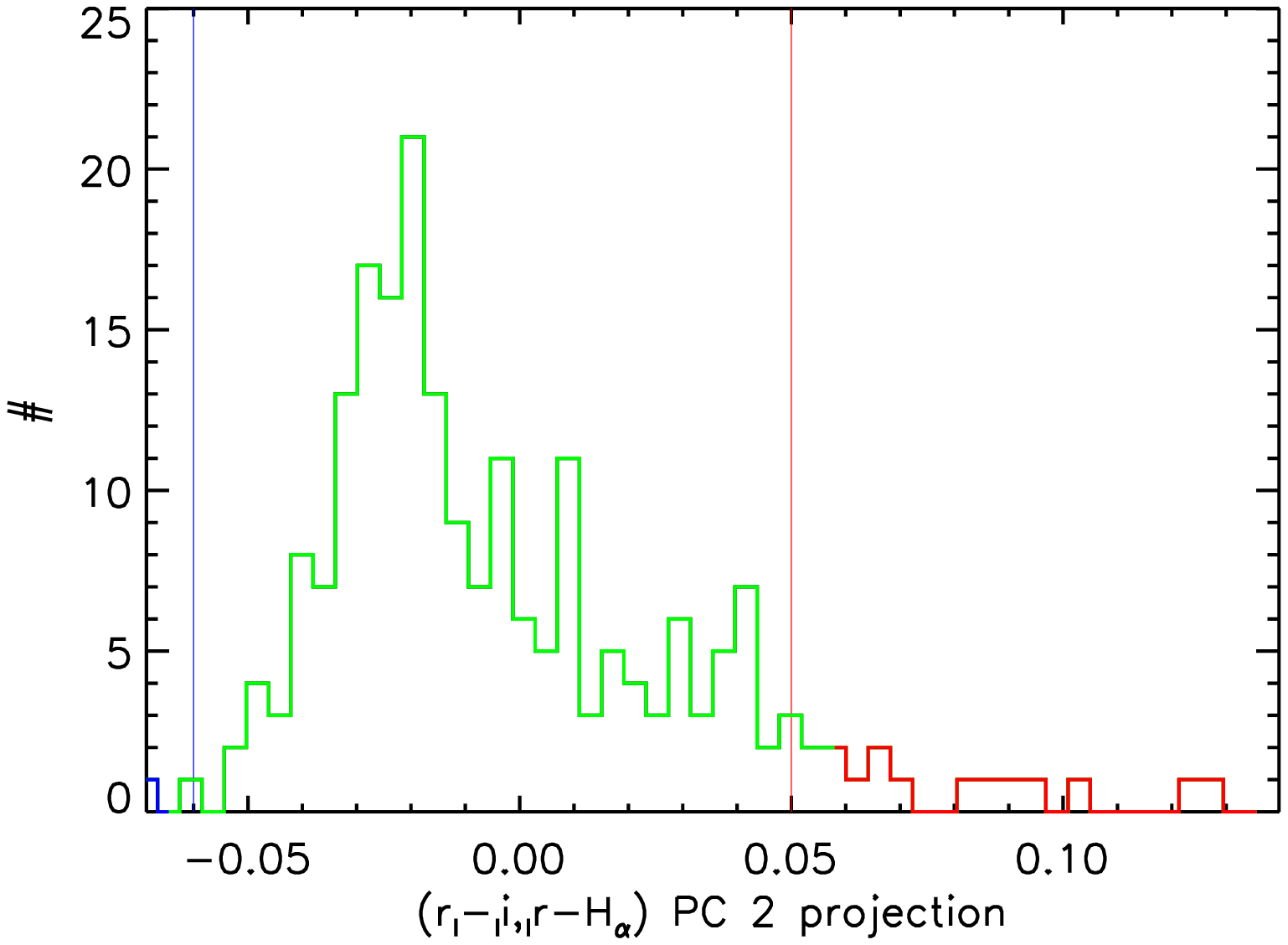}
\includegraphics[width=2.5in]{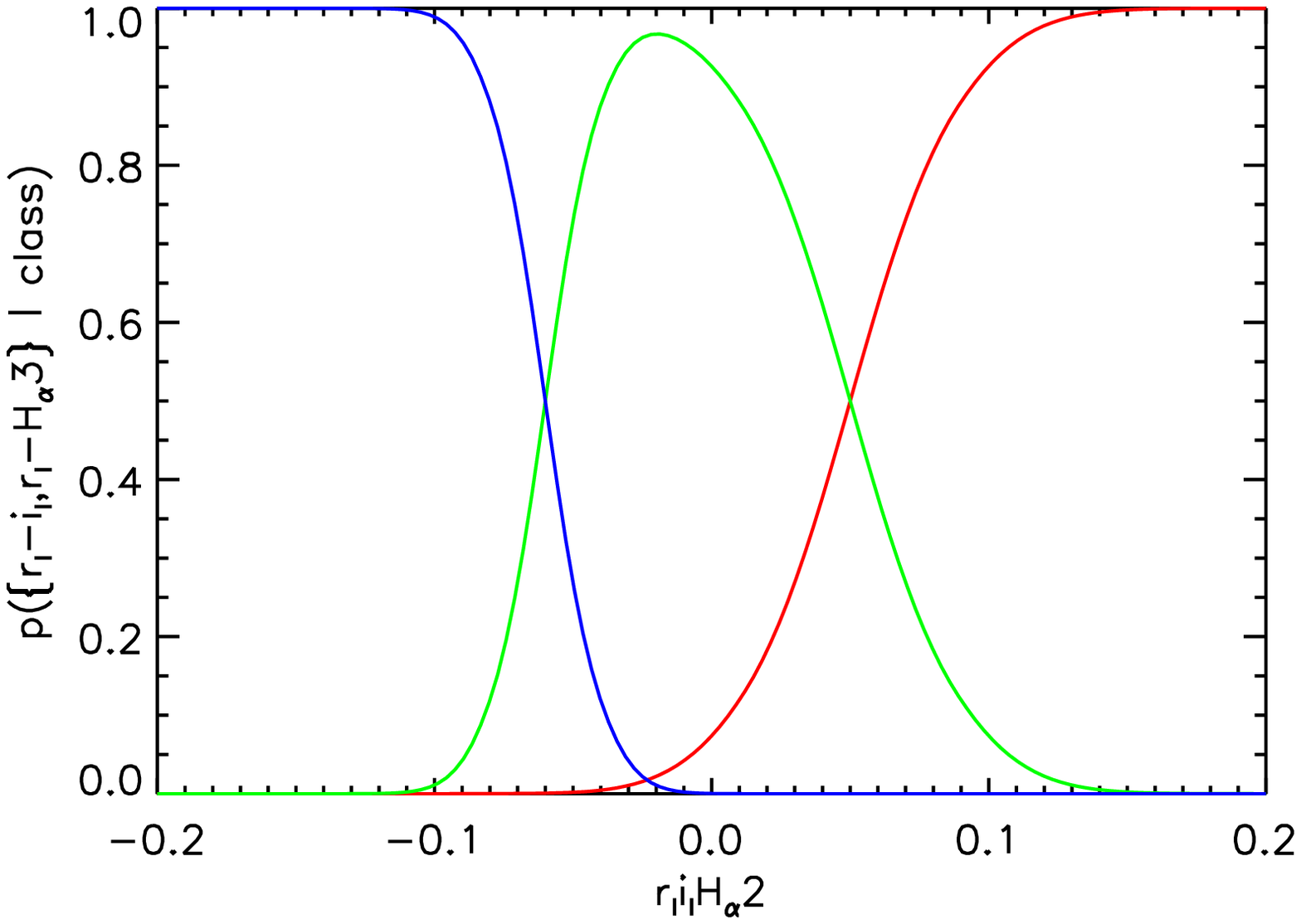}
\includegraphics[width=2.5in]{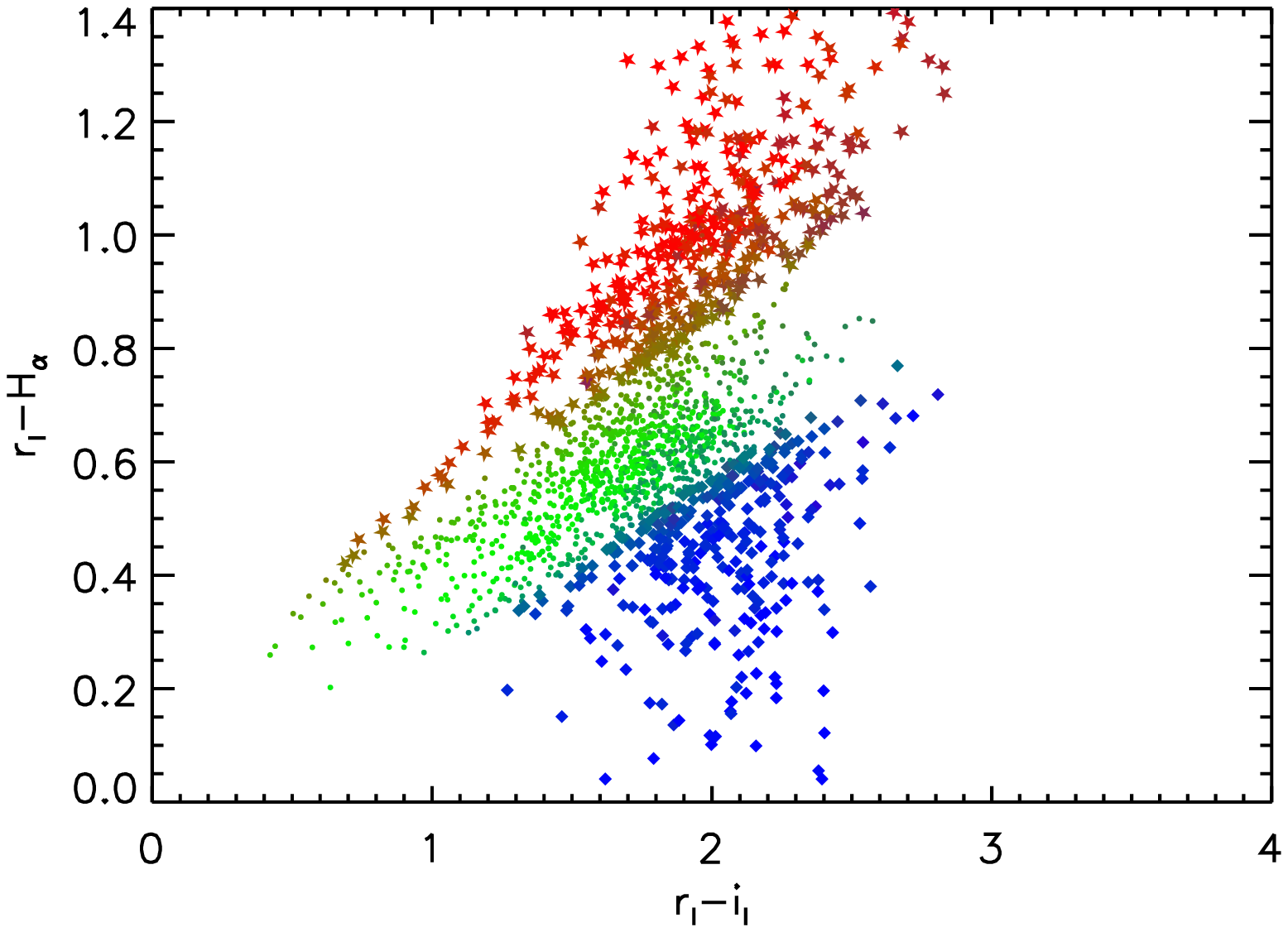}
\caption{
As Figure ~\ref{f:riz2pc}, for the $2^{nd}$ Principal Component of $\{\rp-\ip,\rp-\Ha\}$.
}
\label{f:riHa2pc}
\vskip 0.5in
\end{figure*}

\begin{figure*}[htb!]
\includegraphics[width=2.5in]{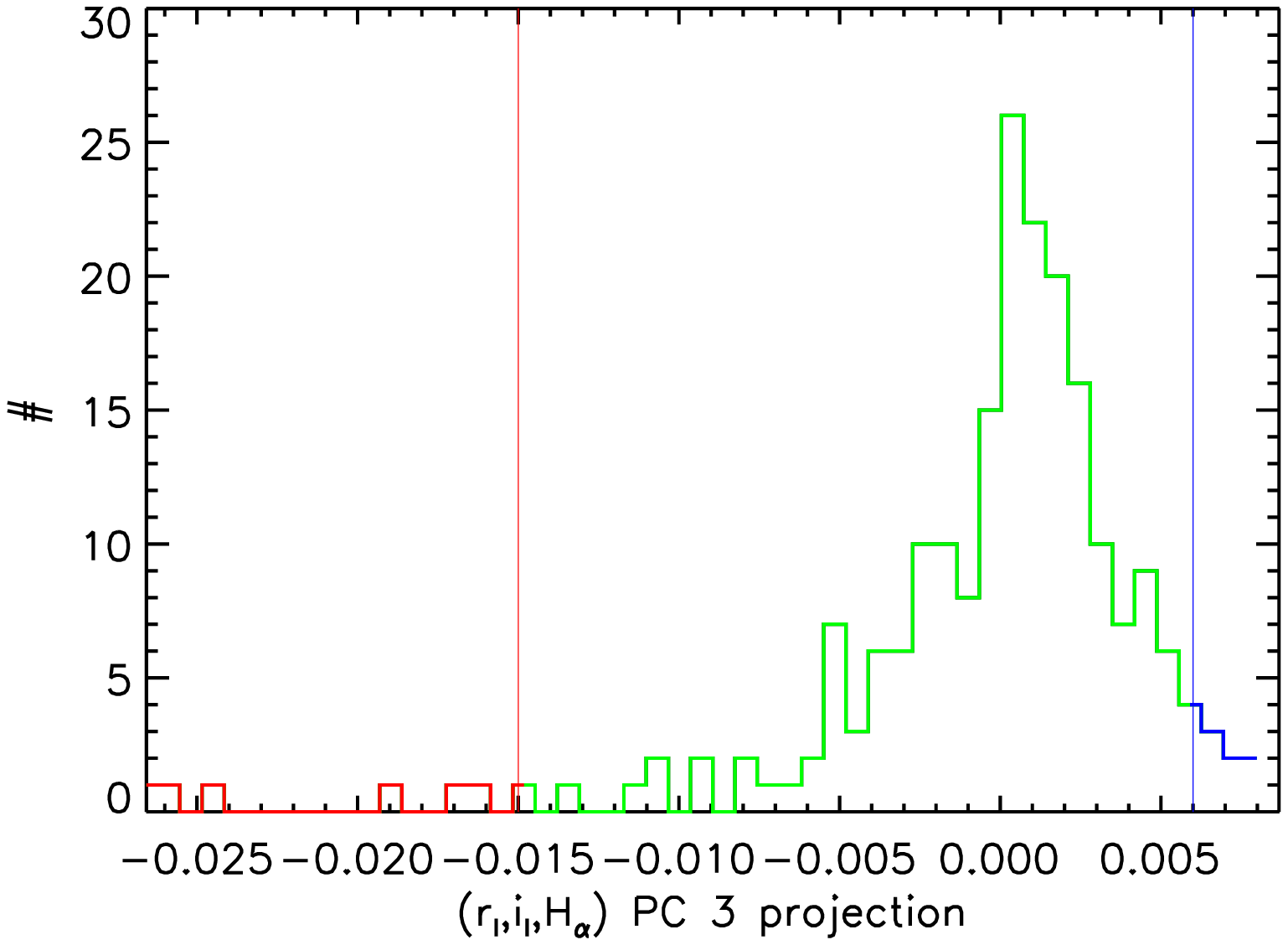}
\includegraphics[width=2.5in]{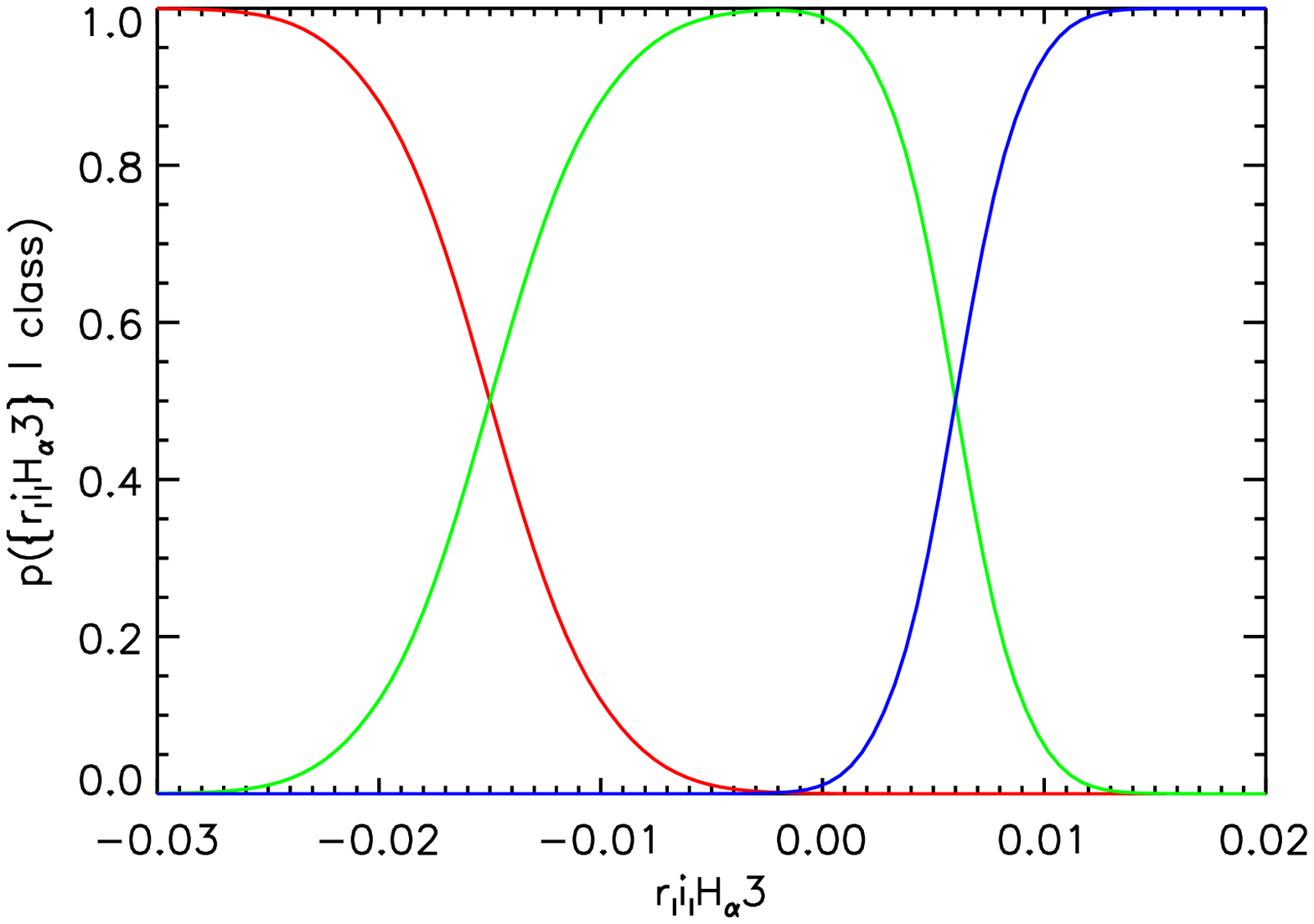}
\includegraphics[width=2.5in]{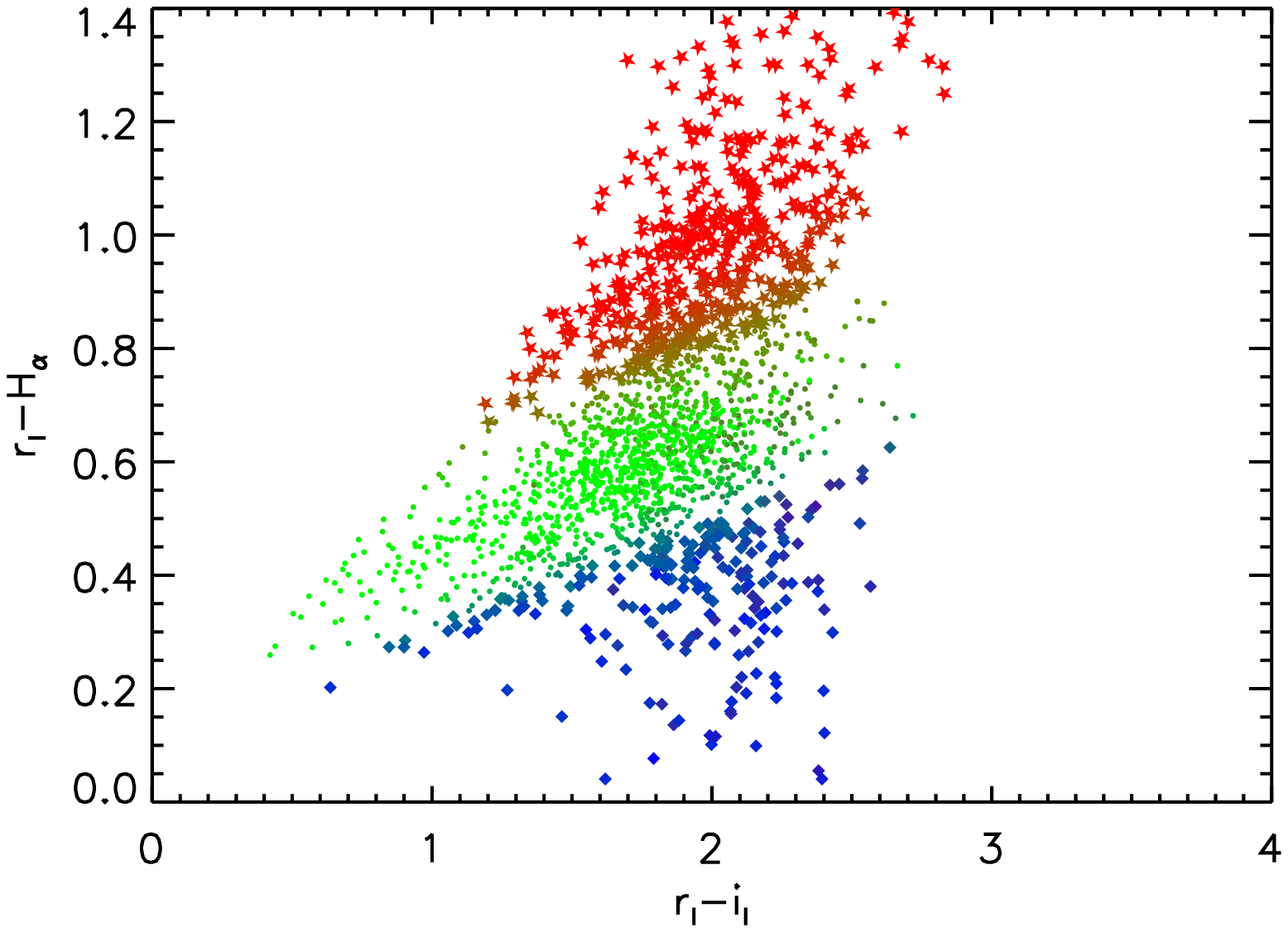}
\caption{
As Figure ~\ref{f:riz2pc}, for the $3^{rd}$ Principal Component of $\{\rp,\ip,\Ha\}$.
}
\label{f:riHa3pc}
\vskip 0.5in
\end{figure*}

\begin{figure*}[htb!]
\includegraphics[width=2.5in]{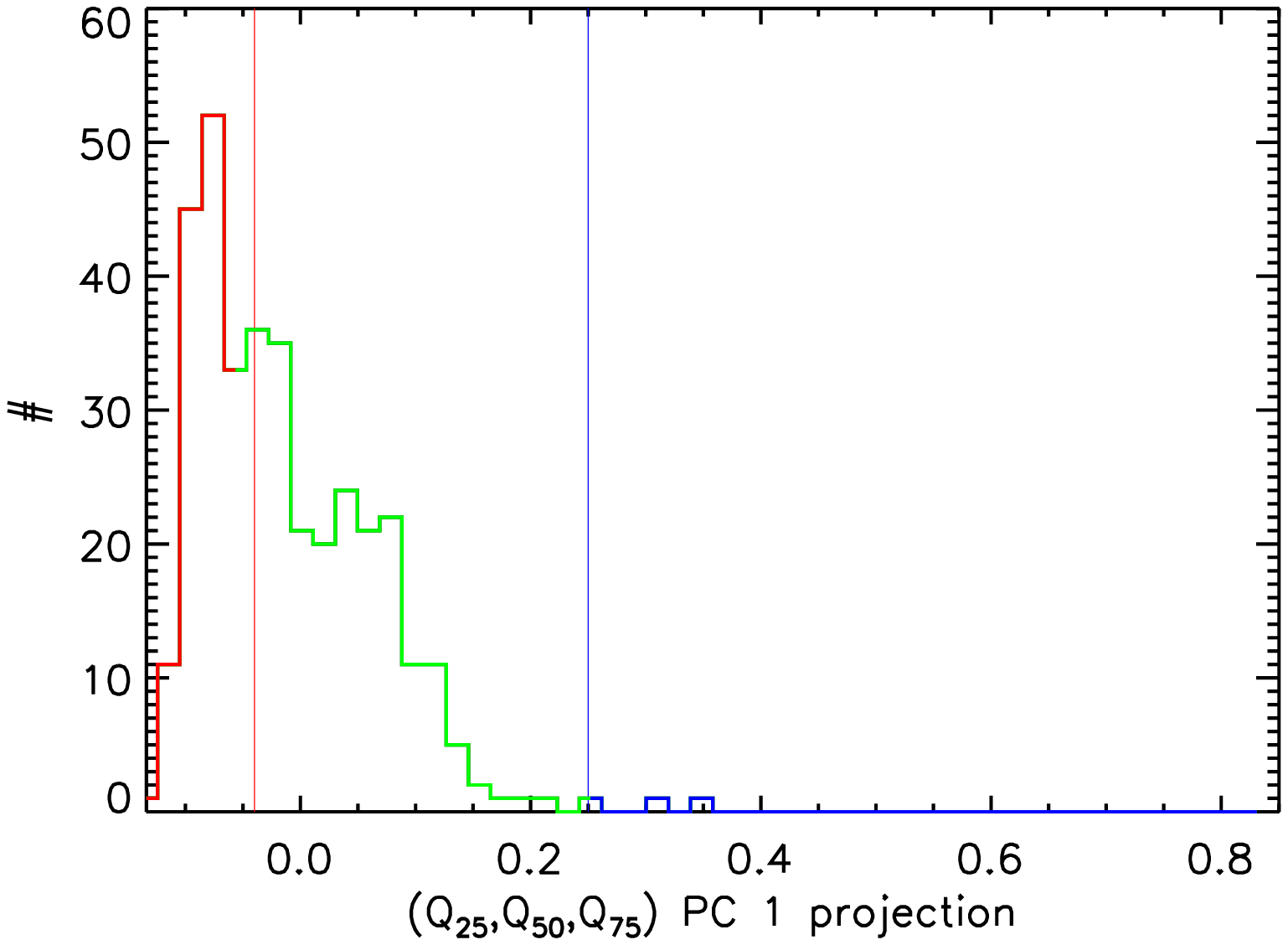}
\includegraphics[width=2.5in]{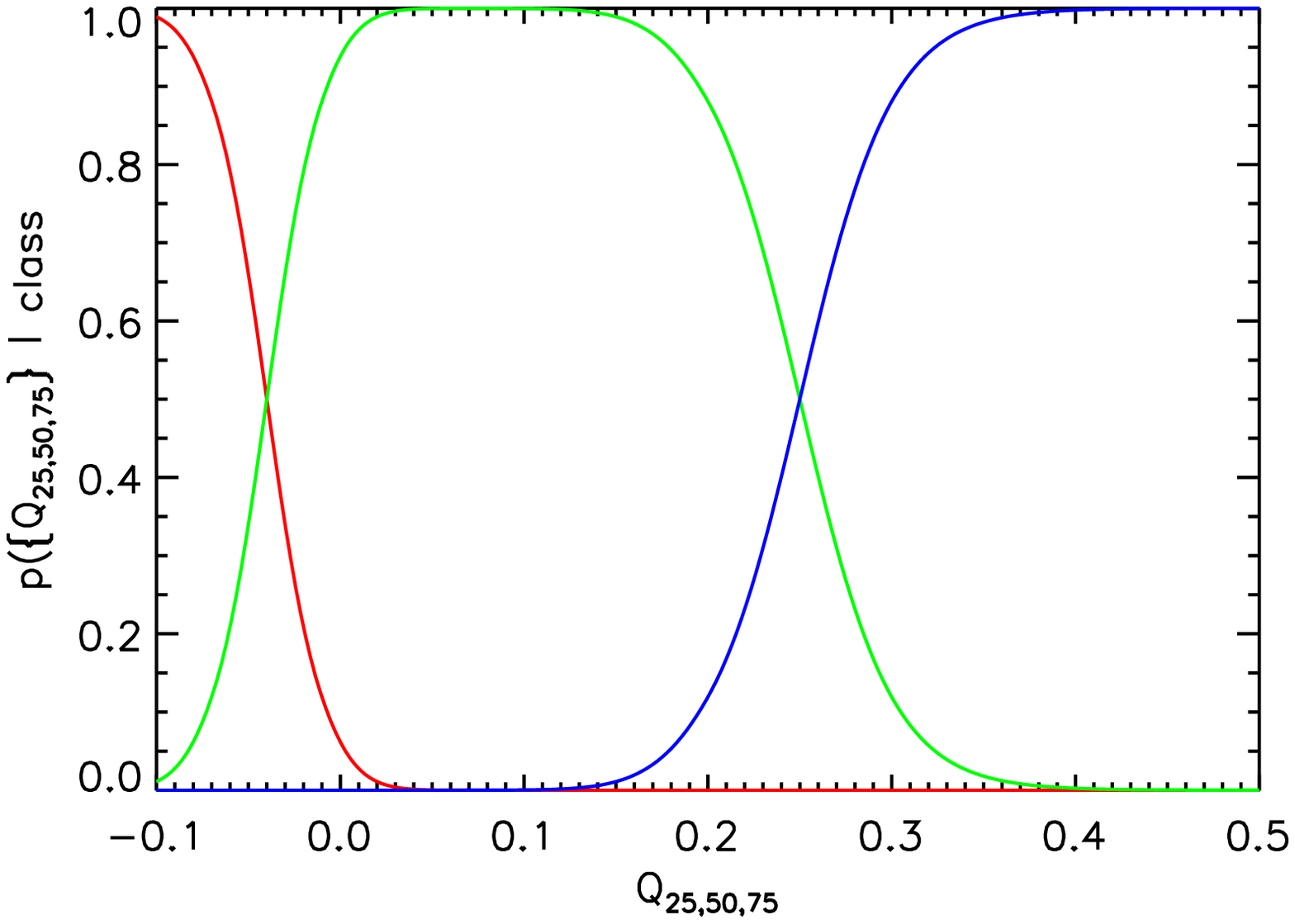}
\includegraphics[width=2.5in]{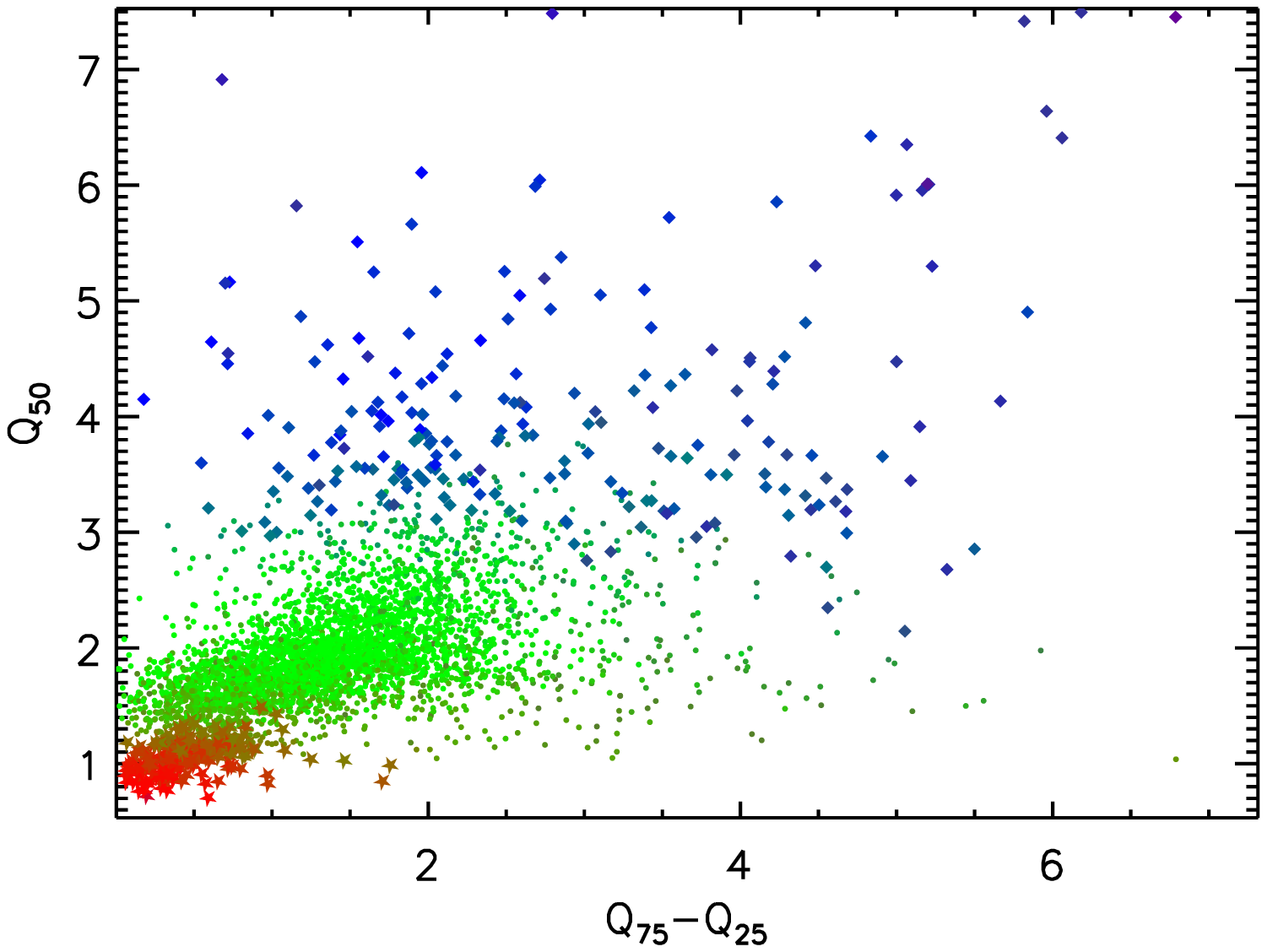}
\caption{
As Figure~\ref{f:riz2pc}, for the $1^{st}$ Principal Component of $\{\qs,\qm,\qh\}$.
}
\label{f:q1231pc}
\vskip 0.5in
\end{figure*}

Here we show the various combinations of magnitudes that were used to define the likelihoods for classification in Section~\ref{s:likeli}.
The combinations PCA(1)$\{\jj,\kk\}$, PCA(2)$\{\jj,\kk\}$, PCA(2)$\{\hh,\jj,\kk\}$, PCA(2)$\{\rp-\ip,\rp-\Ha\}$, PCA(3)$\{\rp,\ip,\Ha\}$, and PCA(1)$\{\qs,\qm,\qh\}$ are shown in Figures~\ref{f:JK1pc}-\ref{f:q1231pc}.

\section{The Effect of Reclassification}

\begin{figure}[htb!]
\centering
\includegraphics[width=3in]{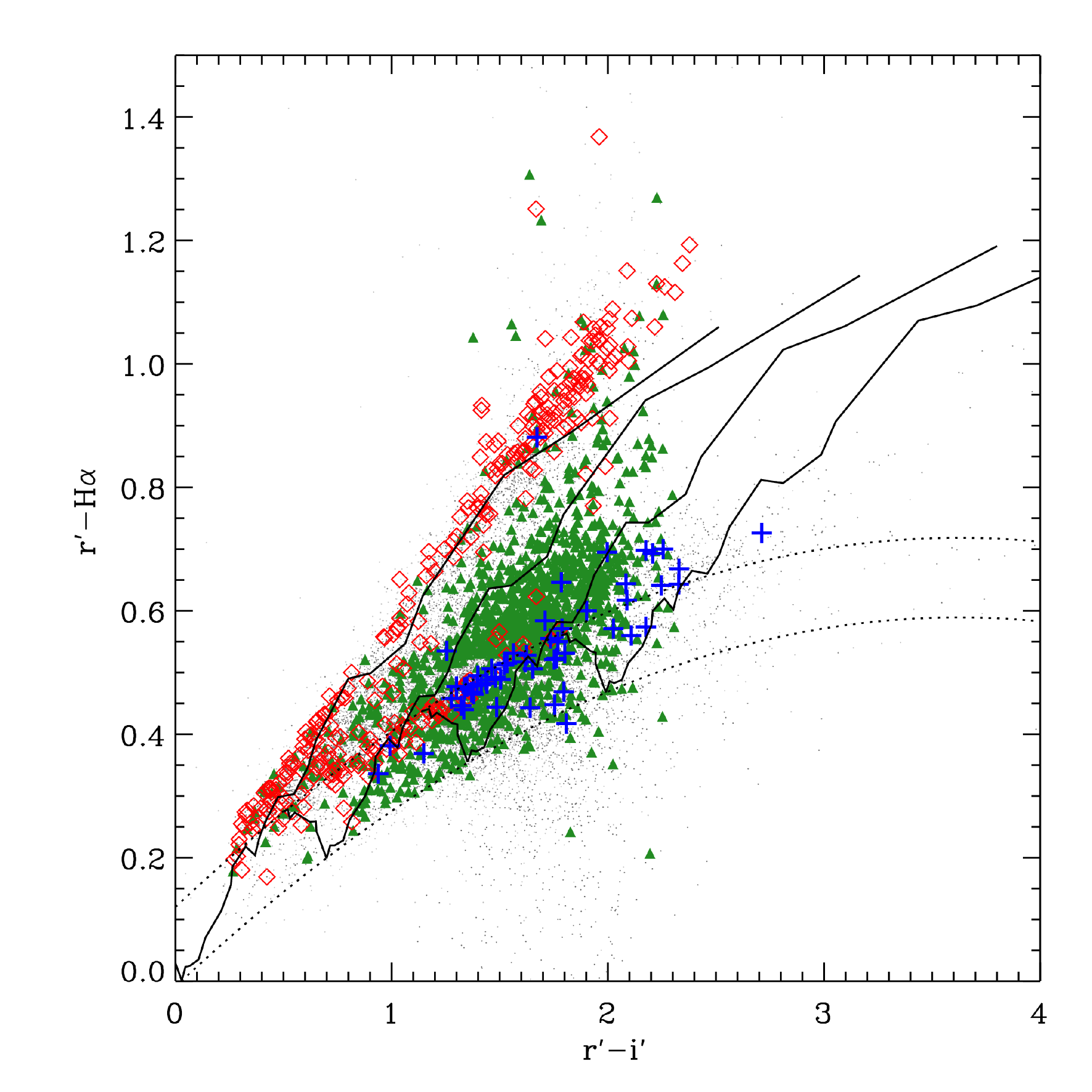}
\includegraphics[width=3in]{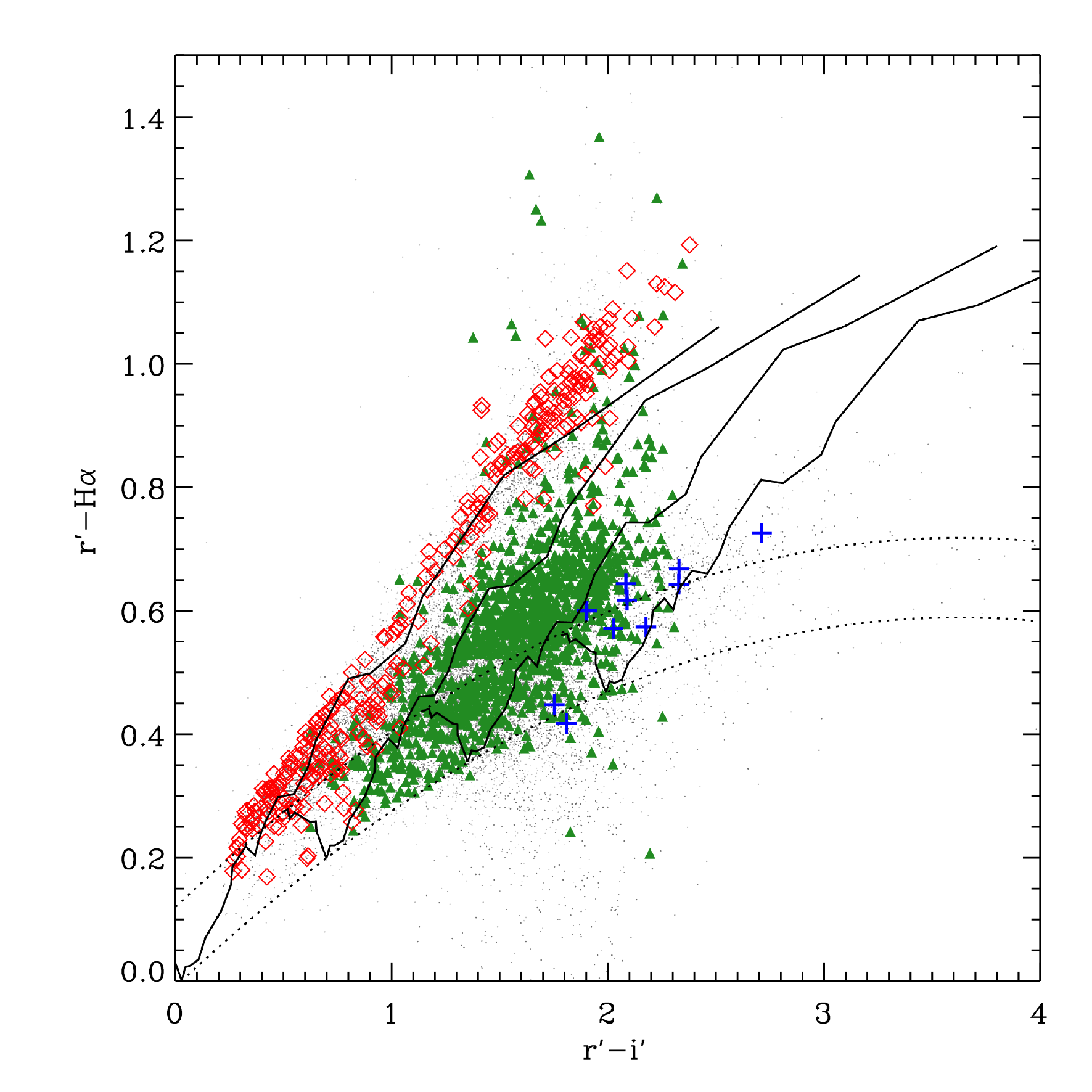}
\caption{As Figure~\ref{b:1}, for $\rp-\Ha$~vs.~$\rp-\ip$.
The black lines are ZAMS with increasing extinction: from E$_{B-V} = 1^m$ to E$_{B-V} = 4^m$ from 
\citet{Drew+05}.
The curved dotted lines mark the locus typically populated by A stars.
}
\label{b:2}
\vskip 0.5in
\end{figure}

\begin{figure}[htb!]
\centering
\includegraphics[width=3in]{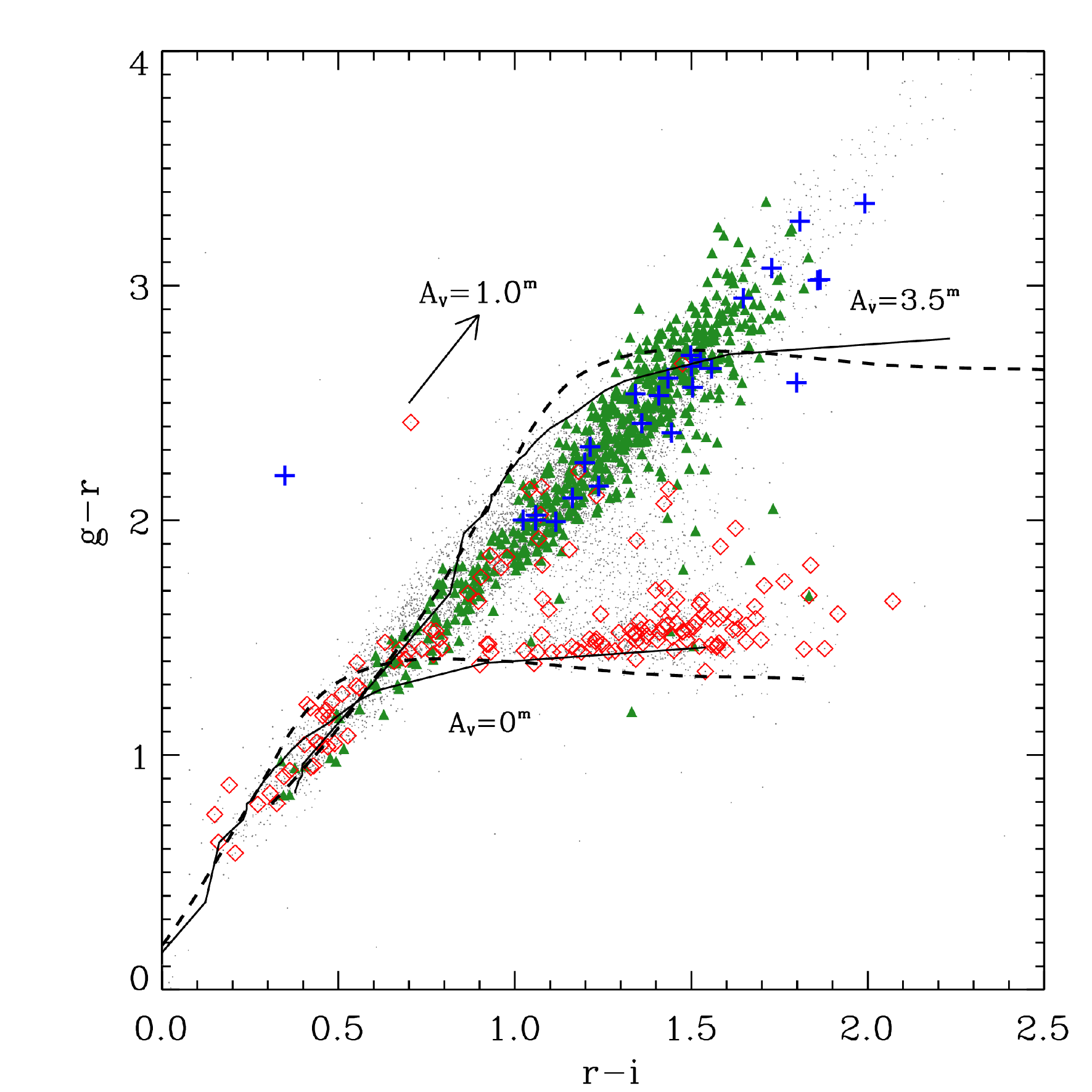}
\includegraphics[width=3in]{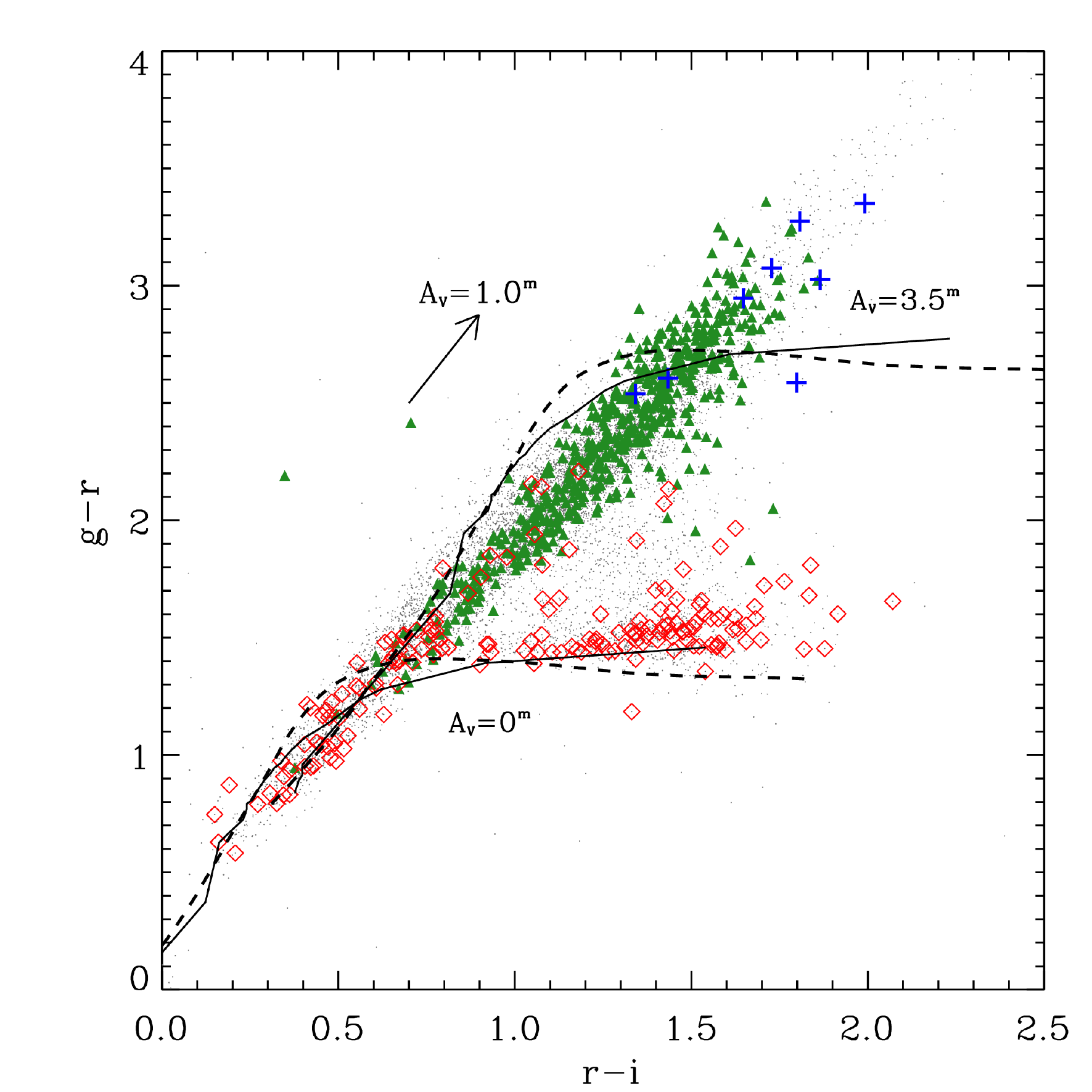}
\caption{As Figure~\ref{b:1}, for $\gp-\rp$~vs.~$\rp - \ip$ of sources with good quality SDSS photometry.
The 2.5 Myrs isochrones from 
\citet{Siess+00}
with $\Av=0$ and $\Av=3.5$ are shown as solid lines; the 2.5 Myrs MIST isochrones as dashed lines.
}
\label{b:3}
\vskip 0.5in
\end{figure}

\begin{figure}[htb!]
\centering
\includegraphics[width=3in]{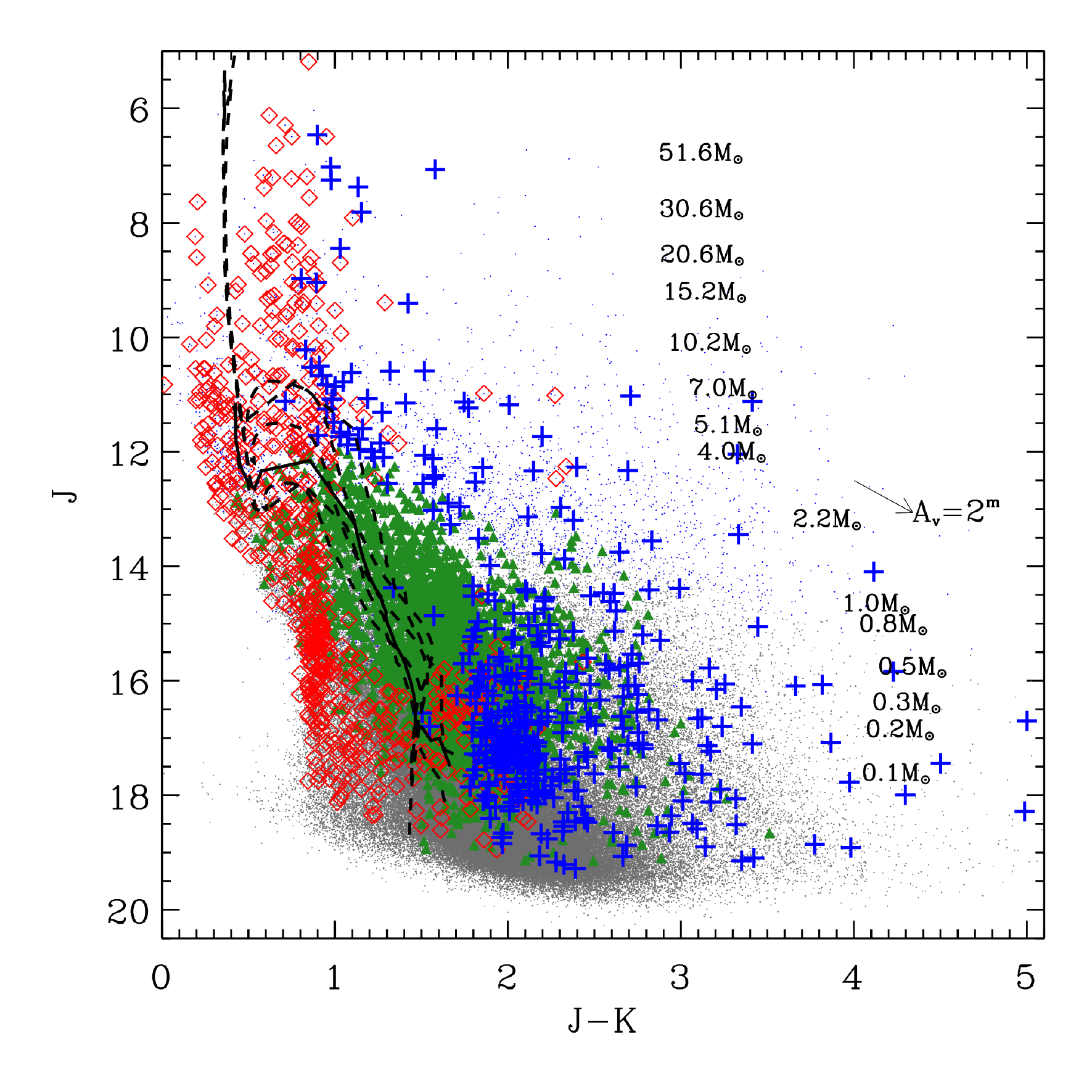}
\includegraphics[width=3in]{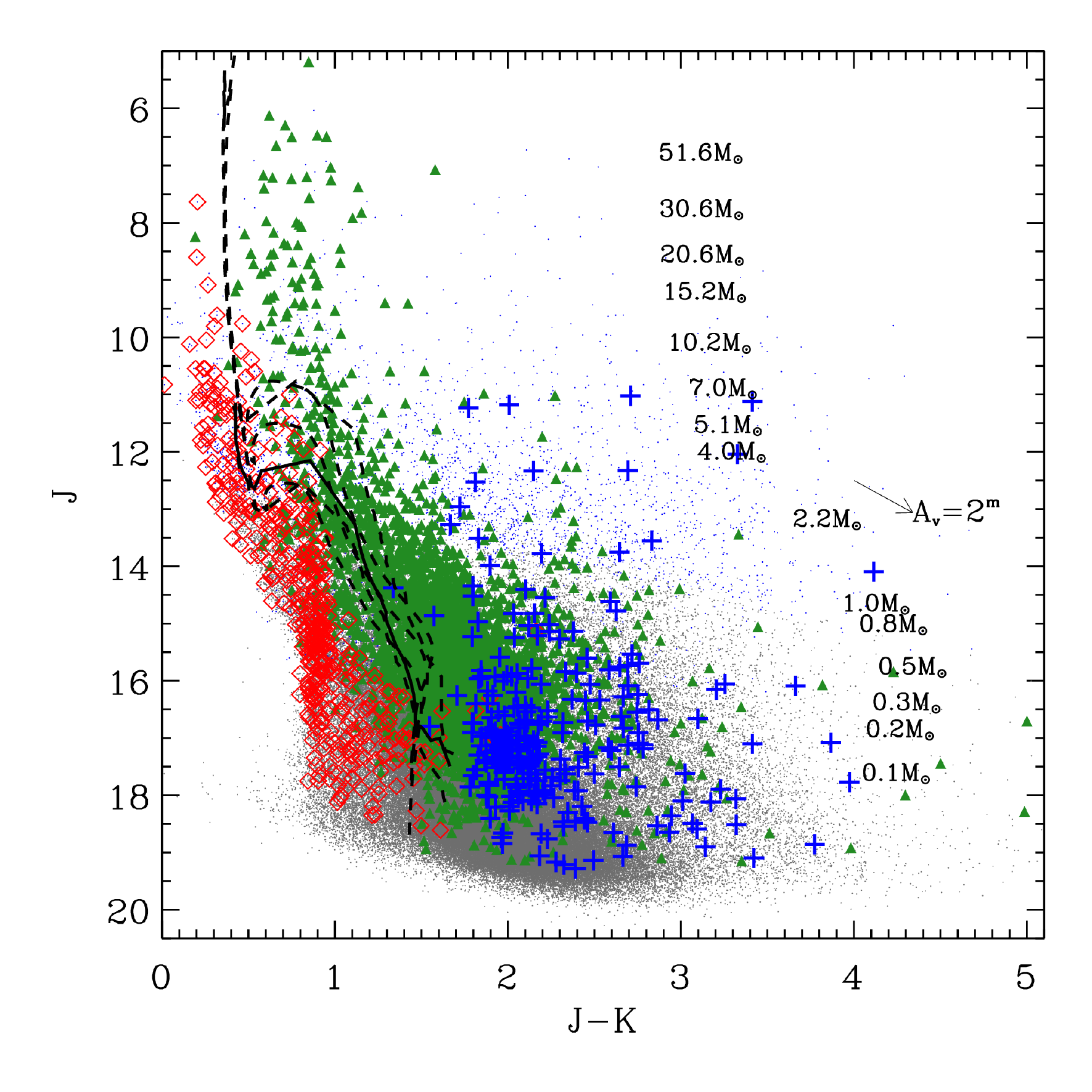}
\caption{As Figure~\ref{b:1}, for $\jj$~vs.~$(\jj-\kk)$ for sources with good 2MASS or UKIDSS photometry.
The solid line is the 2.5 Myrs isochrone with $\Av=3.5$ from 
\citet{Siess+00};
while the dashed line marks the MIST isochrone.
}
\label{b:4}
\vskip 0.5in
\end{figure}

\begin{figure}[htb!]
\centering
\includegraphics[width=3in]{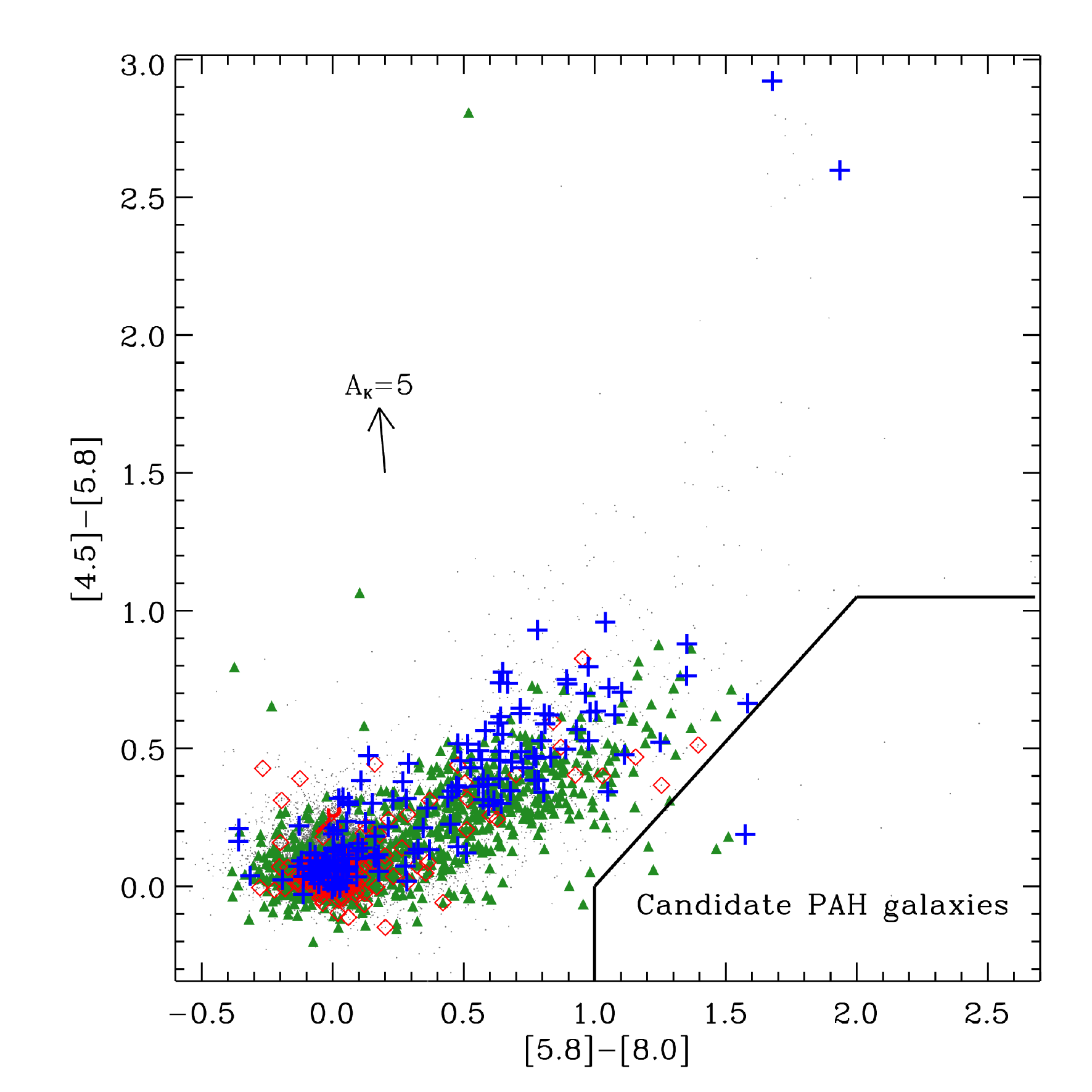}
\includegraphics[width=3in]{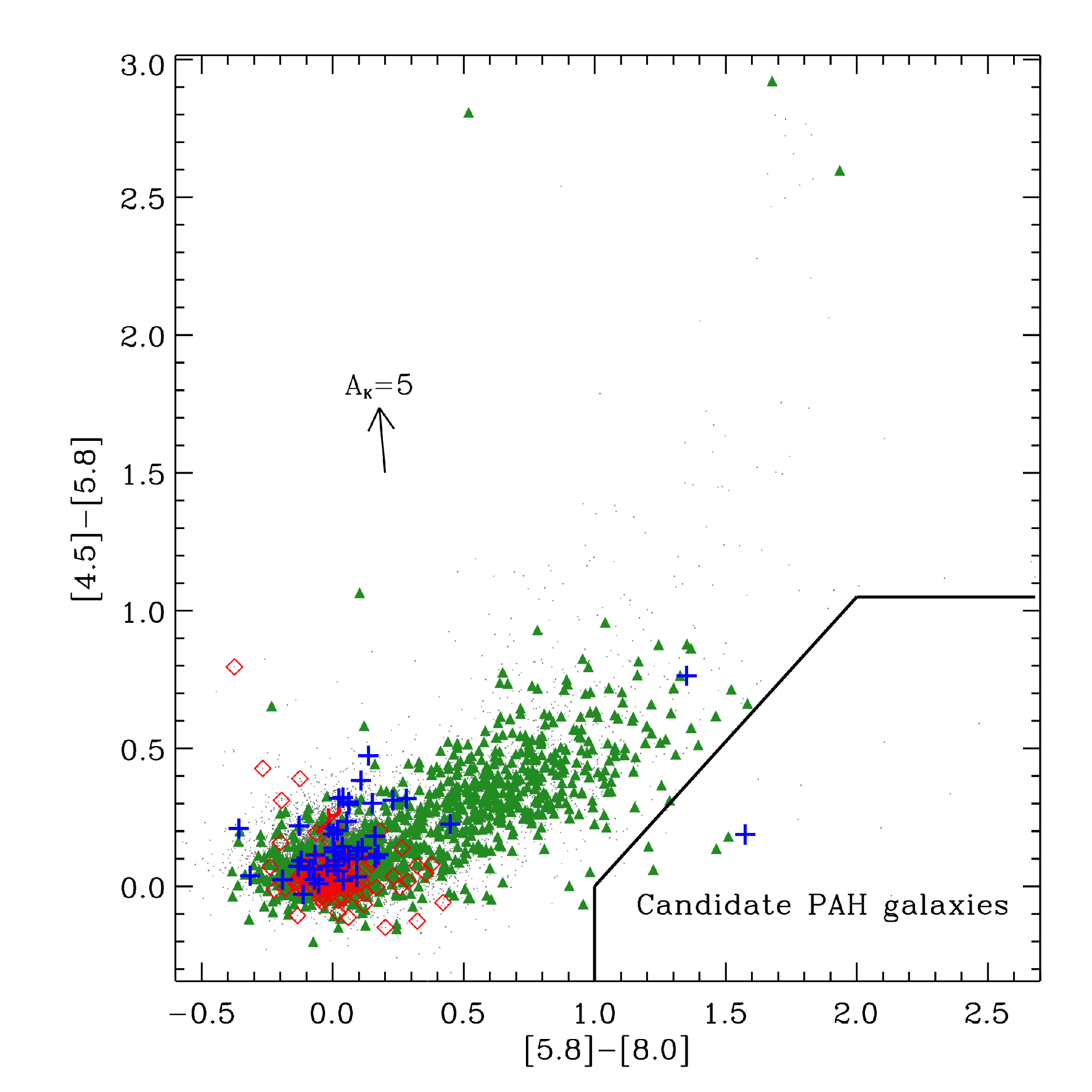}
\caption{As in Figure~\ref{b:1}, for Spitzer $[4.5]-[5.8]$~vs.~$[5.8]-[8.0]$ colors of sources with good IRAC photometry.
The solid line delimits the locus typically populated by {\bf PAH} galaxies.
}
\label{b:5}
\vskip 0.5in
\end{figure}

\begin{figure}[htb!]
\centering
\includegraphics[width=3in]{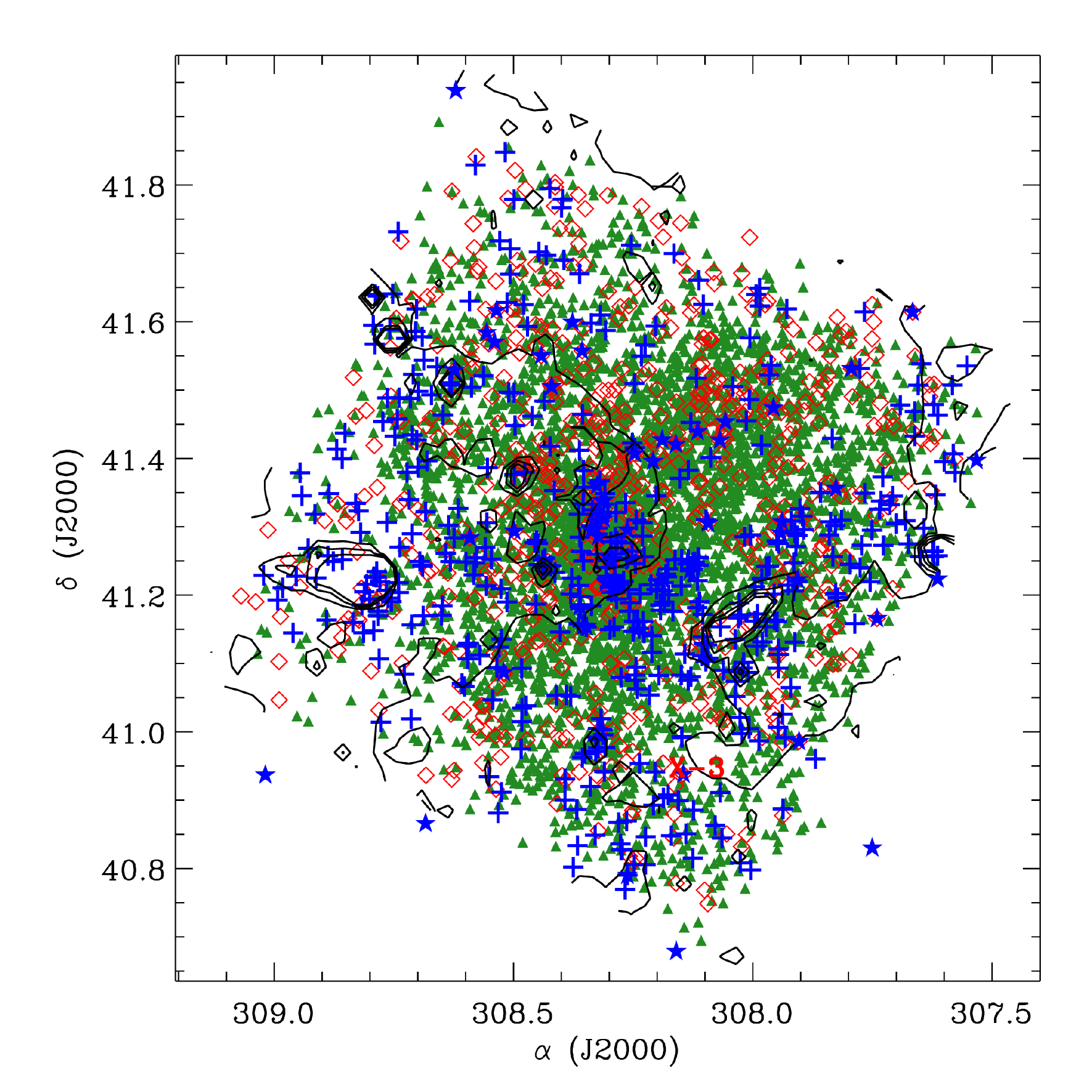}
\includegraphics[width=3in]{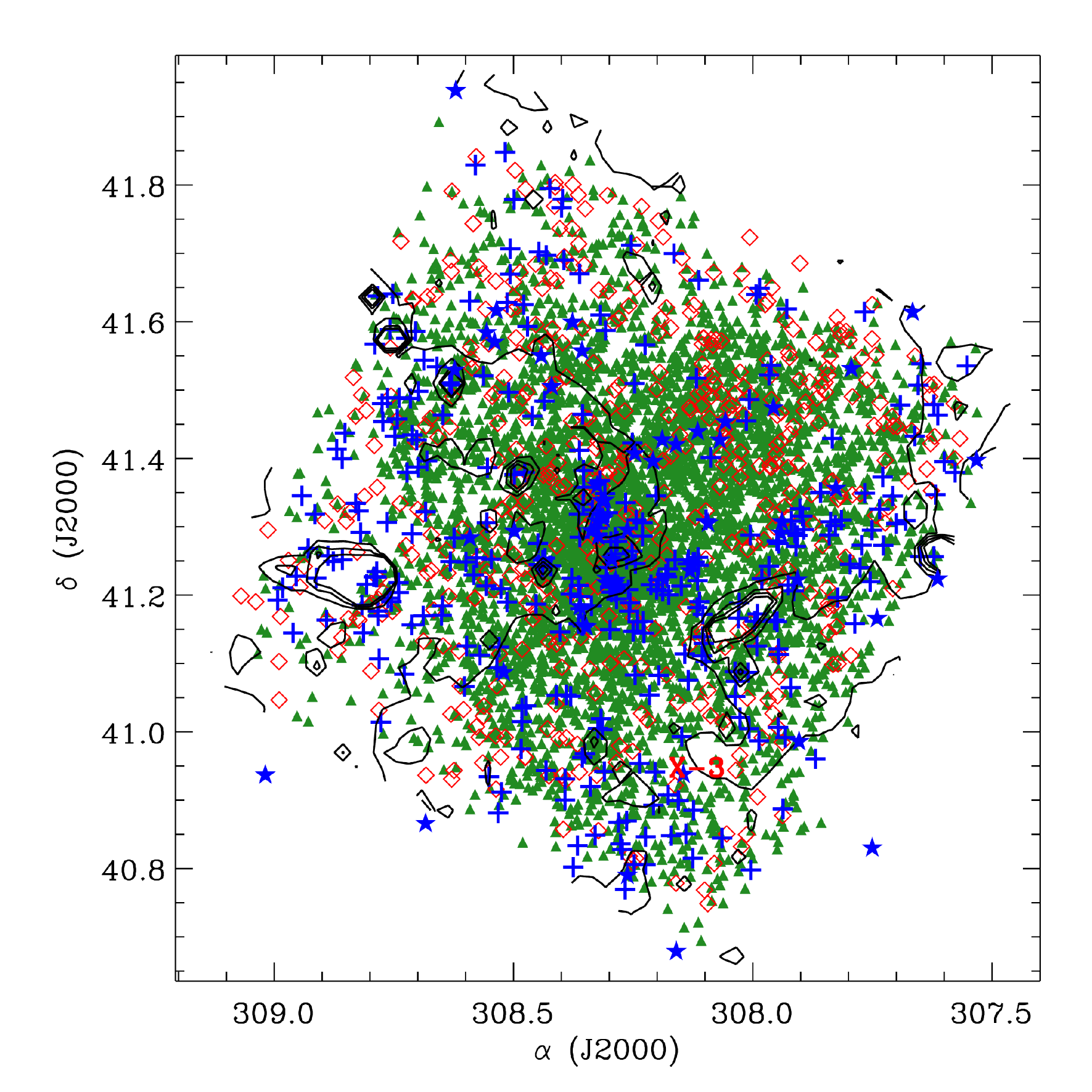}
\caption{Spatial distribution of the X-ray sources detected in our survey, and classified according to the automated NBC scheme ({\sl left}), and after manual reclassification ({\sl right}).
The black contours delimit continuum emission levels at $[8.0]\mu$m Spitzer band.
The colors and symbols are as in Figure~\ref{b:1}.
}
\label{b:6}
\vskip 0.5in
\end{figure}

Here we show various plots of the properties of the catalogued sources comparing the classes as derived from the automated Naive Bayes approach, and after manual reclassification (see Section~\ref{s:reclassify}).  In all figures, red points mark sources classified as foreground, green points mark those classified as members of the association, and blue points mark background objects.
The diagrams for $\rp-\Ha$~vs.~$\rp-\ip$, $\gp-\rp$~vs.~$\rp - \ip$, $\jj$~vs.~$\jj-\kk$, Spitzer $[4.5]-[5.8]$~vs.~$[5.8]-[8.0]$, and their spatial distributions are shown in Figures~\ref{b:2}-\ref{b:6}.

\end{document}